\documentclass[prd,aps,showpacs,epsf,floats]{revtex4}%
\usepackage{amssymb}
\usepackage{amsfonts}
\usepackage{amsmath}
\usepackage{graphicx}%
\providecommand{\U}[1]{\protect\rule{.1in}{.1in}}
\begin{document}
\title{\textbf{Scheme for constructing graphs associated with stabilizer quantum
codes}}
\author{Carlo Cafaro$^{1,2,3}$, Damian Markham$^{4}$, and Peter van Loock$^{2}$}
\affiliation{$^{1}$Max-Planck Institute for the Science of Light,
Gunther-Scharowsky-Str.1/Bau 26, 91058 Erlangen, Germany }
\affiliation{$^{2}$Institute of Physics, University of Mainz, Staudingerweg 7, 55128 Mainz, Germany}
\affiliation{$^{3}$Department of Mathematics, Clarkson University, 8 Clarkson Ave, Potsdam,
NY 13699-5815, USA}
\affiliation{$^{4}$CNRS LTCI, D\'{e}partement Informatique et R\'{e}seaux, Telecom
ParisTech, 23 avenue d'Italie, CS 51327, 75214 Paris CEDEX 13, France}

\begin{abstract}
We propose a systematic scheme for the construction of graphs associated with
binary stabilizer codes. The scheme is characterized by three main steps:
first, the stabilizer code is realized as a codeword-stabilized (CWS) quantum
code; second, the canonical form of the CWS\ code is uncovered; third, the
input vertices are attached to the graphs. To check the effectiveness of the
scheme, we discuss several graphical constructions of various useful
stabilizer codes characterized by single and multi-qubit encoding operators.
In particular, the error-correcting capabilities of such quantum codes are
verified in graph-theoretic terms as originally advocated by Schlingemann and
Werner. Finally, possible generalizations of our scheme for the graphical
construction of both (stabilizer and nonadditive) nonbinary and
continuous-variable quantum codes are briefly addressed.

\end{abstract}

\maketitle

\section{Introduction}

Classical graphs \cite{die, west, wilson} are closely related to quantum error
correcting codes (QECCs) \cite{gotty}. The first construction of QECCs based
upon the use of graphs and finite Abelian groups appears in \cite{werner} and
is provided by Schlingemann and Werner (SW-work). However, while in
\cite{werner} it is proved that all codes constructed from graphs are
stabilizer codes, it remains unclear how to embed the usual stabilizer code
constructions into the proposed graphical scheme. Therefore, although
necessary and sufficient conditions are uncovered for the graph such that the
resulting code corrects a certain number of errors, the power of the graphical
approach to quantum coding for stabilizer codes cannot be fully exploited
unless this embedding issue is resolved. In \cite{dirk}, Schlingemann (S-work)
clarifies this issue by establishing that each quantum stabilizer code (both
binary and nonbinary) could be realized as a graph code and vice-versa. Almost
at the same time, inspired by the work presented in \cite{werner}, the
equivalence of graphical quantum codes and stabilizer codes is also
established by Grassl et \textit{al}. in \cite{markus}. Despite being very
important, the works in \cite{dirk} and \cite{markus} still suffer from the
fact that no systematic scheme for constructing a graph of a stabilizer code
or the stabilizer of a graphical quantum code is available. The solution of
this point is especially important in view of the fact that although \ any
stabilizer code over a finite field has an equivalent representation as a
graphical quantum code, unfortunately, this representation is not unique.
Furthermore, the chosen representation does not reflect all the properties of
the quantum code. A crucial step forward for the description and understanding
of the interplay between properties of graphs and stabilizer codes is achieved
thanks to the introduction of the notion of graph states (and cluster states,
\cite{hans}) into the graphical construction of QECCs as presented by Hein et
\textit{al}. in \cite{hein}. In this last work, it is shown how graph states
are in correspondence to graphs and special focus is devoted to the question
of how the entanglement in a graph state is related to the topology of its
underlying graph. In \cite{hein}, it is also pointed out that codewords of
various QECCs could be regarded as special instances of graph states and
criteria for the equivalence of graph states under local unitary
transformations entirely on the level of the underlying graphs are presented.
Similar findings are uncovered by Van den Nest et \textit{al}. in \cite{bart}
(VdN-work) where a constructive scheme showing that each stabilizer state is
equivalent to a graph state under local Clifford operations is discussed.
Thus, the main finding of Schlingemann in \cite{dirk} is re-obtained in
\cite{bart} for the special case of binary quantum states. Most importantly,
in \cite{bart}, an algorithmic procedure for transforming any binary quantum
stabilizer code into a graph code appears. However, to the best of our
knowledge, nobody has fully and jointly exploited the results provided by
either Schlingemann in \cite{dirk} or Van den Nest et \textit{al}. in
\cite{bart} to provide a more systematic procedure for constructing graphs
associated with arbitrary binary stabilizer codes with special emphasis on the
verification of their error-correcting capabilities. We emphasize that this
last point constitutes one of the original motivations for introducing the
concept of a graph into quantum error correction (QEC) \cite{werner}.

The CWS quantum code formalism presents a unifying approach for constructing
both additive and nonadditive QECCs, for both binary \cite{cross} (CWS-work)
and nonbinary states \cite{chen}. Furthermore, every CWS code in its canonical
form can be fully characterized by a graph and a classical code. In
particular, any CWS code is locally Clifford equivalent to a CWS code with a
graph state stabilizer and word operators consistent only of $Z$s
\cite{cross}. Since the notion of stabilizer codes, graph codes and graph
states can be recast into the CWS formalism, it seems natural to investigate
the graphical depiction of stabilizer codes as originally thought by
Schlingemann and Werner within this generalized framework where stabilizer
codes are realized as CWS\ codes. Proceeding along this line of investigation,
we shall observe that the notion of graph state in QEC as presented in
\cite{hein} emerges naturally. Furthermore, the algorithmic procedure for
transforming any (binary) quantum stabilizer code into a graph code advocated
in \cite{bart} can be exploited and jointly used with the results in
\cite{dirk} where the notions of both coincidence and adjacency matrices of a
classical graphs are introduced. For the sake of completeness, we point out
that the CWS formalism has been already employed into the literature for the
graphical construction of both binary \cite{yu1} and nonbinary \cite{yu2}
(both additive/stabilizer and nonadditive) QECCs. For instance in \cite{yu1},
regarding stabilizer codes as CWS codes and employing a graphical approach to
quantum coding, a classification of all the extremal stabilizer codes up to
eight qubits and the construction of the optimal $\left(  \left(  10\text{,
}24\text{, }3\right)  \right)  $ code together with a family of $1$-error
detecting nonadditive codes with the highest encoding rate so far is
presented. With a leap of imagination, in \cite{yu1} it is also envisioned a
graphical quantum computation based directly on graphical objects. Indeed,
this vision became recently more realistic in the work of Beigi et
\textit{al}. \cite{beigi}. Here, being essentially within the CWS framework, a
systematic method for constructing both binary and nonbinary concatenated
quantum codes based on graph concatenation is developed. Graphs representing
the inner and the outer codes are concatenated via a simple graph operation
(the so-called generalized local complementation, \cite{beigi}). Despite their
very illuminating findings, in \cite{beigi} it is emphasized that the elusive
role played by graphs in QEC is still not well-understood. In neither
\cite{yu1} nor \cite{beigi}, the Authors are concerned with the joint
exploitation of the results provided by either Van den Nest et \textit{al.} in
\cite{bart} (algorithmic procedure for transforming any binary quantum
stabilizer code into a graph code) or Schlingemann in \cite{dirk} (use of both
the coincidence and adjacency matrices of a classical graphs in QEC) in order
to provide a more systematic procedure for constructing graphs associated with
arbitrary binary stabilizer codes with special emphasis on the verification of
their error correcting capabilities which, as pointed out earlier, constituted
a major driving motivation for the introduction of graphs in QEC
\cite{werner}. Instead, we aim here at investigating such unexplored topics
and hope to further advance our understanding of the role played by classical
graphs in quantum coding.

In this article, we propose a systematic scheme for the construction of graphs
with both input and output vertices associated with arbitrary binary
stabilizer codes. The scheme is characterized by three main steps: first, the
stabilizer code is realized as a CWS quantum code; second, the canonical form
of the CWS\ code is uncovered; third, the input vertices are attached to the
graphs with only output vertices. To check the effectiveness of the scheme, we
discuss several graphical constructions of various useful stabilizer codes
characterized by single and multi-qubit encoding operators. In particular, the
error-correcting capabilities of such quantum codes are verified in
graph-theoretic terms as originally advocated by Schlingemann and Werner.
Finally, possible generalizations of our scheme for the graphical construction
of both (stabilizer and nonadditive) nonbinary and continuous variables
quantum codes is briefly addressed.

The layout of the article is as follows. In Section II, we introduce some
preliminary material. First, the notions of graphs, graph states and graph
codes are presented. Second, local Clifford transformations on graph states
and local complementations on graphs are briefly described. Third, the CWS
quantum codes formalism is briefly explained. In Section III, we re-examine
some basic ingredients of the Schlingemann-Werner work (SW-work,
\cite{werner}), the Schlingemann work (S-work, \cite{dirk}) and, finally, the
Van den Nest et al. work (VdN-work, \cite{bart}). We focus on those aspects of
these works that are especially important for our systematic scheme. In this
Section IV, we formally describe our scheme and, for the sake of clarity,
apply it to the graphical construction of the Leung et al. four-qubit quantum
code for the error correction of single amplitude damping errors
\cite{debbie}. Finally, concluding remarks and a brief discussion on possible
extensions of our schematic graphical construction to both (stabilizer and
nonadditive) nonbinary and continuous variables quantum codes appear in
Section V.

Several explicit constructions of graphs for various stabilizer codes
characterized by either single or multi-qubit encoding operators are worked
out in the Appendices. Specifically, we discuss the graphical construction of
the following quantum codes:\ the three-qubit repetition code, the
perfect\textbf{ }$1$-erasure correcting four-qubit code, the perfect\textbf{
}$1$-error correcting five-qubit code,\textbf{ }$1$-error correcting six-qubit
quantum degenerate codes, the CSS seven-qubit stabilizer code, the Shor
nine-qubit stabilizer code, the Gottesman $2$-error correcting eleven-qubit
code,\textbf{ }$\left[  \left[  4\text{, }2\text{, }2\right]  \right]
$\textbf{\ }stabilizer codes, and, finally, the Gottesman $\left[  \left[
8\text{, }3\text{, }3\right]  \right]  $\ stabilizer code.

\section{From graph theory to the CWS formalism}

In this section, we present some preliminary material. First, the notions of
graphs, graph states and graph codes are introduced. Second, local Clifford
transformations on graph states and local complementations on graphs are
briefly presented. Third, the CWS quantum codes formalism is briefly discussed.

\subsection{Graphs, graph states, and graph codes}

A\textbf{\ }graph $G=G\left(  V\text{, }E\right)  $ is characterized by a set
$V$ of $n$ vertices and a set of edges $E$ specified by the adjacency matrix
$\Gamma$ \cite{die, west, wilson}. This matrix is a $n\times n$ symmetric
matrix with vanishing diagonal elements and $\Gamma_{ij}=1$ if vertices $i$,
$j$ are connected and $\Gamma_{ij}=0$ otherwise. The neighborhood of a vertex
$i$ is the set of all vertices $v\in V$ that are connected to $i$ and is
defined by $N_{i}\overset{\text{def}}{=}\left\{  v\in V:\Gamma_{iv}=1\right\}
$. When the vertices $a$, $b\in V$ are the end points of an edge, they are
referred to as being adjacent. An $\left\{  a\text{, }c\right\}  $ path is an
ordered list of vertices $a=a_{1}$, $a_{2}$,..., $a_{n-1}$, $a_{n}=c$, such
that for all $i$, $a_{i}$ and $a_{i+1}$ are adjacent. A connected graph is a
graph that has an $\left\{  a\text{, }c\right\}  $ path for any two $a$, $c\in
V$. Otherwise it is referred to as disconnected. A vertex represents a
physical system, e.g., a qubit (two-dimensional Hilbert space), qudit
($d$-dimensional Hilbert space), or continuous variables (CV)\ (continuous
Hilbert space). An edge between two vertices represents the physical
interaction between the corresponding systems. In what follows, we shall take
into consideration simple graphs only. These are graphs that contain neither
loops (edges connecting vertices with itself) nor multiple edges. Furthermore,
for the time being, we do not make a distinction between different types of
vertices. However, later on we will assign some vertices as inputs, and some
as outputs.

Graph states \cite{hans} are multipartite entangled states that play a
key-role in graphical constructions of QECCs codes and, in addition, are very
important in quantum secret sharing \cite{damian2008} which is, to a certain
extent, equivalent to error correction \cite{anne2013}. For a very recent
experimental demonstration of a graph state quantum error correcting code, we
refer to \cite{damian2014}.

Consider a system of $n$ qubits that are labeled by those $n$ vertices in $V$
and denote by $I^{i}$, $X^{i}$, $Y^{i}$, $Z^{i}$ (or, equivalently,
$X^{i}\equiv\sigma_{x}^{i}$, $Y^{i}\equiv\sigma_{y}^{i}$, $Z^{i}\equiv
\sigma_{z}^{i}$) the identity matrix and the three Pauli operators acting on
the qubit $i\in V$. The $n$-qubit graph state $\left\vert G\right\rangle $
associated with the graph $G$ is defined by \cite{hein},%
\begin{equation}
\left\vert G\right\rangle \overset{\text{def}}{=}%
{\displaystyle\prod\limits_{\Gamma_{ij}=1}}
\mathcal{U}_{ij}\left\vert +\right\rangle _{x}^{V}=\frac{1}{\sqrt{2^{n}}}%
{\displaystyle\sum\limits_{\vec{\mu}=\mathbf{0}}^{\mathbf{1}}}
\left(  -1\right)  ^{\frac{1}{2}\vec{\mu}\cdot\Gamma\cdot\vec{\mu}}\left\vert
\vec{\mu}\right\rangle _{z}\text{,}%
\end{equation}
where $\left\vert +\right\rangle _{x}^{V}$ is the joint $+1$ eigenstate of
$X^{i}$ with $i\in V$, $\mathcal{U}_{ij}$ is the controlled phase gate between
qubits $i$ and $j$ given by,%
\begin{equation}
\mathcal{U}_{ij}\overset{\text{def}}{=}\frac{1}{2}\left[  I+Z_{i}+Z_{j}%
-Z_{i}Z_{j}\right]  \text{,}%
\end{equation}
and $\left\vert \vec{\mu}\right\rangle _{z}$ is the joint eigenstate of
$Z^{i}$ with $i\in V$ and $\left(  -1\right)  ^{\mu_{i}}$ as eigenvalues. The
graph-state basis of the $n$-qubit Hilbert space $\mathcal{H}_{2}^{n}$ is
given by $\left\{  \left\vert G^{C}\right\rangle \overset{\text{def}}{=}%
Z^{C}\left\vert G\right\rangle \right\}  $ where $C$ is an element of the set
of all the subsets of $V$ denoted by $2^{V}$. A collection of subsets
$\left\{  C_{1}\text{,..., }C_{K}\right\}  $ specifies a $K$-dimensional
subspace of $\mathcal{H}_{2}^{n}$ that is spanned by the graph-state basis
$\left\{  \left\vert G^{C_{i}}\right\rangle \right\}  $ with $i=1$,..., $K$.
The graph state $\left\vert G\right\rangle $ is the unique joint $+1$
eigenstate of the $n$-vertex stabilizers $\mathcal{G}_{i}$ with $i\in V$
defined as \cite{hein},%
\begin{equation}
\mathcal{G}_{i}\overset{\text{def}}{=}X^{i}Z^{N_{i}}\overset{\text{def}}%
{=}X^{i}%
{\displaystyle\prod\limits_{j\in N_{i}}}
Z^{j}\text{.}%
\end{equation}
A graph code, first introduced into the realm of QEC in \cite{werner} and
later reformulated into the graph state formalism in \cite{hein}, is defined
to be one in which a graph $G$ is given and the codespace (or, coding space)
is spanned by a subset of the graph state basis. These states are regarded as
codewords, although we recall that what is significant from the point of view
of the QEC properties is the subspace they span, not the codewords themselves
\cite{robert}.

\subsection{Local Clifford transformations and local complementations}

\subsubsection{Transformations on quantum states\textbf{\ }}

The Clifford group $\mathcal{C}_{n}$ is the normalizer of the Pauli group
$\mathcal{P}_{\mathcal{H}_{2}^{n}}$ in $\mathcal{U}\left(  2^{n}\right)  $,
i.e., it is the group of unitary operators $U$ satisfying $U\mathcal{P}%
_{\mathcal{H}_{2}^{n}}U^{\dagger}=\mathcal{P}_{\mathcal{H}_{2}^{n}}$. The
local Clifford group $\mathcal{C}_{n}^{l}$ is the subgroup of $\mathcal{C}%
_{n}$ and consists of all $n$-fold tensor products of elements in
$\mathcal{C}_{1}$. The Clifford group is generated by a simple set of quantum
gates: the Hadamard gate $H$, the phase gate $P$ and the CNOT gate
$U_{\text{CNOT}}$ \cite{gaitan}. Using the well-known representations of the
Pauli matrices in the computational basis, it is straightforward to show that
the action of $H$ on such matrices reads%
\begin{equation}
\sigma_{x}\rightarrow H\sigma_{x}H^{\dagger}=\sigma_{z}\text{, }\sigma
_{y}\rightarrow H\sigma_{y}H^{\dagger}=-\sigma_{y}\text{, }\sigma
_{z}\rightarrow H\sigma_{z}H^{\dagger}=\sigma_{x}\text{.}%
\end{equation}
The action of the phase gate $P$ on $\sigma_{x}$, $\sigma_{y}$ and $\sigma
_{z}$ is given by,%
\begin{equation}
\sigma_{x}\rightarrow P\sigma_{x}^{\dagger}P=\sigma_{y}\text{, }\sigma
_{y}\rightarrow P\sigma_{y}^{\dagger}P=-\sigma_{x}\text{, }\sigma
_{z}\rightarrow P\sigma_{z}^{\dagger}P=\sigma_{z}\text{.}%
\end{equation}
Finally, the CNOT\ gate leads to the following transformations rules,%
\begin{align}
\sigma_{x}\otimes I  &  \rightarrow U_{\text{CNOT}}\left(  \sigma_{x}\otimes
I\right)  U_{\text{CNOT}}^{\dagger}=\sigma_{x}\otimes\sigma_{x}\text{,
}I\otimes\sigma_{x}\rightarrow U_{\text{CNOT}}\left(  I\otimes\sigma
_{x}\right)  U_{\text{CNOT}}^{\dagger}=I\otimes\sigma_{x}\text{,}\nonumber\\
& \nonumber\\
\sigma_{z}\otimes I  &  \rightarrow U_{\text{CNOT}}\left(  \sigma_{z}\otimes
I\right)  U_{\text{CNOT}}^{\dagger}=\sigma_{z}\otimes I\text{, }I\otimes
\sigma_{z}\rightarrow U_{\text{CNOT}}\left(  I\otimes\sigma_{z}\right)
U_{\text{CNOT}}^{\dagger}=\sigma_{z}\otimes\sigma_{z}\text{.}%
\end{align}
Observe that the CNOT gate propagates bit flip errors from the control to the
target, and phase errors from the target to the control. As a side remark, we
stress that another useful two-qubit gate is the controlled-phase gate
$U_{\text{CP}}\overset{\text{def}}{=}\left(  I\otimes H\right)  U_{\text{CNOT}%
}\left(  I\otimes H\right)  $. The controlled-phase gate has the following
action on the generators of $\mathcal{P}_{\mathcal{H}_{2}^{2}}$,%
\begin{align}
\sigma_{x}\otimes I  &  \rightarrow U_{\text{CP}}\left(  \sigma_{x}\otimes
I\right)  U_{\text{CP}}^{\dagger}=\sigma_{x}\otimes\sigma_{z}\text{, }%
I\otimes\sigma_{x}\rightarrow U_{\text{CP}}\left(  I\otimes\sigma_{x}\right)
U_{\text{CP}}^{\dagger}=\sigma_{z}\otimes\sigma_{x}\text{,}\nonumber\\
& \nonumber\\
\sigma_{z}\otimes I  &  \rightarrow U_{\text{CP}}\left(  \sigma_{z}\otimes
I\right)  U_{\text{CP}}^{\dagger}=\sigma_{z}\otimes I\text{, }I\otimes
\sigma_{z}\rightarrow U_{\text{CP}}\left(  I\otimes\sigma_{z}\right)
U_{\text{CP}}^{\dagger}=I\otimes\sigma_{z}\text{.}%
\end{align}
We observe that a controlled-phase gate does not propagate phase errors,
though a bit-flip error on one qubit spreads to a phase error on the other qubit.

We also point out that a unitary operator $U$ that fixes the stabilizer group
$S_{\text{stabilizer}}$ \ (we refer to \cite{daniel-phd} for a detailed
characterization of the quantum stabilizer formalism in QEC) of a quantum
stabilizer code $\mathcal{C}_{\text{stabilizer}}$ under conjugation is an
encoded operation. In other words, $U$ is an encoded operation that maps
codewords to codewords whenever $US_{\text{stabilizer}}U^{\dagger
}=S_{\text{stabilizer}}$. In particular, if $S^{\prime}\overset{\text{def}}%
{=}USU^{\dagger}$ (every element of $S^{\prime}$ can be written as
$UsU^{\dagger}$ for some $s\in S$) and $\left\vert c\right\rangle $ is a
codeword stabilized by every element in $S$, then $\left\vert c^{\prime
}\right\rangle =U\left\vert c\right\rangle $ is stabilized by every stabilizer
element in $S^{\prime}$.

\subsubsection{Transformations on graphs}

If there exists a local unitary (LU) transformation $U$ such that $U\left\vert
G\right\rangle =\left\vert G^{\prime}\right\rangle $, the states $\left\vert
G\right\rangle $ and $\left\vert G^{\prime}\right\rangle $ will have the same
entanglement properties. If $\left\vert G\right\rangle $ and $\left\vert
G^{\prime}\right\rangle $ are graph states, we say that their corresponding
graphs $G$ and $G^{\prime}$ will then represent equivalent quantum codes, with
the same distance, weight distribution, and other properties. Determining
whether two graphs are LU-equivalent is a difficult task, but a sufficient
condition for equivalence was given in \cite{hein}. Let the graphs $G=\left(
V\text{, }E\right)  $ and $G^{\prime}=\left(  V\text{, }E^{\prime}\right)  $
on $n$ vertices correspond to the $n$-qubit graph states $\left\vert
G\right\rangle $ and $\left\vert G^{\prime}\right\rangle $. We define the two
$2\times2$ unitary matrices,%
\begin{equation}
\tau_{x}\overset{\text{def}}{=}\sqrt{-i\sigma_{x}}=\frac{1}{\sqrt{2}}\left(
\begin{array}
[c]{cc}%
-1 & i\\
i & -1
\end{array}
\right)  \text{ and, }\tau_{z}\overset{\text{def}}{=}\sqrt{i\sigma_{z}%
}=\left(
\begin{array}
[c]{cc}%
\omega & 0\\
0 & \omega^{3}%
\end{array}
\right)  \text{,}%
\end{equation}
where $\omega^{4}=i^{2}=-1$, and $\sigma_{x}$ and $\sigma_{z}$ are Pauli
matrices. Given a graph $G=\left(  V=\left\{  0\text{,..., }n-1\right\}
\text{, }E\right)  $, corresponding to the graph state $\left\vert
G\right\rangle $, we define a local unitary transformation $U_{a}$,%
\begin{equation}
U_{a}\overset{\text{def}}{=}%
{\displaystyle\bigotimes\limits_{i\in N_{a}}}
\tau_{x}^{\left(  i\right)  }%
{\displaystyle\bigotimes\limits_{i\notin N_{a}}}
\tau_{z}^{\left(  i\right)  }\text{,}%
\end{equation}
where $a\in V$ is any vertex, $N_{a}\subset V$ is the neighborhood of $a$, and
$\tau_{x}^{\left(  i\right)  }$ means that the transform $\tau_{x}$ should be
applied to the qubit corresponding to vertex $i$. Given a graph $G$, if there
exists a finite sequence of vertices $\left(  u_{0}\text{,..., }%
u_{k-1}\right)  $ such that $U_{u_{k-1}}$...$U_{u_{0}}\left\vert
G\right\rangle =\left\vert G^{\prime}\right\rangle $, then $G$ and $G^{\prime
}$ are LU-equivalent \cite{hein}. It was discovered by Hein et \textit{al}.
and by Van den Nest et \textit{al}. that the sequence of transformations
taking $\left\vert G\right\rangle $ to $\left\vert G^{\prime}\right\rangle $
can equivalently be expressed as a sequence of simple graph operations taking
$G$ to $G^{\prime}$. In particular, it was shown in \cite{bart} that a graph
$G$ determines uniquely a graph state $\left\vert G\right\rangle $ and two
graph states ($\left\vert G_{1}\right\rangle $ and $\left\vert G_{2}%
\right\rangle $) determined by two graphs ($G_{1}$ and $G_{2}$) are equivalent
up to some local Clifford transformations iff these two graphs are related to
each other by local complementations (LCs). The concept of LC was originally
introduced by Bouchet in \cite{france}. A LC of a graph on a vertex $v$ refers
to the operation that in the neighborhood of $v$ we connect all the
disconnected vertices and disconnect all the connected vertices. All the
graphs on up to $12$ vertices have been classified under LCs and graph
isomorphisms \cite{parker}. In summary, the relation between graphs and
quantum codes can be rather complicated since one graph may provide
inequivalent codes and different graphs may provide equivalent codes. However,
it has been established that the family of codes given by a graph is
equivalent to the family of codes given by a local complementation of that graph.

As pointed out earlier, unitary operations $U$ in the local Clifford group
$\mathcal{C}_{n}^{l}$ act on graph states $\left\vert G\right\rangle $.
However, there exists also graph theoretical rules, transformations acting on
graphs, which correspond to local Clifford operations. These operations
generate the orbit of any graph state under local Clifford operations. The LC
orbit of a graph $G$ is the set of all non-isomorphic graphs, including $G$
itself, that can be transformed into $G$ by any sequence of local
complementations and vertex permutations. The transformation laws for a graph
state$\left\vert G\right\rangle $ and a graph stabilizer under local unitary
transformations $U$ read,%
\begin{equation}
\left\vert G\right\rangle \rightarrow\left\vert G^{\prime}\right\rangle
=U\left\vert G\right\rangle \text{ and, }S_{\Gamma}\rightarrow S_{\Gamma
^{\prime}}=US_{\Gamma}U^{\dagger}\text{,} \label{oggi}%
\end{equation}
respectively. Neglecting overall phases, it turns out that local Clifford
operations $U\in\mathcal{C}_{n}^{l}$ are just the symplectic transformations
$Q$ of $%
\mathbb{Z}
_{2}^{2n}$ which preserve the symplectic inner product \cite{moor}. Therefore,
the $\left(  2n\times2n\right)  $-matrices $Q$ satisfy the relation
$Q^{\text{T}}PQ=P$ where T denotes the transpose operation and $P$ is the
$\left(  2n\times2n\right)  $-matrix that defines a symplectic inner product
in $%
\mathbb{Z}
_{2}^{2n}$,%
\begin{equation}
P\overset{\text{def}}{=}\left(
\begin{array}
[c]{cc}%
0 & I\\
I & 0
\end{array}
\right)  \text{.}%
\end{equation}
Furthermore, since local Clifford operations act on each qubit separately,
they have the additional block structure%
\begin{equation}
Q\overset{\text{def}}{=}\left(
\begin{array}
[c]{cc}%
A & B\\
C & D
\end{array}
\right)  \text{,}%
\end{equation}
where the $\left(  n\times n\right)  $-blocks $A$, $B$, $C$, $D$ are diagonal.
It was shown in \cite{bart} that each binary stabilizer code is equivalent to
a graph code. In particular, each graph code characterized by the adjacency
matrix $\Gamma$ corresponds to a stabilizer matrix $\mathcal{S}_{b}%
\overset{\text{def}}{=}\left(  \Gamma\left\vert I\right.  \right)  $ and
transpose stabilizer (generator matrix) $\mathcal{T}\overset{\text{def}}%
{=}\mathcal{S}_{b}^{T}=\binom{\Gamma}{I}$. The generator matrix $\binom
{\Gamma^{\prime}}{I}$ for a graph state with adjacency matrix $\Gamma^{\prime
}$ reads,%
\begin{equation}
\binom{\Gamma}{I}\rightarrow\binom{\Gamma^{\prime}}{I}=\left(
\begin{array}
[c]{cc}%
A & B\\
C & D
\end{array}
\right)  \binom{\Gamma}{I}\left(  C\Gamma+D\right)  ^{-1}\text{,} \label{nons}%
\end{equation}
where,%
\begin{equation}
\Gamma^{\prime}\overset{\text{def}}{=}Q\left(  \Gamma\right)  =\left(
A\Gamma+B\right)  \left(  C\Gamma+D\right)  ^{-1}\text{.} \label{tl1}%
\end{equation}
Observe that in order to have properly defined generators matrices in Eq.
(\ref{nons}), $C\Gamma+D$ must be nonsingular and $\Gamma^{\prime}$ must have
vanishing diagonal elements. The graphical analog of the transformation law in
Eq. (\ref{tl1}) was provided in \cite{bart}. Before stating this result, some
additional terminology awaits to be introduced.

Two vertices $i$ and $j$ of a graph $G=\left(  V\text{, }E\right)  $ are
called adjacent vertices, or neighbors, if $\left\{  i\text{, }j\right\}  \in
E $. The neighborhood $N\left(  i\right)  \subseteq V$ of a vertex $i$ is the
set of all neighbors of $i$. A graph $G^{\prime}=\left(  V^{\prime}\text{,
}E^{\prime}\right)  $ which satisfies $V^{\prime}\subseteq V$ and $E^{\prime
}\subseteq E$ is a subgraph of $G$ and one writes $G^{\prime}\subseteq G$. For
a subset $A\subseteq V$ of vertices, the induced subgraph $G\left[  A\right]
\subseteq G$ is the graph with vertex set $A$ and edge set $\left\{  \left\{
i\text{, }j\right\}  \in E:i\text{, }j\in A\right\}  $. If $G$ has an
adjacency matrix $\Gamma$, its complement $G^{\text{c}}$ is the graph with
adjacency matrix $\Gamma+\mathbf{I}$, where $\mathbf{I}$ is the $\left(
n\times n\right)  $-matrix which has all ones, except for the diagonal entries
which are zero. For each vertex $i=1$,..., $n$, a local complementation
$g_{i}$ sends the $n$-vertex graph $G$ to the graph $g_{i}\left(  G\right)  $
which is obtained by replacing the induced subgraph $G\left[  N\left(
i\right)  \right]  $ by its complement. In other words,%
\begin{equation}
\Gamma\rightarrow\Gamma^{\prime}\equiv g_{i}(\Gamma)\overset{\text{def}}%
{=}\Gamma+\Gamma\Lambda_{i}\Gamma+\Lambda^{\left(  i\right)  }\text{,}
\label{n1}%
\end{equation}
where $\Lambda_{i}$ has a $1$ on the $i$th diagonal entry and zeros elsewhere
and $\Lambda^{\left(  i\right)  }$ is a diagonal matrix such that yields zeros
on the diagonal of $g_{i}(\Gamma)$. Finally, the graphical analog of Eq.
(\ref{tl1}) becomes,%
\begin{equation}
Q_{i}\left(  \Gamma\right)  =g_{i}(\Gamma)\text{,} \label{tl2}%
\end{equation}
with,%
\begin{equation}
Q_{i}\overset{\text{def}}{=}\left(
\begin{array}
[c]{cc}%
I & \text{diag}\left(  \Gamma_{i}\right) \\
\Lambda_{i} & I
\end{array}
\right)  \text{,} \label{n2}%
\end{equation}
and diag$\left(  \Gamma_{i}\right)  \overset{\text{def}}{=}$diag$\left(
\Gamma_{i1}\text{,..., }\Gamma_{in}\right)  $. Observe that substituting
(\ref{n2}) in (\ref{tl1}) and using (\ref{n1}), Eq. (\ref{tl2}) gives%
\begin{equation}
Q_{i}\left(  \Gamma\right)  =g_{i}(\Gamma)\Leftrightarrow\Gamma+\Gamma
\Lambda_{i}\Gamma+\Lambda^{\left(  i\right)  }=\Gamma+\Gamma\Lambda_{i}%
\Gamma+\left[  \text{diag}\left(  \Gamma_{i}\right)  +\text{diag}\left(
\Gamma_{i}\right)  \Lambda_{i}\Gamma\right]  \text{,}%
\end{equation}
that is,%
\begin{equation}
\Lambda^{\left(  i\right)  }=\text{diag}\left(  \Gamma_{i}\right)
+\text{diag}\left(  \Gamma_{i}\right)  \Lambda_{i}\Gamma\text{.}%
\end{equation}
The translation of the action of local Clifford operations on graph states
into the action of local complementations on graphs as presented in Eq.
(\ref{tl2}) is a major achievement of \cite{bart}.

\subsection{The CWS-work}

CWS codes include all stabilizer codes as well as several nonadditive codes.
However, for the sake of completeness, we point out that there are indeed
quantum codes that cannot be recast within the CWS framework as pointed out in
\cite{cross} and shown in \cite{ruskai}. CWS codes in standard form can be
specified by a graph $G$ and a (nonadditive, in general) classical binary code
$\mathcal{C}_{^{\text{classical}}}$. The $n$ vertices of the graph $G$
correspond to the $n$ qubits of the code and its adjacency matrix is $\Gamma$.
Given the graph state $\left\vert G\right\rangle $ and the binary code
$\mathcal{C}_{^{\text{classical}}}$, a unique base state $\left\vert
S\right\rangle $ and a set of word operators $\left\{  w_{k}\right\}  $ are
specified. The base state $\left\vert S\right\rangle $ is a single stabilizer
state stabilized by the word stabilizer $\mathcal{S}_{\text{CWS}}$, a maximal
Abelian subgroup of the Pauli group $\mathcal{P}_{\mathcal{H}_{2}^{n}}$.

Let $\left(  \left(  n\text{, }K\text{, }d\right)  \right)  $ denote a quantum
code on $n$ qubits that encodes $K$ dimensions with distance $d$. Following
\cite{cross}, it can be shown that a $\left(  \left(  n\text{, }K\text{,
}d\right)  \right)  $ codeword stabilized code with word operators
$\mathcal{W}=\left\{  w_{l}\right\}  $ with $l\in\left\{  1\text{,...,
}K\right\}  $ and codeword stabilizer $\mathcal{S}_{\text{CWS}}$ is locally
Clifford equivalent to a codeword stabilized code with word operators
$\mathcal{W}^{\prime}$,%
\begin{equation}
\mathcal{W}^{\prime}\overset{\text{def}}{=}\left\{  w_{l}^{\prime
}=Z^{\mathbf{c}_{l}}\right\}  \text{,} \label{can1}%
\end{equation}
and codeword stabilizer $\mathcal{S}_{\text{CWS}}^{\prime}$,%
\begin{equation}
\mathcal{S}_{\text{CWS}}^{\prime}\overset{\text{def}}{=}\left\langle
S_{l}^{\prime}\right\rangle =\left\langle X_{l}Z^{\mathbf{r}_{l}}\right\rangle
\text{,} \label{can2}%
\end{equation}
where $\mathbf{c}_{l}$s are codewords defining the classical binary code
$\mathcal{C}_{\text{classical}}$ and $\mathbf{r}_{l}$ is the $l$th row vector
of the adjacency matrix $\Gamma$ of the graph $G$. For the sake of clarity, we
stress that $Z^{\mathbf{v}}$ in Eq. (\ref{can2}) is the notational shorthand
for%
\begin{equation}
Z^{\mathbf{v}}\overset{\text{def}}{=}Z^{v_{1}}\otimes\text{...}\otimes
Z^{v_{n}}\text{,}%
\end{equation}
where $\mathbf{v=}\left(  v_{1}\text{,..., }v_{n}\right)  \in F_{2}^{n}$ is a
binary $n$-vector. Thus, any CWS code is locally Clifford equivalent to a CWS
code with a graph-state stabilizer and word operators consisting only of $Z$s.
Moreover, the word operators can always be chosen to include the identity.
Eqs. (\ref{can1})\ and (\ref{can2}) characterize the so-called standard form
of a CWS quantum code. For a CWS code in standard form, the base state
$\left\vert S\right\rangle $ is a graph state. Furthermore, the codespace of a
CWS\ code is spanned by a set of basis vectors which result from applying the
word operators $w_{k}$ on the base state $\left\vert S\right\rangle $,%
\begin{equation}
\mathcal{C}_{\text{CWS}}\overset{\text{def}}{=}\text{Span}\left\{  \left\vert
w_{l}\right\rangle \right\}  \text{ with, }\left\vert w_{l}\right\rangle
\overset{\text{def}}{=}w_{l}\left\vert S\right\rangle \text{.}%
\end{equation}
Therefore, the dimension of the codespace equals the number of word operators.
These operators are Pauli operators in $\mathcal{P}_{\mathcal{H}_{2}^{n}}$
that anticommute with one or more of the stabilizer generators for the base
state. Thus, word operators map the base state onto an orthogonal state. The
only exception is that in general the set of word operators also includes the
identity operator so that the base state is a codeword of the quantum code as
well. These basis states are also eigenstates of the stabilizer generators,
but with some of the eigenvalues differing from $+1$. In addition, it turns
out that a single qubit Pauli error $X$, $Z$ or $ZX$ acting on a codeword
$\omega\left\vert S\right\rangle $ of a CWS code in standard form is
equivalent up to a sign to another multi-qubit error consisting of $Z$s.
Therefore, since all errors become $Z$s, the original quantum error model is
transformed into a classical (induced by the CWS\ formalism) error model
characterized, in general, by multi-qubit errors. The map $\mathcal{C}%
l_{\mathcal{S}_{\text{CWS}}}$ that defines this transformation reads,%
\begin{equation}
\mathcal{C}l_{\mathcal{S}_{\text{CWS}}}:\mathcal{E}\ni E\equiv\pm
Z^{\mathbf{v}}X^{\mathbf{u}}\mapsto\mathcal{C}l_{\mathcal{S}_{\text{CWS}}%
}\left(  \pm Z^{\mathbf{v}}X^{\mathbf{u}}\right)  \overset{\text{def}}%
{=}\mathbf{v\oplus}%
{\displaystyle\bigoplus\limits_{l=1}^{n}}
u_{l}\mathbf{r}_{l}\in\left\{  0\text{, }1\right\}  ^{n}\text{,}%
\end{equation}
where $\mathcal{E}$ denotes the set of Pauli errors $E$, $\mathbf{r}_{l}$ is
the $l$th\textit{\ }row of the adjacency matrix $\Gamma$ for the graph $G$ and
$u_{l}$ is the $l$\textit{th }bit of the vector $\mathbf{u}$.
Finally,\textbf{\ }it was shown in \cite{cross} that any stabilizer code is a
CWS code. Specifically, a quantum stabilizer code $\left[  \left[
n,k,d\right]  \right]  $ (where the parameters $n$, $k$, $d$ denote the
length, the dimension and the distance of the quantum code, respectively) with
stabilizer $\mathcal{S}\overset{\text{def}}{=}\left\langle S_{1}\text{,...,
}S_{n-k}\right\rangle $ where $S_{j}$ with $j\in\left\{  1\text{,...,
}n-k\right\}  $ denote the stabilizer generators and logical operations
$\bar{X}_{1}$,..., $\bar{X}_{k}$ and $\bar{Z}_{1}$,..., $\bar{Z}_{k}$ is
equivalent to a CWS code defined by,%
\begin{equation}
\mathcal{S}_{\text{CWS}}\overset{\text{def}}{=}\left\langle S_{1}\text{, ...,
}S_{n-k}\text{, }\bar{Z}_{1}\text{,..., }\bar{Z}_{k}\right\rangle \text{,}%
\end{equation}
and word operators $\omega_{\mathbf{v}}$,%
\begin{equation}
\omega_{\mathbf{v}}=\bar{X}_{1}^{\left(  \mathbf{v}\right)  _{1}}%
\otimes\text{...}\otimes\bar{X}_{k}^{\left(  \mathbf{v}\right)  _{k}}\text{.}%
\end{equation}
The vector $\mathbf{v}$ denotes a $k$-bit string and $\left(  \mathbf{v}%
\right)  _{l}\equiv v_{l}$ with $l\in\left\{  1\text{,..., }k\right\}  $ is
the $l$th bit of the vector $\mathbf{v}$. For further details on binary
CWS\ quantum codes, we refer to \cite{cross}. Finally, for a very recent
investigation on the symmetries of CWS codes, we refer to \cite{tqc}.

\section{From graphs to stabilizer codes and vice-versa}

In this section, we revisit some basic ingredients of the Schlingemann-Werner
work (SW-work, \cite{werner}), the Schlingemann work (S-work, \cite{dirk})
and, finally, the Van den Nest et \textit{al}. work (VdN-work, \cite{bart}).
We focus on those aspects of these works that will be especially relevant for
our proposed scheme.

\subsection{The Schlingemann-Werner work}

The basic graphical construction of quantum codes within the SW-work
\cite{werner} can be described as follows. Quantum codes are completely
characterized by a unidirected graph $G$ $=G\left(  V\text{, }E\right)  $
characterized by a set $V$ of $n$ vertices and a set of edges $E$ specified by
the coincidence matrix $\Xi$ with both input and output vertices and a finite
Abelian group $\mathcal{G}$ with a nondegenerate symmetric bicharacter $\chi$.
We remark that there are various types of matrices that can be used to specify
a given graph (for instance, incidence and adjacency matrices \cite{die}). The
coincidence matrix introduced in \cite{werner} is simply the adjacency matrix
of a graph with both input and output vertices (and, it should not be confused
with the so-called incidence matrix of a graph). The sets of input and output
vertices will be denoted by $X$ and $Y$, respectively. Let $\mathcal{G}$ be
any finite additive Abelian group of cardinality $\left\vert \mathcal{G}%
\right\vert =n$ with the addition operation denoted by $+$ and null element
$0$. A nondegenerate symmetric bicharacter is a map $\chi:\mathcal{G}%
\times\mathcal{G}\ni\left(  g\text{, }h\right)  \mapsto\chi\left(  g\text{,
}h\right)  \equiv\left\langle g\text{, }h\right\rangle \in%
\mathbb{C}
$ satisfying the following properties \cite{partha}: (i) $\left\langle
g\text{, }h\right\rangle =\left\langle h\text{, }g\right\rangle $, $\forall
g$, $h\in\mathcal{G}$; (ii) $\left\langle g\text{, }h_{1}+h_{2}\right\rangle
=\left\langle g\text{, }h_{1}\right\rangle \left\langle g\text{, }%
h_{2}\right\rangle $, $\forall g$, $h_{1}$, $h_{2}\in\mathcal{G}$; (iii)
$\left\langle g\text{, }h\right\rangle =1$ $\forall h\in\mathcal{G}%
\Leftrightarrow g=0$. If $\mathcal{G}=%
\mathbb{Z}
_{n}\overset{\text{def}}{=}\left\{  0\text{,..., }n-1\right\}  $ (the cyclic
group of order $n$) with addition modulo $n$ as the group operation, the
bicharacter $\chi$ can be chosen as%
\begin{equation}
\chi\left(  g\text{, }h\right)  \equiv\left\langle g\text{, }h\right\rangle
\overset{\text{def}}{=}e^{i\frac{2\pi}{n}gh}\text{, }%
\end{equation}
with $g$, $h\in%
\mathbb{Z}
_{n}$. The encoding operator $\mathbf{v}_{G}$ of an error correcting code is
an isometry (a bijective map between two metric spaces that preserve
distances),%
\begin{equation}
\mathbf{v}_{G}:L^{2}\left(  \mathcal{G}^{X}\right)  \rightarrow L^{2}\left(
\mathcal{G}^{Y}\right)  \text{,}%
\end{equation}
where $L^{2}\left(  \mathcal{G}^{X}\right)  $ is the $\left\vert X\right\vert
$-fold tensor product $\mathcal{H}^{\otimes X}$ with $\mathcal{H}=L^{2}\left(
\mathcal{G}\right)  $ (the Hilbert space $\mathcal{H}$ is realized as the
space of integrable functions over $\mathcal{G}$) and $\mathcal{G}%
\overset{\text{def}}{=}%
\mathbb{Z}
_{2}$ in the qubit case. Similarly, $L^{2}\left(  \mathcal{G}^{Y}\right)  $ is
the $\left\vert Y\right\vert $-fold tensor product $\mathcal{H}^{\otimes Y}$.
The Hilbert space $L^{2}\left(  \mathcal{G}\right)  $ is defined as,%
\begin{equation}
L^{2}\left(  \mathcal{G}\right)  \overset{\text{def}}{=}\left\{
\psi\left\vert \psi:\mathcal{G}\rightarrow%
\mathbb{C}
\right.  \right\}  \text{,}%
\end{equation}
with scalar product between two elements $\psi_{1}$ and $\psi_{2}$ in
$L^{2}\left(  \mathcal{G}\right)  $ given by,%
\begin{equation}
\left\langle \psi_{1}\text{, }\psi_{2}\right\rangle \overset{\text{def}}%
{=}\frac{1}{\left\vert \mathcal{G}\right\vert }\sum_{g}\bar{\psi}_{1}\left(
g\right)  \psi_{2}\left(  g\right)  \text{.}%
\end{equation}
The action of $\mathbf{v}_{G}$ on $L^{2}\left(  \mathcal{G}^{X}\right)  $ is
defined as \cite{werner},%
\begin{equation}
\left(  \mathbf{v}_{G}\psi\right)  \left(  g^{Y}\right)  \overset{\text{def}%
}{=}\int dg^{X}\mathbf{v}_{G}\left[  g^{X\cup Y}\right]  \psi\left(
g^{X}\right)  \text{,} \label{g12}%
\end{equation}
where $\mathbf{v}_{_{G}}\left[  g^{X\cup Y}\right]  $, the integral kernel of
the isometry $\mathbf{v}_{_{G}}$, is given by \cite{werner},%
\begin{align}
\mathbf{v}_{G}\left[  g^{X\cup Y}\right]   &  =\left\vert \mathcal{G}%
\right\vert ^{\frac{\left\vert X\right\vert }{2}}\prod\limits_{\left\{
z\text{, }z^{\prime}\right\}  }\chi\left(  g_{z}\text{, }g_{z^{\prime}%
}\right)  ^{\Xi\left(  z\text{, }z^{\prime}\right)  }=\left\vert
\mathcal{G}\right\vert ^{\frac{\left\vert X\right\vert }{2}}\prod
\limits_{\left\{  z\text{, }z^{\prime}\right\}  }\left[  \exp\left(
\frac{2\pi i}{p}g_{z}g_{z^{\prime}}\right)  \right]  ^{\Xi\left(  z\text{,
}z^{\prime}\right)  }\nonumber\\
& \nonumber\\
&  =\left\vert \mathcal{G}\right\vert ^{\frac{\left\vert X\right\vert }{2}%
}\prod\limits_{\left\{  z\text{, }z^{\prime}\right\}  }\left[  \exp\left(
\frac{2\pi i}{p}g_{z}\Xi\left(  z\text{, }z^{\prime}\right)  g_{z^{\prime}%
}\right)  \right]  =\left\vert \mathcal{G}\right\vert ^{\frac{\left\vert
X\right\vert }{2}}\exp\left(  \frac{\pi i}{p}g^{X\cup Y}\cdot\Xi\cdot g^{X\cup
Y}\right)  \text{.} \label{g11}%
\end{align}
The product in Eq. (\ref{g11}) must be taken over each two elementary subsets
$\left\{  z\text{, }z^{\prime}\right\}  $ in $X\cup Y$. Substituting Eq.
(\ref{g11}) into Eq. (\ref{g12}), the action of $\mathbf{v}_{G}$ on
$L^{2}\left(  \mathcal{G}^{X}\right)  $ finally becomes,%
\begin{equation}
\left(  \mathbf{v}_{G}\psi\right)  \left(  g^{Y}\right)  =\int dg^{X}%
\left\vert \mathcal{G}\right\vert ^{\frac{\left\vert X\right\vert }{2}}%
\exp\left(  \frac{\pi i}{p}g^{X\cup Y}\cdot\Xi\cdot g^{X\cup Y}\right)
\psi\left(  g^{X}\right)  \text{.} \label{g13}%
\end{equation}
We recall that the sequential steps of a QEC cycle can be described as
follows,%
\begin{equation}
\rho\overset{\text{coding}}{\longrightarrow}\mathbf{v}\rho\mathbf{v}^{\ast
}\equiv\rho^{\prime}\text{, }\rho^{\prime}\overset{\text{noise}}%
{\longrightarrow}\mathbf{T}\left(  \rho^{\prime}\right)  =\sum\limits_{\alpha
}F_{\alpha}\rho^{\prime}F_{\alpha}^{\ast}\equiv\rho^{\prime\prime}\text{,
}\rho^{\prime\prime}\overset{\text{recovery}}{\longrightarrow}\mathbf{R}%
\left(  \rho^{\prime\prime}\right)  =\rho\text{,}%
\end{equation}
that is,%
\begin{equation}
\mathbf{R}\left(  \mathbf{T}\left(  \mathbf{v}\rho\mathbf{v}^{\ast}\right)
\right)  =\rho\text{.}%
\end{equation}
Furthermore, the traditional Knill-Laflamme error-correction conditions read,%
\begin{equation}
\left\langle \mathbf{v}\psi_{1}\text{, }F_{\alpha}^{\ast}F_{\beta}%
\mathbf{v}\psi_{2}\right\rangle =\omega\left(  F_{\alpha}^{\ast}F_{\beta
}\right)  \left\langle \psi_{1}\text{, }\psi_{2}\right\rangle \text{,}
\label{kl}%
\end{equation}
where the multiplicative factor $\omega\left(  F_{\alpha}^{\ast}F_{\beta
}\right)  $ does not depend on the states $\psi_{1}$ and $\psi_{2}$. The
graphical analog of Eq. (\ref{kl}) is given by,%
\begin{equation}
\left\langle \mathbf{v}\psi_{1}\text{, }F\mathbf{v}\psi_{2}\right\rangle
=\omega\left(  F\right)  \left\langle \psi_{1}\text{, }\psi_{2}\right\rangle
\text{,} \label{g14}%
\end{equation}
for all operators in $\mathcal{U}(E)$, the set of all operators in
$L^{2}(\mathcal{G}^{Y})$ which are localized in $E\subset Y$. Thus, operators
in $\mathcal{U}(E)$ are given by the tensor product of an arbitrary operator
on $\mathcal{H}^{\otimes E}$ with the identity on $\mathcal{H}^{\otimes
Y\backslash E}$. A graph code corrects $e$ errors if and only if it detects
all error configurations $E\subset Y$ with $\left\vert E\right\vert \leq2e$.
Given this graphical construction of the encoding operator $\mathbf{v}_{G}$ in
Eq. (\ref{g13}) and the graphical quantum error-correction conditions in Eq.
(\ref{g14}), the main finding provided by Schlingemann and Werner can be
restated as follows: given a finite Abelian group $\mathcal{G}$ and a weighted
graph $G$, an error configuration $E\subset Y$ is detected by the quantum code
$\mathbf{v}_{G}$ if and only if given that%
\begin{equation}
d^{X}=0\text{ and, }\Xi_{E}^{X}d^{E}=0\text{,} \label{wc1}%
\end{equation}
then,%
\begin{equation}
\Xi_{X\cup E}^{I}d^{X\cup E}=0\Rightarrow d^{X\cup E}=0\text{,} \label{wc2}%
\end{equation}
with $I=Y\backslash E$. In general, the condition\textbf{\ }$\Xi_{B}^{A}%
d^{B}=0$\ is a set of equations, one for each integration vertex\textbf{\ }%
$a\in A$\textbf{: }for each vertex\textbf{\ }$a\in A$\textbf{, }we have to sum
the\textbf{\ }$d_{b}$\textbf{\ }for all vertices\textbf{\ }$b\in B$%
\textbf{\ }connected to\textbf{\ }$a$\textbf{, }and equate it to zero\textbf{.
}Furthermore, we underline that the fact that\textbf{\ }$v_{G}$\ is
an\textbf{\ }isometry is equivalent to the detection of zero errors. In
graph-theoretic terms, the detection of zero errors requires that\textbf{\ }%
$\Xi_{X}^{Y}d^{X}=0$\textbf{\ }implies\textbf{\ }$d^{X}=0$. A code
that\textbf{\ }satisfies Eq. (\ref{wc2}) given Eq. (\ref{wc1})\textbf{\ }can
be either nondegenerate or degenerate.\ We shall assume that Eq. (\ref{wc2})
with the additional constraints in Eq. (\ref{wc1}) denotes the weak version
(necessary and sufficient conditions) of the graph-theoretic error detection
conditions. However, sufficient graph-theoretic error detection conditions can
be introduced as well. Specifically, an error configuration $E$ is detectable
by a quantum code if,%
\begin{equation}
\Xi_{X\cup E}^{I}d^{X\cup E}=0\Rightarrow d^{X\cup E}=0\text{.} \label{sc}%
\end{equation}
We shall denote conditions in Eq. (\ref{sc}) without any additional set of
graph-theoretic constraints (like the ones provided in Eq. (\ref{wc1})) the
strong version (sufficient conditions) of the graph-theoretic error detection
conditions. We finally emphasize, as originally pointed out in \cite{werner},
that a code that satisfies Eq. (\ref{sc}) is nondegenerate.

\subsection{The Schlingemann-work}

Schlingemann was able to show that stabilizer codes, either binary or
nonbinary, are equivalent to graph codes (and vice-versa). However, as far as
our proposed scheme concerns, the main finding uncovered in the S-work
\cite{dirk} may be stated as follows. Consider a graph code with only one
input and $\left(  n-1\right)  $-output vertices. Its corresponding
coincidence matrix $\Xi_{n\times n}$ can be written as,%
\begin{equation}
\Xi_{n\times n}\overset{\text{def}}{=}\left(
\begin{array}
[c]{cc}%
0_{1\times1} & B_{1\times\left(  n-1\right)  }^{\dagger}\\
B_{\left(  n-1\right)  \times\left(  1\right)  } & A_{\left(  n-1\right)
\times\left(  n-1\right)  }%
\end{array}
\right)  \text{,} \label{gammac}%
\end{equation}
where $A_{\left(  n-1\right)  \times\left(  n-1\right)  }$ denotes the
$\left(  n-1\right)  \times\left(  n-1\right)  $-symmetric adjacency matrix
$\Gamma_{\left(  n-1\right)  \times\left(  n-1\right)  }$. Then, the graph
code with symmetric coincidence matrix $\Xi_{n\times n}$ in Eq. (\ref{gammac})
is equivalent to stabilizer codes being associated with the isotropic subspace
$\mathcal{S}_{\text{isotropic}}$ defined as,%
\begin{equation}
\mathcal{S}_{\text{isotropic}}\overset{\text{def}}{=}\left\{  \left(
Ak\left\vert k\right.  \right)  :k\in\ker B^{\dagger}\right\}  \text{,}%
\end{equation}
that is, omitting unimportant phase factors, with the binary stabilizer group
$\mathcal{S}_{\text{binary}}$,%
\begin{equation}
\mathcal{S}_{\text{binary}}\overset{\text{def}}{=}\left\{  g_{k}=X^{k}%
Z^{Ak}:k\in\ker B^{\dagger}\right\}  \text{.}%
\end{equation}
Observe that a stabilizer operator $g_{k}\in\mathcal{S}_{\text{binary}}$ for
an $n$-vertex graph has a $2n$-dimensional binary vector space representation
such that $g_{k}\leftrightarrow v_{g_{k}}\overset{\text{def}}{=}\left(
Ak\left\vert k\right.  \right)  $.

More generally, consider a $\left[  \left[  n,k,d\right]  \right]  $ binary
quantum stabilizer code associated with a graph $G=\left(  V\text{, }E\right)
$ characterized by the $\left(  n+k\right)  \times\left(  n+k\right)  $
symmetric coincidence matrix $\Xi_{\left(  n+k\right)  \times\left(
n+k\right)  }$,%
\begin{equation}
\Xi_{\left(  n+k\right)  \times\left(  n+k\right)  }\overset{\text{def}}%
{=}\left(
\begin{array}
[c]{cc}%
0_{k\times k} & B_{k\times n}^{\dagger}\\
B_{n\times k} & \Gamma_{n\times n}%
\end{array}
\right)  \text{.} \label{losai-2}%
\end{equation}
To attach the input vertices, $\Xi$ has to be constructed in such a manner
that the following conditions are satisfied: i) first, $\det\Gamma_{n\times
n}=0$ ($\operatorname{mod}2$); ii) second,\ the matrix $B_{k\times n}%
^{\dagger}$ must define a $k$-dimensional subspace in $\mathbf{F}_{2}^{n}$
spanned by $k$ linearly independent binary vectors of length $n$ not included
in the Span of the raw-vectors defining the symmetric adjacency matrix
$\Gamma_{n\times n}$,%
\begin{equation}
\text{Span}\left\{  \vec{v}_{1}\text{,..., }\vec{v}_{k}\right\}  \cap\text{
Span}\left\{  \vec{v}_{\Gamma}^{\left(  1\right)  }\text{,..., }\vec
{v}_{\Gamma}^{\left(  n\right)  }\text{ }\right\}  =\left\{  \emptyset
\right\}  \text{,}%
\end{equation}
where $\vec{v}_{j}\in\mathbf{F}_{2}^{n}$ for $j\in\left\{  1\text{,...,
}k\right\}  $ and $\vec{v}_{\Gamma}^{\left(  i\right)  }\in\mathbf{F}_{2}^{n}$
for $i\in\left\{  1\text{,..., }n\right\}  $; iii) third, Span$\left\{
\vec{v}_{1}\text{,..., }\vec{v}_{k}\right\}  $ contains a vector $\vec{v}%
_{B}\in\mathbf{F}_{2}^{n}$ such that $\vec{v}_{B}\cdot\vec{v}_{\Gamma
}^{\left(  i\right)  }=0$ for any $i\in\left\{  1\text{,..., }n\right\}  $.
Condition i) is needed to avoid disconnected graphs. Condition ii) is required
to have a properly defined isometry capable of detecting zero errors. Finally,
condition iii) is needed to generate an isotropic subspace (or, in other
words, an Abelian subgroup of the Pauli group, the so-called stabilizer group)
with,%
\begin{equation}
\left(  \Gamma\vec{v}_{\Gamma}^{\left(  l\right)  }\text{, }\vec{v}_{\Gamma
}^{\left(  l\right)  }\right)  \odot\left(  \Gamma\vec{v}_{\Gamma}^{\left(
m\right)  }\text{, }\vec{v}_{\Gamma}^{\left(  m\right)  }\right)  =0\text{, }%
\end{equation}
for any pair $\left(  \vec{v}_{\Gamma}^{\left(  l\right)  }\text{, }\vec
{v}_{\Gamma}^{\left(  m\right)  }\right)  $ in $\left\{  \vec{v}_{\Gamma
}^{\left(  1\right)  }\text{,..., }\vec{v}_{\Gamma}^{\left(  n\right)  }\text{
}\right\}  $ where the symbol $\odot$ denotes the symplectic product
\cite{gaitan}.

As a final remark, we point out that in a more general framework like the one
presented in \cite{dirk}, we could consider three types of vertices: input,
auxiliary and output vertices. The input vertices label the input systems and
are used for encoding. The auxiliary vertices are inputs used as auxiliary
degrees of freedom for implementing additional constraints for the protected
code subspace. Finally, output vertices simply label the output quantum systems.

\subsection{The Van den Nest-work}

The main achievement of the VdN-work in \cite{bart} is the construction of a
very useful algorithmic procedure for transforming any binary quantum
stabilizer code into a graph code. Before describing this procedure, we remark
that it is straightforward to check that a graph code given by the adjacency
matrix $\Gamma$ corresponds to a stabilizer matrix $\mathcal{S}_{b}%
\overset{\text{def}}{=}\left(  \Gamma\left\vert I\right.  \right)  $ and
transpose stabilizer $\mathcal{T}\overset{\text{def}}{=}\mathcal{S}%
_{b}^{\text{T}}=\binom{\Gamma}{I}$. That said, consider a quantum stabilizer
code with stabilizer matrix,%
\begin{equation}
\mathcal{S}_{b}\overset{\text{def}}{=}\left(  Z\left\vert X\right.  \right)
\text{,} \label{ssss}%
\end{equation}
and transpose stabilizer $\mathcal{T}$ given by,%
\begin{equation}
\mathcal{T}\overset{\text{def}}{=}\mathcal{S}_{b}^{\text{T}}=\binom
{Z^{\text{T}}}{X^{\text{T}}}\equiv\binom{A}{B}\text{.} \label{t}%
\end{equation}
Let us define\textbf{ }$\mathcal{S}_{b}$ in Eq. (\ref{ssss}). Given a set of
generators of the stabilizer, the stabilizer matrix\textbf{ }$\mathcal{S}_{b}%
$\textbf{\ }is constructed by assembling the binary representations of the
generators as the rows of a full rank\textbf{ }$\left(  n\times2n\right)
$-matrix. The transpose of the binary stabilizer matrix (i.e., the transpose
stabilizer)\textbf{ }$\mathcal{T}$\textbf{\ \ }is simply the full rank\textbf{
}$\left(  2n\times n\right)  $-matrix obtained from\textbf{ }$\mathcal{S}_{b}%
$\textbf{\ }after exchanging rows with columns\textbf{. }The goal of the
algorithmic procedure is to convert the transpose stabilizer $\mathcal{T}$ in
Eq. (\ref{t}) of a given stabilizer code into the transpose stabilizer
$\mathcal{T}^{\prime}=$ $\binom{A^{\prime}}{B^{\prime}}$ of an equivalent
graph code. Then, the matrix $A^{\prime}$ will represent the adjacency matrix
of the corresponding graph. Two scenarios may occur: i) $B$ is a $n\times n$
invertible matrix; ii) $B$ is not an invertible matrix. In the first scenario
where $B$ is invertible, a right-multiplication of the transpose stabilizer
$\mathcal{T}=$ $\binom{A}{B}$ by $B^{-1}$ will perform a basis change, an
operation that provides us with an equivalent stabilizer code,%
\begin{equation}
\mathcal{T}B^{-1}=\binom{A}{B}B^{-1}=\binom{AB^{-1}}{I}\text{.}%
\end{equation}
Then, the matrix $AB^{-1}$ will denote the resulting adjacency matrix of the
corresponding graph. Furthermore, if the matrix $AB^{-1}$ has nonzero diagonal
elements, we can simply set these elements to zero in order to satisfy the
standard requirements for a correct definition of an adjacency matrix of
simple graphs. In the second scenario where $B$ is not invertible, we can
always find a suitable local Clifford unitary transformation $U$ such that
\cite{bart},%
\begin{equation}
\mathcal{S}_{b}\overset{\text{def}}{=}\left(  Z\left\vert X\right.  \right)
\overset{U}{\rightarrow}\mathcal{S}_{b}^{\prime}\overset{\text{def}}{=}\left(
Z^{\prime}\left\vert X^{\prime}\right.  \right)  \text{,}%
\end{equation}
and,%
\begin{equation}
\mathcal{T}\overset{\text{def}}{=}\mathcal{S}_{b}^{\text{T}}=\binom
{Z^{\text{T}}}{X^{\text{T}}}\equiv\binom{A}{B}\overset{U}{\rightarrow
}\mathcal{T}^{\prime}\overset{\text{def}}{=}\mathcal{S}_{b}^{\prime\text{T}%
}=\binom{Z^{^{\prime}\text{T}}}{X^{\prime\text{T}}}\equiv\binom{A^{\prime}%
}{B^{\prime}}\text{,}%
\end{equation}
with $\det B^{\prime}\neq0$. Therefore, right-multiplying $\mathcal{T}%
^{\prime}$ with $B^{\prime-1}$, we get%
\begin{equation}
\mathcal{T}^{\prime}B^{\prime-1}=\binom{A^{\prime}}{B^{\prime}}B^{\prime
-1}=\binom{A^{\prime}B^{\prime-1}}{I}\text{.}%
\end{equation}
Thus, the adjacency matrix of the corresponding graph becomes $A^{\prime
}B^{\prime-1}$.

The above-described algorithmic procedure for transforming any binary quantum
stabilizer code into a graph code is very important for our proposed scheme as
it will become clear in the next section.

\section{The scheme}

In this section, we formally describe our scheme and apply it to the graphical
construction of the Leung et \textit{al}. four-qubit quantum code for the
error correction of single amplitude damping errors.

\subsection{Description of the scheme}

We emphasize that our ultimate goal is the construction of classical graphs
$G\left(  V\text{, }E\right)  $ with both input and output vertices defined by
the coincidence matrix $\Xi$ in order to verify the error-correcting
capabilities of the corresponding quantum stabilizer codes via the
graph-theoretic error correction conditions advocated in the SW-work. To
achieve this goal, we propose a systematic scheme based on a very simple idea.
The CWS-, VdN- and S-works must be combined in such a manner that, with
respect to our ultimate goal, the weak-points of one method should be
compensated by the strong-points of another method.

\subsubsection{Step one}

The CWS formalism offers a very general framework where both binary/nonbinary
and/or additive/nonadditive quantum codes can be described. For this reason,
the starting point of our scheme is the realization of binary stabilizer codes
as CWS quantum codes. Although this is a relatively straightforward step, the
CWS code that one obtains is not, in general, in the standard canonical form.
From the CWS-work in \cite{cross}, it is known that there does exist a local
(unitary) Clifford operations that allows in principle to write down the
CWS\ code that realizes the binary stabilizer code in standard form. However,
the CWS-work does not suggest any algorithmic procedure to achieve this
standard form. In the absence of a systematic procedure, uncovering a local
Clifford unitary $U$ such that $\mathcal{S}_{\text{CWS}}^{\prime}%
\overset{\text{def}}{=}U\mathcal{S}_{\text{CWS}}U^{\dagger}$ (every element
$s^{\prime}\in\mathcal{S}_{\text{CWS}}^{\prime}$ can be written as
$UsU^{\dagger}$ for some $s\in\mathcal{S}_{\text{CWS}}$) may constitute a very
tedious challenge. Fortunately, we can avoid this. Before explaining how, let
us introduce the codeword stabilizer matrix $\mathcal{H}_{\mathcal{S}%
_{\text{CWS}}}\overset{\text{def}}{=}\left(  Z\left\vert X\right.  \right)  $
corresponding to the codeword stabilizer $\mathcal{S}_{\text{CWS}}$.

\subsubsection{Step two}

Two main achievements of the VdN-work in \cite{bart} are the following: first,
each stabilizer state is equivalent to a graph state under local Clifford
operations; second, an algorithmic procedure for transforming any binary
quantum stabilizer code into a graph code is provided. Observe that a
stabilizer state can be regarded as a quantum code with parameters $\left[
\left[  n\text{, }0\text{, }d\right]  \right]  $. Our idea is to exploit the
algorithmic procedure provided by the VdN-work by translating the starting
point of the algorithmic procedure in the CWS language. To achieve this, we
replace the generator matrix of the stabilizer state with the codeword
stabilizer matrix $\mathcal{H}_{\mathcal{S}_{\text{CWS}}}$ corresponding to
the codeword stabilizer $\mathcal{S}_{\text{CWS}}$ of the CWS code that
realizes the binary stabilizer code whose graphical depiction is being sought.
This way, we can simply apply the VdN algorithmic procedure to uncover the
standard form of the CWS code and, if necessary, the explicit expression for
the local (unitary) Clifford operation that links the non-standard to the
standard forms of the CWS code. After applying this VdN algorithmic procedure
adapted to the CWS formalism, we can construct a graph characterized by a
symmetric adjacency matrix $\Gamma$ with only output vertices. How do we
attach possible input vertices to this graph associated with the $\left[
\left[  n\text{, }k\text{, }d\right]  \right]  $ binary stabilizer codes with
$k\neq0$?

\subsubsection{Step three}

Unlike the VdN-work whose findings are limited to the binary quantum states,
the S-work extends its applicability to both binary and nonbinary quantum
codes. In particular, in \cite{dirk} it was shown that any stabilizer code is
a graph code and vice-versa. However, in the S-work an analog of the
algorithmic procedure for transforming any binary quantum stabilizer code into
a graph code is missing. Despite this fact, the S-work does provide a very
useful result for our proposed scheme. Namely, it is shown that a graph code
with associated graph $G\left(  V,E\right)  $ with both input and output
vertices and corresponding symmetric coincidence matrix $\Xi$ is equivalent to
stabilizer codes being associated with a suitable isotropic subspace space
$\mathcal{S}_{\text{isotropic}}$. Recall that at the end of the
above-mentioned step two, we are basically given both the isotropic subspace
and the graph without input vertices, that is the symmetric adjacency matrix
$\Gamma$ embedded in the more general coincidence matrix $\Xi$. Therefore, by
exploiting the just mentioned very useful specific finding of the S-work in a
reverse direction (we are allowed to do so since a graph code is equivalent to
a stabilizer code and vice-versa), in some sense, we can construct the full
coincidence matrix $\Xi$ and finally attach the input vertices to the graph.
What can we do with a graphical depiction of a binary stabilizer code?

\subsubsection{Step three+one}

In the SW-work, outstanding graphical QEC conditions were introduced
\cite{werner}. However, these conditions were only partially employed for
quantum codes associated with graphs and the codes needed not be necessarily
stabilizer codes. By logically combining the CWS-, VdN- and S-works, the power
of the graphical QEC conditions in \cite{werner} can be fully exploited in a
systematic manner in both directions: from graph codes to stabilizer codes and vice-versa.

\medskip

In summary, given a binary quantum stabilizer code $\mathcal{C}%
_{\text{stabilizer}}$, the systematic procedure that we propose can be
described in $3+1=4$ points as follows:

\begin{itemize}
\item Realize the stabilizer code $\mathcal{C}_{\text{stabilizer}}$ as a CWS
quantum code $\mathcal{C}_{\text{CWS}}$;

\item Apply the VdN-work adapted to the CWS formalism to identify the standard
form of the CWS code that realizes the stabilizer code whose graphical
depiction is being sought. In other words, find the graph $G$ with only output
vertices characterized by the symmetric adjacency matrix $\Gamma$ associated
with $\mathcal{C}_{\text{CWS}}$ in the standard form;

\item Exploit the S-work as explained to identify the extended graph with both
input and output vertices characterized by the symmetric coincidence matrix
$\Xi$ associated with the isometric encoding map that defines $\mathcal{C}%
_{\text{CWS}}$;

\item Use the SW-work to apply the graph-theoretic error-correction conditions
to the extended graph in order to explicitly verify the error-correcting
capabilities of the corresponding $\mathcal{C}_{\text{stabilizer}}$ realized
as a $\mathcal{C}_{\text{CWS}}$ quantum code.
\end{itemize}

\subsection{Application of the scheme}

We think there is no better way to describe and understand the effectiveness
of our proposed scheme than by simply working out in detail a simple
illustrative example. In what follows, we wish to uncover the graph\textbf{
}associated with the Leung et \textit{al}. $\left[  \left[  4\text{,}1\right]
\right]  $ four-qubit stabilizer (nondegenerate) quantum code \cite{debbie}.
Several explicit constructions of graphs for various stabilizer codes
characterized by either single or multi-qubit encoding operators are added in
the Appendices: the three-qubit repetition code, the perfect $1$-erasure
correcting four-qubit code, the perfect\textbf{ }$1$-error correcting
five-qubit code,\textbf{ }$1$-error correcting six-qubit quantum degenerate
codes, the CSS seven-qubit stabilizer code, the Shor nine-qubit stabilizer
code, the Gottesman\textbf{ }$2$\textbf{-}error correcting eleven-qubit
code\textbf{, }$\left[  \left[  4\text{, }2\text{, }2\right]  \right]
$\textbf{\ }stabilizer codes, and, finally, the Gottesman\textbf{ }$\left[
\left[  8\text{, }3\text{, }3\right]  \right]  $\textbf{\ }stabilizer code.

\subsubsection{Step one}

Recall that the stabilizer $\mathcal{S}_{\text{b}}^{\text{Leung}}$ of the
Leung et \textit{al}. $\left[  \left[  4\text{, }1\right]  \right]  $ code is
given by \cite{fletcher},%
\begin{equation}
\mathcal{S}_{\text{b}}^{\text{Leung}}\overset{\text{def}}{=}\left\langle
X^{1}X^{2}X^{3}X^{4}\text{, }Z^{1}Z^{2}\text{, }Z^{3}Z^{4}\right\rangle
\text{,}%
\end{equation}
with a suitable logical $\bar{Z}$ operation given by $\bar{Z}=Z^{1}Z^{3}$.
Therefore, when regarded within the CWS framework \cite{cross}, the Leung et
\textit{al}. code is equivalent to a CWS\ code defined with codeword
stabilizer,%
\begin{equation}
\mathcal{S}_{\text{CWS}}^{\text{Leung}}\overset{\text{def}}{=}\left\langle
X^{1}X^{2}X^{3}X^{4}\text{, }Z^{1}Z^{2}\text{, }Z^{3}Z^{4}\text{, }Z^{1}%
Z^{3}\right\rangle \text{.} \label{scws}%
\end{equation}

\subsubsection{Step two}

Taking into consideration Eq. (\ref{scws}), we observe that $\mathcal{S}%
_{\text{Leung}}^{\text{CWS}}$ is local Clifford equivalent to $S_{\text{Leung}%
}^{\prime\text{CWS}}$ given by,%
\begin{equation}
\mathcal{S}_{\text{CWS}}^{\prime\text{Leung}}\overset{\text{def}}%
{=}U\mathcal{S}_{\text{CWS}}^{\text{Leung}}U^{\dagger}\text{,}%
\end{equation}
with $U\overset{\text{def}}{=}I^{1}\otimes H^{2}\otimes H^{3}\otimes H^{4}$
where $H$ denotes the Hadamard transformation. We notice that the codeword
stabilizer matrix $\mathcal{H}_{\mathcal{S}_{\text{CWS}}^{\prime\text{Leung}}%
}$ associated with the codeword stabilizer $\mathcal{S}_{\text{CWS}}%
^{\prime\text{Leung}}$ reads,%
\begin{equation}
\mathcal{H}_{\mathcal{S}_{\text{CWS}}^{\prime\text{Leung}}}\overset
{\text{def}}{=}\left(  Z^{\prime}\left\vert X^{\prime}\right.  \right)
=\left(
\begin{array}
[c]{cccc}%
0 & 1 & 1 & 1\\
1 & 0 & 0 & 0\\
0 & 0 & 0 & 0\\
1 & 0 & 0 & 0
\end{array}
\left\vert
\begin{array}
[c]{cccc}%
1 & 0 & 0 & 0\\
0 & 1 & 0 & 0\\
0 & 0 & 1 & 1\\
0 & 0 & 1 & 0
\end{array}
\right.  \right)  \text{,}%
\end{equation}
with $\det X^{\prime}\neq0$. Therefore, we can find a suitable graph with
output vertices only that is associated with the Leung et \textit{al}. code by
applying the VdN algorithmic procedure. The transpose of $\mathcal{H}%
_{\mathcal{S}_{\text{CWS}}^{\prime\text{Leung}}}$ becomes,%
\begin{equation}
\mathcal{T}^{\prime}\overset{\text{def}}{=}\mathcal{H}_{\mathcal{S}%
_{\text{CWS}}^{\prime\text{Leung}}}^{\text{T}}\equiv\binom{A^{\prime}%
}{B^{\prime}}=\left(
\begin{array}
[c]{c}%
\underline{%
\begin{array}
[c]{cccc}%
0 & 1 & 0 & 1\\
1 & 0 & 0 & 0\\
1 & 0 & 0 & 0\\
1 & 0 & 0 & 0
\end{array}
}\\%
\begin{array}
[c]{cccc}%
1 & 0 & 0 & 0\\
0 & 1 & 0 & 0\\
0 & 0 & 1 & 1\\
0 & 0 & 1 & 0
\end{array}
\end{array}
\right)  \text{.} \label{bprimo}%
\end{equation}
From Eq. (\ref{bprimo}) it turns out that $B^{\prime}$ is a $4\times4$
invertible matrix with inverse given by,%
\begin{equation}
B^{\prime-1}=\left(
\begin{array}
[c]{cccc}%
1 & 0 & 0 & 0\\
0 & 1 & 0 & 0\\
0 & 0 & 0 & 1\\
0 & 0 & 1 & 1
\end{array}
\right)  \text{.}%
\end{equation}
Finally, the adjacency matrix $\Gamma$ of a graph that realizes the Leung et
\textit{al}. code is given by $\Gamma=A^{\prime}B^{\prime-1}$, that is%
\begin{equation}
\Gamma=A^{\prime}B^{\prime-1}=\left(
\begin{array}
[c]{cccc}%
0 & 1 & 0 & 1\\
1 & 0 & 0 & 0\\
1 & 0 & 0 & 0\\
1 & 0 & 0 & 0
\end{array}
\right)  \left(
\begin{array}
[c]{cccc}%
1 & 0 & 0 & 0\\
0 & 1 & 0 & 0\\
0 & 0 & 0 & 1\\
0 & 0 & 1 & 1
\end{array}
\right)  =\left(
\begin{array}
[c]{cccc}%
0 & 1 & 1 & 1\\
1 & 0 & 0 & 0\\
1 & 0 & 0 & 0\\
1 & 0 & 0 & 0
\end{array}
\right)  \overset{\text{def}}{=}\Gamma_{\text{Leung}}\text{.} \label{gl}%
\end{equation}
As a side remark, we recall that a graph determines uniquely a graph state and
two graph states determined by two graphs are equivalent up to some\emph{\ }%
local Clifford transformations if and only if these two graphs are related to
each other via local complementations\emph{\ }(LC) \cite{bart}. Avoiding
unnecessary formalities, we recall that a local complementation of a graph on
a vertex $v$ can be regarded as the the operation where in the neighborhood of
$v$ we connect all the disconnected vertices and disconnect all the connected
vertices. For instance, applying a local complementation on vertex $v=1$ on
the graph with adjacency matrix $\Gamma$ in Eq. (\ref{gl}), we obtain%
\begin{equation}
\Gamma_{\text{Leung}}\overset{\text{def}}{=}\left(
\begin{array}
[c]{cccc}%
0 & 1 & 1 & 1\\
1 & 0 & 0 & 0\\
1 & 0 & 0 & 0\\
1 & 0 & 0 & 0
\end{array}
\right)  \overset{\text{LC}_{v=1}}{\longrightarrow}\Gamma_{\text{Leung}%
}^{\prime}\overset{\text{def}}{=}\left(
\begin{array}
[c]{cccc}%
0 & 1 & 1 & 1\\
1 & 0 & 1 & 1\\
1 & 1 & 0 & 1\\
1 & 1 & 1 & 0
\end{array}
\right)  \text{.} \label{tt}%
\end{equation}
It turns out that $\Gamma_{\text{Leung}}$ and $\Gamma_{\text{Leung}}^{\prime}$
are the only two adjacency matrices corresponding to the only two connected
graphs, up to graph isomorphisms, that realize\textbf{\ }the Leung et
\textit{al}. $\left[  \left[  4\text{, }1\right]  \right]  $ code. As a matter
of fact, recall that the LC orbit $\mathbf{L=}\left[  G\right]  $ of a graph
$G$ is the set of all non-isomorphic graphs, including $G$ itself, that can be
transformed into $G$ by any sequence of local complementations and vertex
permutations. Let $\mathcal{G}_{n}$ denote the set of all non-isomorphic
simple unidirected connected graphs on $n$ vertices. Let $\mathcal{L}%
_{n}\overset{\text{def}}{=}\left\{  \mathbf{L}_{1}\text{,..., }\mathbf{L}%
_{k}\right\}  $ be the set of all distinct orbits of graphs in $\mathcal{G}%
_{n}$. All $\mathbf{L}\in\mathcal{L}_{n}$ are disjoint and $\mathcal{L}_{n}$
constitutes a partitioning of $\mathcal{G}_{n}$, that is to say%
\begin{equation}
\mathcal{G}_{n}\overset{\text{def}}{=}%
{\displaystyle\bigcup\limits_{i=1}^{k}}
\mathbf{L}_{i}\text{.}%
\end{equation}
Two graphs, $G_{1}$ and $G_{2}$, are equivalent with respect to local
complementations and vertex permutations if one of the graphs is in the LC
orbit of the other, for instance $G_{2}\in\left[  G_{1}\right]  $. In
\cite{danielsen1}, the set $\mathcal{L}_{4}$ of all LC orbits on $4$ vertices
was generated. It was shown that despite the fact that there are $2^{\binom
{4}{2}}=64$ unidirected simple graphs on $4$ vertices, the number of
non-isomorphic connected graphs is only $\left\vert \mathcal{G}_{4}\right\vert
=6$. Furthermore, it was shown that there are only $\left\vert \mathcal{L}%
_{4}\right\vert =2$ distinct LC orbits on $4$ vertices, $\mathcal{L}%
_{4}=\left\{  \mathbf{L}_{1}\text{, }\mathbf{L}_{2}\right\}  $ with,%
\begin{equation}
\Gamma_{\mathbf{L}_{1}}^{\left(  1\right)  }\overset{\text{def}}{=}\left(
\begin{array}
[c]{cccc}%
0 & 1 & 1 & 1\\
1 & 0 & 0 & 1\\
1 & 0 & 0 & 1\\
1 & 1 & 1 & 0
\end{array}
\right)  \text{, }\Gamma_{\mathbf{L}_{1}}^{\left(  2\right)  }\overset
{\text{def}}{=}\left(
\begin{array}
[c]{cccc}%
0 & 1 & 1 & 1\\
1 & 0 & 1 & 0\\
1 & 1 & 0 & 0\\
1 & 0 & 0 & 0
\end{array}
\right)  \text{, }\Gamma_{\mathbf{L}_{1}}^{\left(  3\right)  }\overset
{\text{def}}{=}\left(
\begin{array}
[c]{cccc}%
0 & 1 & 0 & 1\\
1 & 0 & 1 & 0\\
0 & 1 & 0 & 0\\
1 & 0 & 0 & 0
\end{array}
\right)  \text{, }\Gamma_{\mathbf{L}_{1}}^{\left(  4\right)  }\overset
{\text{def}}{=}\left(
\begin{array}
[c]{cccc}%
0 & 1 & 0 & 1\\
1 & 0 & 1 & 0\\
0 & 1 & 0 & 1\\
1 & 0 & 1 & 0
\end{array}
\right)  \text{, }%
\end{equation}
and,%
\begin{equation}
\Gamma_{\mathbf{L}_{2}}^{\left(  5\right)  }\overset{\text{def}}{=}\left(
\begin{array}
[c]{cccc}%
0 & 1 & 1 & 1\\
1 & 0 & 1 & 1\\
1 & 1 & 0 & 1\\
1 & 1 & 1 & 0
\end{array}
\right)  \text{, }\Gamma_{\mathbf{L}_{2}}^{\left(  6\right)  }\overset
{\text{def}}{=}\left(
\begin{array}
[c]{cccc}%
0 & 1 & 1 & 1\\
1 & 0 & 0 & 0\\
1 & 0 & 0 & 0\\
1 & 0 & 0 & 0
\end{array}
\right)  \text{. }%
\end{equation}
We stress that $\Gamma_{\mathbf{L}_{2}}^{\left(  5\right)  }$ and
$\Gamma_{\mathbf{L}_{2}}^{\left(  6\right)  }$ correspond to $\Gamma
_{\text{Leung}}^{\prime}$ and $\Gamma_{\text{Leung}}$, respectively.
Therefore, we have uncovered that the Leung et \textit{al}. $\left[  \left[
4\text{, }1\right]  \right]  $ code can be realized by graphs that belong to
the orbit $\mathbf{L}_{2}$ of $\mathcal{L}_{4}$ in $\mathcal{G}_{4}%
=\mathbf{L}_{1}\cup\mathbf{L}_{2}$, the set of all non-isomorphic unidirected
connected graphs on $4$ vertices.

For the sake of completeness, we also point out that all graphs on up to $12$
vertices have been classified under LCs and graph isomorphisms \cite{parker}.
Furthermore, the number of graphs on $n$ unlabeled vertices or the number of
connected graphs with $n$ vertices can be found in \cite{sloane}. Finally, a
very recent database of interesting graphs appears in \cite{dam}.

\subsubsection{Step three}

Let us consider the symmetric adjacency matrix $\Gamma_{\text{Leung}}$ as
given in Eq. (\ref{tt}). How do we find the enlarged \textbf{graph }with
corresponding symmetric coincidence matrix $\Xi_{\text{Leung}}$ given
$\Gamma_{\text{Leung}}?$ Recall that the graph related to $\Gamma
_{\text{Leung}}$ realizes a stabilizer code which is locally Clifford
equivalent to the Leung et \textit{al}. code with standard binary stabilizer
matrix $\mathcal{S}_{\text{b}}^{\prime}$ given by%
\begin{equation}
\mathcal{S}_{\text{b}}^{\prime}\overset{\text{def}}{=}\left\langle X^{1}%
Z^{2}Z^{3}Z^{4}\text{, }Z^{1}X^{2}\text{, }X^{3}X^{4}\right\rangle \text{.}%
\end{equation}
Putting $g_{1}\overset{\text{def}}{=}$ $X^{1}Z^{2}Z^{3}Z^{4}$, $g_{2}%
\overset{\text{def}}{=}Z^{1}X^{2}$ and $g_{3}\overset{\text{def}}{=}X^{3}%
X^{4}$, we have%
\begin{equation}
\mathcal{S}_{b}^{\prime}=\left\langle g_{1}\text{, }g_{2}\text{, }%
g_{3}\right\rangle =\left\{  I\text{, }g_{1}\text{, }g_{2}\text{, }%
g_{3}\text{, }g_{1}g_{2}\text{, }g_{1}g_{3}\text{, }g_{2}g_{3}\text{, }%
g_{1}g_{2}g_{3}\right\}  \text{.} \label{sbp}%
\end{equation}
The $8$-dimensional binary vector representation of these\textbf{\ }stabilizer
operators is given by,%
\begin{align}
I  &  \leftrightarrow v_{I}=\left(  0000\left\vert 0000\right.  \right)
\text{, }g_{1}\leftrightarrow v_{g_{1}}=\left(  0111\left\vert 1000\right.
\right)  \text{, }g_{2}\leftrightarrow v_{g_{2}}=\left(  1000\left\vert
0100\right.  \right)  \text{,}\nonumber\\
& \nonumber\\
\text{ }g_{3}  &  \leftrightarrow v_{g_{3}}=\left(  0000\left\vert
0011\right.  \right)  \text{, }g_{1}g_{2}\leftrightarrow v_{g_{1}g_{2}%
}=\left(  1111\left\vert 1100\right.  \right)  \text{, }g_{1}g_{3}%
\leftrightarrow v_{g_{1}g_{3}}=\left(  0111\left\vert 1011\right.  \right)
\text{,}\nonumber\\
& \nonumber\\
\text{ }g_{2}g_{3}  &  \leftrightarrow v_{g_{2}g_{3}}=\left(  1000\left\vert
0111\right.  \right)  \text{, }g_{1}g_{2}g_{3}\leftrightarrow v_{g_{1}%
g_{2}g_{3}}=\left(  1111\left\vert 1111\right.  \right)  \text{.}%
\end{align}
Recall that for a graph code with both $1$-input and $n$-output vertices, its
corresponding coincidence matrix $\Xi_{\left(  n+1\right)  \times\left(
n+1\right)  }$ has the form expressed in Eq. (\ref{losai-2}). The graph code
with symmetric coincidence matrix $\Xi_{\left(  n+1\right)  \times\left(
n+1\right)  }$ is equivalent to stabilizer codes being associated with the
isotropic subspace $\mathcal{S}_{\text{isotropic}}$ defined as,%
\begin{equation}
\mathcal{S}_{\text{isotropic}}\overset{\text{def}}{=}\left\{  \left(
Ak\left\vert k\right.  \right)  :k\in\ker B^{\dagger}\right\}  \text{,}%
\end{equation}
that is, omitting unimportant phase factors, with the binary stabilizer group
$\mathcal{S}_{b}$,%
\begin{equation}
\mathcal{S}_{b}\overset{\text{def}}{=}\left\{  g_{k}=X^{k}Z^{Ak}:k\in\ker
B^{\dagger}\right\}  \text{.}%
\end{equation}
In our case,\ in agreement with the four conditions for attaching input
vertices as outlined in the S-work paragraph, it turns out that%
\begin{equation}
B_{4\times1}\overset{\text{def}}{=}\left(
\begin{array}
[c]{c}%
0\\
0\\
1\\
1
\end{array}
\right)  \text{and, }A\overset{\text{def}}{=}\left(
\begin{array}
[c]{cccc}%
0 & 1 & 1 & 1\\
1 & 0 & 0 & 0\\
1 & 0 & 0 & 0\\
1 & 0 & 0 & 0
\end{array}
\right)  \equiv\Gamma_{\text{Leung}}\text{.} \label{b}%
\end{equation}
Finally, the enlarged graph is defined by the following symmetric coincidence
matrix $\Xi_{\text{Leung}}$,%
\begin{equation}
\Xi_{\text{Leung}}\overset{\text{def}}{=}\left(
\begin{array}
[c]{ccccc}%
0 & 0 & 0 & 1 & 1\\
0 & 0 & 1 & 1 & 1\\
0 & 1 & 0 & 0 & 0\\
1 & 1 & 0 & 0 & 0\\
1 & 1 & 0 & 0 & 0
\end{array}
\right)  \text{.} \label{glb}%
\end{equation}
An additional self-consistency check that substantiates the correctness
of\textbf{\ }$\Xi_{\text{Leung}}$\textbf{\ }in Eq. (\ref{glb}) is represented
by the fact that any $g_{k}$ in $\mathcal{S}_{b}^{\prime}$ in Eq. (\ref{sbp})
has a $8$-dimensional binary vector representation of the form $v_{g_{k}}$
with $v_{g_{k}}\overset{\text{def}}{=}\left(  \Gamma_{\text{Leung}}k_{g_{k}%
}\left\vert k_{g_{k}}\right.  \right)  $ with $k_{g_{k}}\in\ker B^{\dagger}$
with $B$ given in Eq. (\ref{b}).

\section{Final remarks}

In this article, we proposed a systematic scheme for the construction of
graphs\textbf{ }with both input and output vertices associated with arbitrary
binary stabilizer codes. The scheme is characterized by three main steps:
first, the stabilizer code is realized as a codeword-stabilized (CWS) quantum
code; second, the canonical form of the CWS\ code is uncovered; third, the
input vertices are attached to the graphs. To check the effectiveness of the
scheme, we discussed several graphical constructions of various useful
stabilizer codes characterized by single and multi-qubit encoding operators
(for details, see appendices). In particular, the error correction
capabilities of such quantum codes are verified in graph-theoretic terms as
originally advocated by Schlingemann and Werner (for details, see appendices).

Finally, in what follows, possible generalizations of our scheme for the
graphical construction of both (stabilizer and nonadditive) nonbinary and
continuous-variable quantum codes will be briefly addressed.

The scheme proposed is limited to binary stabilizer codes. How about nonbinary
and continuous-variable (CV) codes? How about nonadditive codes? We point out
the following three points:

\medskip

\begin{itemize}
\item \emph{From additive to nonadditive case}. The codeword-stabilized
quantum code formalism presents a unifying approach to both additive and
nonadditive quantum error-correcting codes, for both binary and nonbinary
states \cite{chen}.

\item \emph{From binary to nonbinary case}. The stabilizer formalism, graph
states and quantum error correcting codes for $d$-dimensional quantum systems
have been extensively considered in \cite{dirk2} and \cite{dirk3}. However, as
pointed out in \cite{hein2}, no straightforward extension of the stabilizer
formalism in terms of generators within the Pauli group is possible for
$d$-level systems. As a consequence, it is possible that results obtained
within the binary framework are no longer valid when taking into consideration
weighted graph states. The generalizations of the Pauli group, the Clifford
group, and the stabilizer states for qudits in a Hilbert space of arbitrary
dimension $d$ appears in \cite{bart2}. When moving into the nonbinary case,
new features emerge. For instance, the symmetric adjacency matrix does not
contain any longer binary entries as in the case of a simple graph as in qubit
systems. The generalization of the Pauli operators, the so-called Weyl
operators, are no longer Hermitian. The finite field $\mathbf{F}_{2}$ is
replaced by the finite field of prime order $d$ and all arithmetic operations
are defined modulo $d$. The dimension $d$ can be naturally generalized to
prime power dimension $d=p^{r}$ with $p$ being prime and $r$ being an integer.
If, however, the underlying integer ring is no longer a field, one loses the
vector space structure of $\mathbf{F}_{d}$, which demands some caution with
respect to the concept of a basis. If $d$ contains multiple prime factors, the
stabilizer, consisting of $d^{N}$ different elements, is in general no longer
generated by a set of only $N$ generators. For the minimal generating set,
more elements $N\leq m\leq2N$ of the stabilizer might be needed as pointed out
in \cite{bart2}. Furthermore, it is possible to show that the action of the
local (generalized) Clifford group on nonbinary stabilizer states can be
translated into operations on graphs. However, unlike the binary case, the
single local complementation is replaced by a pair of two different
graph-theoretic operations. Furthermore, an efficient polynomial time
algorithm to verify whether two graph states, in the non-binary case, are
locally Clifford equivalent is available \cite{beigi2}. Despite these
challenges, new important advances have been recently achieved. For instance,
an explicit method of going from qudit CSS codes to qudit graph codes,
including all the encoding and decoding procedures, has been presented in
\cite{anne}.

\item \emph{From discrete to continuous case}. A remarkable difference between
discrete and continuous variables (DV and CV, respectively) quantum
information is that while quantum states and unitary transformations involved
are described by integer-valued parameters in the DV case, they are
characterized by \emph{real}-valued parameters in the CV case. The
continuous-variable analog of the Pauli and Clifford algebras and groups
together with sets of gates that can efficiently simulate any arbitrary
unitary transformation in these groups were defined in \cite{sam2}. The
standard Pauli group for CV quantum computation on $n$ coupled oscillator is
the Heisenberg-Weyl group $\mathcal{HW}\left(  n\right)  $ which consists of
phase-space displacement operators for the $n$ oscillators. Unlike the
discrete Pauli group for qubits, the group $\mathcal{HW}\left(  n\right)  $ is
a continuous Lie group, and can therefore only be generalized by a set of
continuously parametrized operators. Furthermore, the Clifford group for CV is
the semidirect product group of the symplectic group and Heisenberg-Weyl
group, $Sp\left(  2n\text{, }%
\mathbb{R}
\right)  \ltimes\mathcal{HW}\left(  n\right)  $, consisting of all phase-space
translations along with all one-mode and two-mode squeezing transformations
\cite{sam2}. This group is generated by inhomogeneous quadratic polynomials in
the canonical operators. For DV, it is possible to generate the Clifford group
using only the CNOT, Hadamard and phase gates. However, in the CV\ case, the
analog of these gates (namely, the $SUM$, the Fourier $F$ and the phase
$P\left(  \eta\right)  $ gates with $\eta\in%
\mathbb{R}
$) are all elements of $Sp\left(  2n\text{, }%
\mathbb{R}
\right)  $. They are generated by homogeneous quadratic Hamiltonians only.
Thus, they are in the subgroup of the Clifford group. In order to generate the
entire Clifford group, one requires a continuous $\mathcal{HW}\left(
1\right)  $ transformation (i.e., a linear Hamiltonian that generates a
one-parameter subgroup of $\mathcal{HW}\left(  1\right)  $) such as the Pauli
operator $X\left(  q\right)  $ with $q\in%
\mathbb{R}
$. Finally, the Clifford group in the CV case is generated by the set
$\left\{  SUM\text{, }F\text{, }P\left(  \eta\right)  \text{, }X\left(
q\right)  :\eta\text{, }q\in%
\mathbb{R}
\right\}  $. Continuous-variable graph states were proposed in \cite{zang2,
peter1}. It is of great relevance understanding the graph-theoretic
transformation rules that describe both local unitary and local Clifford
unitary equivalences of arbitrary CV graph states. For a particular class of
CV graph states, the so-called CV four-mode unweighted graph states, such
transformation rules have been uncovered in \cite{zang3}. It turns out that
even for such restricted class of states, the corresponding local Clifford
unitary cannot exactly mirror that for the qubit case and a greater level of
complexity arises in the CV framework. In addition, the complete
implementation of local complementations for CV weighted graphs (a weighted
graph state is described by a graph $G=\left(  V\text{, }E\right)  $ in which
every edge is specified by a factor $\Omega_{ab}$ corresponding to the
strength of modes $a$ and $b$; for unweighted graph states, all the
interactions have the same strength) remains an open problem. In \cite{zang4},
the graphical description of local Clifford transformations for CV weighted
graph states were considered. In particular, it was shown that unlike qubit
weighted graph states, CV weighted graph states can be expressed by the
stabilizer formalism in terms of generators in the Pauli group. The main
reason for this difference is that the CZ gate for qubit is periodic as a
function of the interaction strength while the CV CZ gate is not.\textbf{ }We
remark that in this context, the CV case is even more subtle, besides the fact
that weighted CV graph states are still stabilizer states unlike weighted
qubit graph states. In particular, the most general form of weighted CV graph
states has a complex adjacency matrix. In fact, all real-valued (with real
adjacency matrix) CV graph states (weighted or unweighted) are unphysical
states (only defined in the limit of infinite squeezing). In order to
represent physical CV graph states, corresponding to pure multi-mode Gaussian
states, the weights become necessarily complex. All this is introduced and
discussed in \cite{peter2} where it is also described how such general,
physical CV graph states transform under local and general Gaussian
transformations. In particular, we emphasize that the general results
presented in \cite{peter2} include Zhang's results in \cite{zang3, zang4} as
the limiting cases of infinite squeezing and real-weighted states. We recall
that in the qubit-case a systematic classification of local Clifford
equivalence of qubit graph states has been executed and an efficient algorithm
with polynomial time complexity in the number of qubits to decide whether two
given stabilizer states are local Clifford equivalent is known. In the
CV\ framework, it can be proved that any CV stabilizer state is equivalent to
a weighted graph state under local Clifford operations, the equivalence
between two stabilizer states under local Clifford operations can be
investigated by studying the equivalence between weighted graph states under
local Clifford operations \cite{zang5}. However, the existence of a universal
method to determine whether two CV stabilizer states with finite modes are
equivalent or not under local Clifford operation has been only partially
addressed in \cite{peter2}\textbf{. }In the CV case, the local-Clifford
equivalence of stabilizer states translates into local-Gaussian unitary
equivalence of (pure) Gaussian states. Furthermore, while a single unifying
definition of complex-weighted CV\ graph states (Gaussian pure states)
together with graph transformation rules for all local Gaussian unitary
operations were presented in \cite{peter2}, no systematic algorithm for
deciding on the local equivalence of two given CV (Gaussian) stabilizer states
was discussed. This issue, however, was recently addressed in \cite{giedke}.
Specifically, necessary and sufficient conditions of Gaussian local unitary
equivalence for arbitrary (mixed or pure) Gaussian states were derived.
Despite such advances, several questions remain to be better understood. For
instance, the relation between local equivalence of CV Gaussian states and
Gaussian local equivalence deserves further investigation \cite{giedke,
fiurasek}. A thorough analysis of this type of questions is not only important
from a theoretical point of view, it can also be of practical use concerning
which states are the most suitable for optical realizations of stabilizer
quantum error correction codes in any dimension \cite{peter-damian}.
\end{itemize}

\smallskip

In view of these considerations, we conclude that the extension of our
proposed scheme to arbitrary nonbinary/CV codes and/or additive/nonadditive
codes might turn out to be nontrivial. However, in light of the recent
advances, we are confident that its generalization could be achieved with a
reasonable effort.\medskip

\begin{acknowledgments}
We thank the ERA-Net CHIST-ERA project HIPERCOM for financial support.
\end{acknowledgments}

\bigskip\pagebreak

\appendix

\section{Single qubit encoding}

Before presenting our illustrative examples, we would like to make few remarks
on graphs in quantum error correction.

The coincidence matrix of a graph characterizes the structural properties of a
graph: number of vertices, number of edges, and, above all, the manner in
which vertices are connected. The structure of graphs associated with
stabilizer quantum codes hides essential information about the graphical error
detection conditions in Eqs. (\ref{wc1}) and (\ref{wc2}). Such graphical
conditions may not be necessarily visible in a direct manner as originally
pointed out in \cite{dirk}. This becomes especially evident when the number of
vertices and edges in the graph increases in the presence of multi-qubit
encodings and/or big code lengths. However, graphs do maintain part of their
appeal in that they provide a\textbf{ }\emph{geometric}\textbf{ }aid in
identifying the explicit\textbf{ }\emph{algebraic}\textbf{ }linear equations
that characterize the graphical error detection conditions without taking into
consideration the explicit form of their corresponding coincidence matrices.
In our opinion, this is no negligible advantage of our graphical approach
since identifying the algebraic equations directly from the coincidence
matrices can become quite tedious without a visual aid provided by graphs.
Clearly, the peculiar advantage of our scheme is that it allows to uncover the
expression of the coincidence matrix of a graph associated with a binary
stabilizer code. We shall further discuss some of these aspects in our
illustrative examples that appear below.

As an additional side remark, we point out that there could be scenarios where
one can exploit the high symmetry of the graph in an efficient manner in order
to check the graphical conditions for error detection \cite{werner}. While
symmetry arguments are elegant and powerful, they require some caution in the
case of graphs in quantum error correction: symmetries of graphs are not
necessarily the same as symmetries of the associated stabilizer codes
\cite{dirk, markus, tqc}. For instance, graphs with different symmetries can
lead to a class of codes that are equivalent to the CSS seven-qubit code as
shown in Ref. \cite{markus}. As recently pointed out in \cite{tqc}, a clear
understanding of the requirements under which a graph can exhibit the same
symmetry as the quantum (CWS, in general) code is still missing. In this
article, we do not address this issue. However, in agreement with the
statement appeared in Ref. \cite{tqc}, we do think that this point is
definitively worth further attention.

\subsection{The $\left[  \left[  3,1,1\right]  \right]  $ stabilizer code}

Before applying our scheme for the construction of the graph\textbf{
}associated with a $\left[  \left[  3,1,1\right]  \right]  $ stabilizer code
\cite{gaitan}, we emphasize how intricate can be finding the explicit
expression of unitary transformations that relate sets of vertex stabilizers
of graphs. For the sake of reasoning, consider the following sets
$\mathcal{S}_{\left\vert \Gamma_{1}\right\rangle }$, $\mathcal{S}_{\left\vert
\Gamma_{2}\right\rangle }$ and $\mathcal{S}_{\left\vert \Gamma_{3}%
\right\rangle }$ defined as
\begin{equation}
\mathcal{S}_{\left\vert \Gamma_{1}\right\rangle }\overset{\text{def}}%
{=}\left\langle X^{1}\text{, }X^{2}\text{, }X^{3}\right\rangle \text{,
}\mathcal{S}_{\left\vert \Gamma_{2}\right\rangle }\overset{\text{def}}%
{=}\left\langle X^{1}Z^{2}Z^{3}\text{, }Z^{1}X^{2}\text{, }Z^{1}%
X^{3}\right\rangle \text{ and, }\mathcal{S}_{\left\vert \Gamma_{3}%
\right\rangle }\overset{\text{def}}{=}\left\langle X^{1}Z^{2}Z^{3}\text{,
}Z^{1}X^{2}Z^{3}\text{, }Z^{1}Z^{2}X^{3}\right\rangle \text{,} \label{A1}%
\end{equation}
respectively. In the canonical basis $\mathcal{B}_{\mathcal{H}_{2}^{3}}$ of
the eight-dimensional \emph{complex} Hilbert space $\mathcal{H}_{2}^{3}$,%
\begin{equation}
\mathcal{B}_{\mathcal{H}_{2}^{3}}\overset{\text{def}}{=}\left\{  \left\vert
000\right\rangle \text{, }\left\vert 001\right\rangle \text{, }\left\vert
010\right\rangle \text{, }\left\vert 011\right\rangle \text{, }\left\vert
100\right\rangle \text{, }\left\vert 101\right\rangle \text{, }\left\vert
110\right\rangle \text{, }\left\vert 111\right\rangle \right\}  \text{,}%
\end{equation}
the graph states $\left\vert \Gamma_{1}\right\rangle $, $\left\vert \Gamma
_{2}\right\rangle $ and $\left\vert \Gamma_{3}\right\rangle $ read,%
\begin{align}
&  \left\vert \Gamma_{1}\right\rangle \overset{\text{def}}{=}\frac{\left\vert
000\right\rangle +\left\vert 001\right\rangle +\left\vert 010\right\rangle
+\left\vert 011\right\rangle +\left\vert 100\right\rangle +\left\vert
101\right\rangle +\left\vert 110\right\rangle +\left\vert 111\right\rangle
}{\sqrt{8}}\text{, }\nonumber\\
&  \text{ }\left\vert \Gamma_{2}\right\rangle \overset{\text{def}}{=}%
\frac{\left\vert 000\right\rangle +\left\vert 001\right\rangle +\left\vert
010\right\rangle +\left\vert 011\right\rangle +\left\vert 100\right\rangle
-\left\vert 101\right\rangle -\left\vert 110\right\rangle +\left\vert
111\right\rangle }{\sqrt{8}}\text{, }\nonumber\\
&  \left\vert \Gamma_{3}\right\rangle \overset{\text{def}}{=}\frac{\left\vert
000\right\rangle +\left\vert 001\right\rangle +\left\vert 010\right\rangle
-\left\vert 011\right\rangle +\left\vert 100\right\rangle -\left\vert
101\right\rangle -\left\vert 110\right\rangle -\left\vert 111\right\rangle
}{\sqrt{8}}\text{.} \label{gammaeq}%
\end{align}
We observe that,%
\begin{equation}
\mathcal{S}_{\left\vert \Gamma_{3}\right\rangle }=\mathcal{U}_{\left\vert
\Gamma_{1}\right\rangle \rightarrow\left\vert \Gamma_{3}\right\rangle
}\mathcal{S}_{\left\vert \Gamma_{1}\right\rangle }\mathcal{U}_{\left\vert
\Gamma_{1}\right\rangle \rightarrow\left\vert \Gamma_{3}\right\rangle
}^{\dagger}\text{,} \label{01}%
\end{equation}
with,%
\begin{equation}
\mathcal{U}_{\left\vert \Gamma_{1}\right\rangle \rightarrow\left\vert
\Gamma_{3}\right\rangle }\overset{\text{def}}{=}\left(  I^{1}\otimes
I^{2}\otimes H^{3}\right)  \cdot\left(  U_{CP}^{12}\otimes I^{3}\right)
\left(  I^{1}\otimes H^{2}\otimes I^{3}\right)  \cdot\left(  I^{1}\otimes
U_{CP}^{23}\right)  \cdot\left(  U_{CP}^{12}\otimes I^{3}\right)  \text{.}
\label{u1}%
\end{equation}
Similarly, it can be shown that%
\begin{equation}
\mathcal{S}_{\left\vert \Gamma_{2}\right\rangle }=\mathcal{U}_{\left\vert
\Gamma_{1}\right\rangle \rightarrow\left\vert \Gamma_{2}\right\rangle
}\mathcal{S}_{\left\vert \Gamma_{1}\right\rangle }\mathcal{U}_{\left\vert
\Gamma_{1}\right\rangle \rightarrow\left\vert \Gamma_{2}\right\rangle
}^{\dagger}\text{,} \label{02}%
\end{equation}
with,%
\begin{equation}
\mathcal{U}_{\left\vert \Gamma_{1}\right\rangle \rightarrow\left\vert
\Gamma_{2}\right\rangle }\overset{\text{def}}{=}\left(  H^{1}\otimes
I^{2}\otimes I^{3}\right)  \cdot\left(  I^{1}\otimes H^{2}\otimes
I^{3}\right)  \cdot\left(  I^{1}\otimes U_{CP}^{23}\right)  \cdot\left(
U_{CP}^{12}\otimes I^{3}\right)  \text{.} \label{u2}%
\end{equation}
\bigskip Finally, combining (\ref{01}) and (\ref{02}), we get%
\begin{equation}
\mathcal{S}_{\left\vert \Gamma_{3}\right\rangle }=\mathcal{U}_{\left\vert
\Gamma_{2}\right\rangle \rightarrow\left\vert \Gamma_{3}\right\rangle
}\mathcal{S}_{\left\vert \Gamma_{2}\right\rangle }\mathcal{U}_{\left\vert
\Gamma_{2}\right\rangle \rightarrow\left\vert \Gamma_{3}\right\rangle
}^{\dagger}=\mathcal{U}_{\left\vert \Gamma_{1}\right\rangle \rightarrow
\left\vert \Gamma_{3}\right\rangle }\mathcal{S}_{\left\vert \Gamma
_{1}\right\rangle }\mathcal{U}_{\left\vert \Gamma_{1}\right\rangle
\rightarrow\left\vert \Gamma_{3}\right\rangle }^{\dagger}=\mathcal{U}%
_{\left\vert \Gamma_{1}\right\rangle \rightarrow\left\vert \Gamma
_{3}\right\rangle }\mathcal{U}_{\left\vert \Gamma_{1}\right\rangle
\rightarrow\left\vert \Gamma_{2}\right\rangle }^{\dagger}\mathcal{S}%
_{\left\vert \Gamma_{2}\right\rangle }\mathcal{U}_{\left\vert \Gamma
_{1}\right\rangle \rightarrow\left\vert \Gamma_{2}\right\rangle }%
\mathcal{U}_{\left\vert \Gamma_{1}\right\rangle \rightarrow\left\vert
\Gamma_{3}\right\rangle }^{\dagger}\text{,}%
\end{equation}
that is,%
\begin{equation}
\mathcal{U}_{\left\vert \Gamma_{2}\right\rangle \rightarrow\left\vert
\Gamma_{3}\right\rangle }=\mathcal{U}_{\left\vert \Gamma_{1}\right\rangle
\rightarrow\left\vert \Gamma_{3}\right\rangle }\mathcal{U}_{\left\vert
\Gamma_{1}\right\rangle \rightarrow\left\vert \Gamma_{2}\right\rangle
}^{\dagger}\text{.} \label{u3}%
\end{equation}
After some algebra, we obtain that the explicit expressions for the Clifford
unitary matrices $\mathcal{U}_{\left\vert \Gamma_{1}\right\rangle
\rightarrow\left\vert \Gamma_{3}\right\rangle }$, $\mathcal{U}_{\left\vert
\Gamma_{1}\right\rangle \rightarrow\left\vert \Gamma_{2}\right\rangle }$, and
$\mathcal{U}_{\left\vert \Gamma_{2}\right\rangle \rightarrow\left\vert
\Gamma_{3}\right\rangle }$ become,%

\begin{align}
&  \mathcal{U}_{\left\vert \Gamma_{1}\right\rangle \rightarrow\left\vert
\Gamma_{3}\right\rangle }\overset{\text{def}}{=}\frac{1}{2}\left(
\begin{array}
[c]{cccccccc}%
1 & 1 & 1 & -1 & 0 & 0 & 0 & 0\\
1 & -1 & 1 & 1 & 0 & 0 & 0 & 0\\
1 & 1 & -1 & 1 & 0 & 0 & 0 & 0\\
1 & -1 & -1 & -1 & 0 & 0 & 0 & 0\\
0 & 0 & 0 & 0 & 1 & 1 & -1 & 1\\
0 & 0 & 0 & 0 & 1 & -1 & -1 & -1\\
0 & 0 & 0 & 0 & -1 & -1 & -1 & 1\\
0 & 0 & 0 & 0 & -1 & 1 & -1 & -1
\end{array}
\right)  \text{, }\mathcal{U}_{\left\vert \Gamma_{1}\right\rangle
\rightarrow\left\vert \Gamma_{2}\right\rangle }\overset{\text{def}}{=}\frac
{1}{2}\left(
\begin{array}
[c]{cccccccc}%
1 & 0 & 1 & 0 & 1 & 0 & -1 & 0\\
0 & 1 & 0 & -1 & 0 & 1 & 0 & 1\\
1 & 0 & -1 & 0 & 1 & 0 & 1 & 0\\
0 & 1 & 0 & 1 & 0 & 1 & 0 & -1\\
1 & 0 & 1 & 0 & -1 & 0 & 1 & 0\\
0 & 1 & 0 & -1 & 0 & -1 & 0 & -1\\
1 & 0 & -1 & 0 & -1 & 0 & -1 & 0\\
0 & 1 & 0 & 1 & 0 & -1 & 0 & 1
\end{array}
\right)  \text{,}\nonumber\\
&  \text{ }\nonumber\\
&  \mathcal{U}_{\left\vert \Gamma_{2}\right\rangle \rightarrow\left\vert
\Gamma_{3}\right\rangle }\overset{\text{def}}{=}\frac{1}{2}\left(
\begin{array}
[c]{cccccccc}%
1 & 1 & 0 & 0 & 1 & 1 & 0 & 0\\
1 & -1 & 0 & 0 & 1 & -1 & 0 & 0\\
0 & 0 & 1 & 1 & 0 & 0 & 1 & 1\\
0 & 0 & 1 & -1 & 0 & 0 & 1 & -1\\
1 & 1 & 0 & 0 & -1 & -1 & 0 & 0\\
1 & -1 & 0 & 0 & -1 & 1 & 0 & 0\\
0 & 0 & -1 & -1 & 0 & 0 & 1 & 1\\
0 & 0 & -1 & 1 & 0 & 0 & 1 & -1
\end{array}
\right)  \text{. }%
\end{align}
A systematic strategy for finding the explicit expressions for the unitary
transformations in Eqs. (\ref{u1}), (\ref{u2})\ and (\ref{u3}) would be very
useful. The VdN-work is especially important in this regard, as we shall see.

Let us consider the three-qubit bit-flip repetition code with codespace
spanned by the codewords $\left\vert 0_{L}\right\rangle \overset{\text{def}%
}{=}$ $\left\vert 000\right\rangle $ and $\left\vert 1_{L}\right\rangle
\overset{\text{def}}{=}$ $\left\vert 111\right\rangle $. The two stabilizer
generators of this code are $g_{1}\overset{\text{def}}{=}Z^{1}Z^{2}$ and
$g_{2}\overset{\text{def}}{=}Z^{1}Z^{3}$ while the logical operations can read
$\bar{Z}\overset{\text{def}}{=}Z^{1}Z^{2}Z^{3}$ and $\bar{X}\overset
{\text{def}}{=}X^{1}X^{2}X^{3}$ with,%
\begin{equation}
\bar{Z}\left\vert 0_{L}\right\rangle =\left\vert 0_{L}\right\rangle \text{,
}\bar{Z}\left\vert 1_{L}\right\rangle =-\left\vert 1_{L}\right\rangle \text{,
}\bar{X}\left\vert 0_{L}\right\rangle =\left\vert 1_{L}\right\rangle \text{
and, }\bar{X}\left\vert 1_{L}\right\rangle =\left\vert 0_{L}\right\rangle
\text{.}%
\end{equation}
This stabilizer code can be regarded as a CWS code with codeword stabilizer
given by,%
\begin{equation}
\mathcal{S}_{\text{CWS}}\overset{\text{def}}{=}\left\langle g_{1}\text{,
}g_{2}\text{, }\bar{Z}\right\rangle =\left\langle Z^{1}Z^{2}\text{, }%
Z^{1}Z^{3}\text{, }Z^{1}Z^{2}Z^{3}\right\rangle =\left\langle Z^{1}\text{,
}Z^{2}\text{, }Z^{3}\right\rangle \text{.}%
\end{equation}
Observe that $\mathcal{S}_{\text{CWS}}=U\mathcal{S}_{\left\vert \Gamma
_{1}\right\rangle }U^{\dagger}$ with $\mathcal{S}_{\left\vert \Gamma
_{1}\right\rangle }$ in Eq. (\ref{A1}) and $U\overset{\text{def}}{=}H^{1}%
H^{2}H^{3}$. Therefore, the graph state associated with $\mathcal{S}%
_{\text{CWS}}$ is locally Clifford equivalent to the graph state $\left\vert
\Gamma_{1}\right\rangle $ in Eq. (\ref{gammaeq}). Let us consider now an
alternative graphical description of the three-qubit bit-flip repetition code
that better fits into our scheme.

Let us consider the codespace of the code spanned by the following new
codewords,%
\begin{equation}
\mathcal{C}\overset{\text{def}}{=}\text{Span }\left\{  \left\vert
0_{L}\right\rangle \overset{\text{def}}{=}\left\vert 000\right\rangle \text{,
}\left\vert 1_{L}\right\rangle \overset{\text{def}}{=}\left\vert
111\right\rangle \right\}  \rightarrow\mathcal{C}^{\prime}\overset{\text{def}%
}{=}\text{Span }\left\{  \left\vert 0_{L}^{\prime}\right\rangle \overset
{\text{def}}{=}\frac{\left\vert 0_{L}\right\rangle +\left\vert 1_{L}%
\right\rangle }{\sqrt{2}}\text{, }\left\vert 1_{L}^{\prime}\right\rangle
\overset{\text{def}}{=}\frac{\left\vert 0_{L}\right\rangle -\left\vert
1_{L}\right\rangle }{\sqrt{2}}\right\}  \text{.}%
\end{equation}
Notice that the codespace of the code does not change since $\mathcal{C}%
^{\prime}=\mathcal{C}$ and we have simply chosen a different orthonormal basis
to describe the code. However, with this alternative choice, the new codeword
stabilizer reads%
\begin{equation}
\mathcal{S}_{\text{CWS}}^{\prime}\overset{\text{def}}{=}\left\langle
g_{1}\text{, }g_{2}\text{, }\bar{Z}^{\prime}\right\rangle =\left\langle
Z^{1}Z^{2}\text{, }Z^{1}Z^{3}\text{, }X^{1}X^{2}X^{3}\right\rangle \text{,}%
\end{equation}
and the remaining logical operation is given by $\bar{X}^{\prime}%
\overset{\text{def}}{=}Z^{1}Z^{2}Z^{3}$. Observe that $\mathcal{S}%
_{\text{CWS}}^{\prime}$ is locally Clifford equivalent to $\mathcal{S}%
_{\text{CWS}}^{^{\prime\prime}}$ with $\mathcal{S}_{\text{CWS}}^{^{\prime
\prime}}\overset{\text{def}}{=}U\mathcal{S}_{\text{CWS}}^{\prime}U$ and
$U\overset{\text{def}}{=}H^{2}H^{3}$. The codeword stabilizer $\mathcal{S}%
_{\text{CWS}}^{^{\prime\prime}}$ reads,%
\begin{equation}
\mathcal{S}_{\text{CWS}}^{^{\prime\prime}}=\left\langle Z^{1}X^{2}\text{,
}Z^{1}X^{3}\text{, }X^{1}Z^{2}Z^{3}\right\rangle \text{.}%
\end{equation}
Observe that $\mathcal{S}_{\text{CWS}}^{^{\prime\prime}}$ equals
$\mathcal{S}_{\left\vert \Gamma_{2}\right\rangle }$ in Eq. (\ref{A1}).
Therefore, the graph state associated with $\mathcal{S}_{\text{CWS}}%
^{\prime\prime}$ is $\left\vert \Gamma_{2}\right\rangle $ in Eq.
(\ref{gammaeq}). We notice that this is such a simple example that we really
do not need to apply our scheme. The adjacency matrix $\Gamma$ of the graph
associated with the CWS code with codeword stabilizer $\mathcal{S}%
_{\text{CWS}}^{^{\prime\prime}}$ reads,%
\begin{equation}
\Gamma=\left(
\begin{array}
[c]{ccc}%
0 & 1 & 1\\
1 & 0 & 0\\
1 & 0 & 0
\end{array}
\right)  \text{.} \label{gamma3}%
\end{equation}
However, acting with a local complementation on the vertex $1$ of the graph
with adjacency matrix $\Gamma$ in (\ref{gamma3}), we get%
\begin{equation}
\Gamma\rightarrow\Gamma^{\prime}=\left(
\begin{array}
[c]{ccc}%
0 & 1 & 1\\
1 & 0 & 1\\
1 & 1 & 0
\end{array}
\right)  \text{.} \label{gammaprime}%
\end{equation}
Furthermore, the new codeword stabilizer becomes $\left[  \mathcal{S}%
_{\text{CWS}}^{^{\prime\prime}}\right]  _{\text{new}}$,%
\begin{equation}
\mathcal{S}_{\text{CWS}}^{^{\prime\prime}}\rightarrow\left[  \mathcal{S}%
_{\text{CWS}}^{^{\prime\prime}}\right]  _{\text{new}}\overset{\text{def}}%
{=}\left\langle X^{1}Z^{2}Z^{3}\text{, }Z^{1}X^{2}Z^{3}\text{, }Z^{1}%
Z^{2}X^{3}\right\rangle \text{.}%
\end{equation}
Note that the graph state associated with $\left[  \mathcal{S}_{\text{CWS}%
}^{^{\prime\prime}}\right]  _{\text{new}}$ is $\left\vert \Gamma
_{3}\right\rangle $ in Eq. (\ref{gammaeq}). We also point out that following
the VdN-work, it turns out that the $6\times6$ local\textbf{\ }unitary
Clifford transformation $Q$ that links the six-dimensional binary vector
representation of the operators in $\mathcal{S}_{\text{CWS}}^{^{\prime\prime}%
}$ and $\left[  \mathcal{S}_{\text{CWS}}^{^{\prime\prime}}\right]
_{\text{new}}$ is given by,%
\begin{equation}
Q\overset{\text{def}}{=}\left(
\begin{array}
[c]{cccccc}%
1 & 0 & 0 & 0 & 0 & 0\\
0 & 1 & 0 & 0 & 1 & 0\\
0 & 0 & 1 & 0 & 0 & 1\\
1 & 0 & 0 & 1 & 0 & 0\\
0 & 0 & 0 & 0 & 1 & 0\\
0 & 0 & 0 & 0 & 0 & 1
\end{array}
\right)  \text{.} \label{q1}%
\end{equation}
For the sake of clarity, consider%
\begin{equation}
\left[  \mathcal{S}_{\text{CWS}}^{^{\prime\prime}}\right]  _{\text{new}%
}\overset{\text{def}}{=}\left\{  I\text{, }g_{1}^{\prime}\text{, }%
g_{2}^{\prime}\text{, }g_{3}^{\prime}\text{, }g_{1}^{\prime}g_{2}^{\prime
}\text{, }g_{1}^{\prime}g_{3}^{\prime}\text{, }g_{2}^{\prime}g_{3}^{\prime
}\text{, }g_{1}^{\prime}g_{2}^{\prime}g_{3}^{\prime}\right\}  \text{,}%
\end{equation}
with $g_{1}^{\prime}\overset{\text{def}}{=}X^{1}Z^{2}Z^{3}$, $g_{2}^{\prime
}\overset{\text{def}}{=}Z^{1}X^{2}Z^{3}$, $g_{3}^{\prime}\overset{\text{def}%
}{=}Z^{1}Z^{2}X^{3}$ and,%
\begin{align}
v_{I}  &  =\left(  000\left\vert 000\right.  \right)  \text{, }v_{g_{1}%
^{\prime}}=\left(  011\left\vert 100\right.  \right)  \text{, }v_{g_{2}%
^{\prime}}=\left(  101\left\vert 010\right.  \right)  \text{, }v_{g_{3}%
^{\prime}}=\left(  110\left\vert 001\right.  \right)  \text{, }v_{g_{1}%
^{\prime}g_{2}^{\prime}}=\left(  110\left\vert 110\right.  \right)  \text{,
}\nonumber\\
& \nonumber\\
v_{g_{1}^{\prime}g_{3}^{\prime}}  &  =\left(  101\left\vert 101\right.
\right)  \text{, }v_{g_{2}^{\prime}g_{3}^{\prime}}=\left(  011\left\vert
011\right.  \right)  \text{, }v_{g_{1}^{\prime}g_{2}^{\prime}g_{3}^{\prime}%
}=\left(  000\left\vert 111\right.  \right)  \text{,} \label{seta}%
\end{align}
where $v_{g^{\prime}}$ denotes the binary vectorial representation of the
Pauli operators. Furthermore,%
\begin{equation}
\mathcal{S}_{\text{CWS}}^{^{\prime\prime}}\overset{\text{def}}{=}\left\{
I\text{, }g_{1}\text{, }g_{2}\text{, }g_{3}\text{, }g_{1}g_{2}\text{, }%
g_{1}g_{3}\text{, }g_{2}g_{3}\text{, }g_{1}g_{2}g_{3}\right\}  \text{,}%
\end{equation}

\begin{figure}[ptb]
\centering
\includegraphics[width=0.35\textwidth]{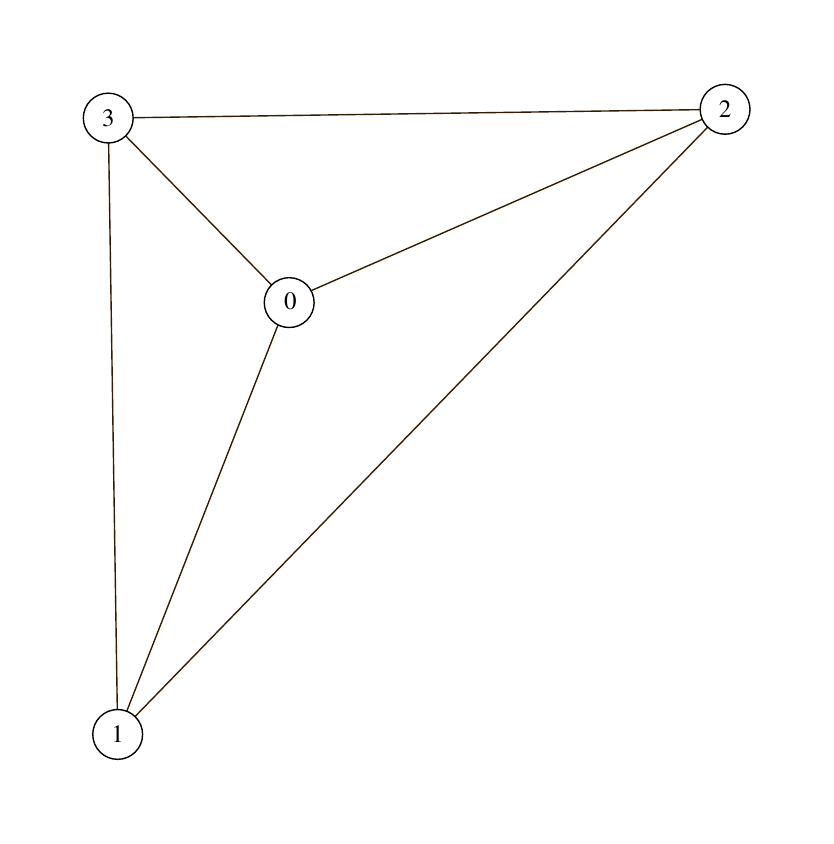}\caption{Graph for a quantum code
that is locally Clifford equivalent to the [[3,1,1]]-code.}%
\label{fig1}%
\end{figure}

with $g_{1}\overset{\text{def}}{=}Z^{1}X^{2}$, $g_{2}\overset{\text{def}}%
{=}Z^{1}X^{3}$, $g_{3}\overset{\text{def}}{=}X^{1}Z^{2}Z^{3}$. Using $Q$ in
Eq. (\ref{q1}), we have%
\begin{align}
I  &  \rightarrow v_{QI}=\left(  000\left\vert 000\right.  \right)  \text{,
}g_{1}\rightarrow v_{Qg_{1}}=\left(  110\left\vert 110\right.  \right)
\text{, }g_{2}\rightarrow v_{Qg_{2}}=\left(  101\left\vert 101\right.
\right)  \text{,}\nonumber\\
& \nonumber\\
\text{ }g_{3}  &  \rightarrow v_{Qg_{3}}=\left(  011\left\vert 100\right.
\right)  \text{, }g_{1}g_{2}\rightarrow v_{Qg_{1}g_{2}}=\left(  011\left\vert
011\right.  \right)  \text{, }g_{1}g_{3}\rightarrow v_{Qg_{1}g_{3}}=\left(
101\left\vert 010\right.  \right)  \text{, }\nonumber\\
& \nonumber\\
g_{2}g_{3}  &  \rightarrow v_{Qg_{2}g_{3}}=\left(  110\left\vert 001\right.
\right)  \text{, }g_{1}g_{2}g_{3}\rightarrow v_{Qg_{1}g_{2}g_{3}}=\left(
000\left\vert 111\right.  \right)  \text{.} \label{setb}%
\end{align}
From Eqs. (\ref{seta}) and (\ref{setb}), we arrive at%
\begin{align}
v_{I}  &  =v_{QI}\text{, }v_{g_{1}^{\prime}}=v_{Qg_{3}}\text{, }%
v_{g_{2}^{\prime}}=v_{Qg_{1}g_{3}}\text{, }v_{g_{3}^{\prime}}=v_{Qg_{2}g_{3}%
}\text{, }\nonumber\\
& \nonumber\\
v_{g_{1}^{\prime}g_{2}^{\prime}}  &  =v_{Qg_{1}}\text{, }v_{g_{1}^{\prime
}g_{3}^{\prime}}=v_{Qg_{2}}\text{, }v_{g_{2}^{\prime}g_{3}^{\prime}}%
=v_{Qg_{1}g_{2}}\text{, }v_{g_{1}^{\prime}g_{2}^{\prime}g_{3}^{\prime}%
}=v_{Qg_{1}g_{2}g_{3}}\text{.}%
\end{align}
Finally, given $\Gamma^{\prime}$ in Eq. (\ref{gammaprime}) and applying the
S-work, the coincidence matrix for a graph associated with a $\left[  \left[
3,1,1\right]  \right]  $ stabilizer code becomes,%
\begin{equation}
\Xi_{\left[  \left[  3,1,1\right]  \right]  }\overset{\text{def}}{=}\left(
\begin{array}
[c]{cccc}%
0 & 1 & 1 & 1\\
1 & 0 & 1 & 1\\
1 & 1 & 0 & 1\\
1 & 1 & 1 & 0
\end{array}
\right)  \text{.} \label{coincidencethree}%
\end{equation}
It is not that difficult to use the graphical quantum error correction
conditions presented in the SW-work and verify that the $\left[  \left[
3,1,1\right]  \right]  $ code with associated coincidence matrix in\ Eq.
(\ref{coincidencethree}) is not a $1$-error correcting quantum code.

\subsection{The $\left[  \left[  4,1\right]  \right]  $ stabilizer code}

Let us consider\textbf{\ }the Grassl et\textbf{\ }\textit{al.} perfect
$1$-erasure correcting four-qubit code with codespace spanned by the following
codewords \cite{markus2},%
\begin{equation}
\left\vert 0_{L}\right\rangle \overset{\text{def}}{=}\frac{\left\vert
0000\right\rangle +\left\vert 1111\right\rangle }{\sqrt{2}}\text{ and,
}\left\vert 1_{L}\right\rangle \overset{\text{def}}{=}\frac{\left\vert
1001\right\rangle +\left\vert 0110\right\rangle }{\sqrt{2}}\text{.}%
\end{equation}
The three stabilizer generators of such a code are given by $g_{1}%
\overset{\text{def}}{=}X^{1}X^{2}X^{3}X^{4}$, $g_{2}\overset{\text{def}}%
{=}Z^{1}Z^{4}$ and $g_{3}\overset{\text{def}}{=}Z^{2}Z^{3}$. Furthermore, the
logical operations are $\bar{X}\overset{\text{def}}{=}X^{1}X^{4}$ and $\bar
{Z}\overset{\text{def}}{=}Z^{1}Z^{3}$. We notice that such a code, just like
the four-qubit code provided by Leung et \textit{al}., is also a $1$-error
detecting code and can be used for the error correction of single amplitude
damping errors. When viewed as a CWS code, the codeword stabilizer reads%
\begin{equation}
\mathcal{S}_{\text{CWS}}\overset{\text{def}}{=}\left\langle g_{1}\text{,
}g_{2}\text{, }g_{3}\text{, }\bar{Z}\right\rangle =\left\langle X^{1}%
X^{2}X^{3}X^{4}\text{, }Z^{1}Z^{4}\text{, }Z^{2}Z^{3}\text{, }Z^{1}%
Z^{3}\right\rangle \text{.}%
\end{equation}
Observe that $\mathcal{S}_{\text{CWS}}$\ is local Clifford equivalent to
$\mathcal{S}_{\text{CWS}}^{\prime}$ with $\mathcal{S}_{\text{CWS}}^{\prime
}\overset{\text{def}}{=}U\mathcal{S}_{\text{CWS}}U^{\dagger}$ and
$U\overset{\text{def}}{=}H^{2}H^{3}H^{4}$. Therefore, $\mathcal{S}%
_{\text{CWS}}^{\prime}$ is given by%
\begin{equation}
\mathcal{S}_{\text{CWS}}^{\prime}=\left\langle X^{1}Z^{2}Z^{3}Z^{4}\text{,
}Z^{1}X^{4}\text{, }X^{2}X^{3}\text{, }Z^{1}X^{3}\right\rangle \text{.}%
\end{equation}
The codeword stabilizer matrix $\mathcal{H}_{\mathcal{S}_{\text{CWS}}^{\prime
}}$ associated with $\mathcal{S}_{\text{CWS}}^{\prime}$ is given by,%
\begin{equation}
\mathcal{H}_{\mathcal{S}_{\text{CWS}}^{\prime}}\overset{\text{def}}{=}\left(
Z^{\prime}\left\vert X^{\prime}\right.  \right)  =\left(
\begin{array}
[c]{cccc}%
0 & 1 & 1 & 1\\
1 & 0 & 0 & 0\\
0 & 0 & 0 & 0\\
1 & 0 & 0 & 0
\end{array}
\left\vert
\begin{array}
[c]{cccc}%
1 & 0 & 0 & 0\\
0 & 0 & 0 & 1\\
0 & 1 & 1 & 0\\
0 & 0 & 1 & 0
\end{array}
\right.  \right)  \text{.}%
\end{equation}
We observe that $\det X^{\prime\text{T}}\neq0$ and, applying the VdN-work, the
symmetric adjacency matrix $\Gamma$ reads\textbf{,}%
\begin{equation}
\Gamma\overset{\text{def}}{=}Z^{\prime\text{T}}\cdot\left(  X^{\prime\text{T}%
}\right)  ^{-1}=\left(
\begin{array}
[c]{cccc}%
0 & 1 & 1 & 1\\
1 & 0 & 0 & 0\\
1 & 0 & 0 & 0\\
1 & 0 & 0 & 0
\end{array}
\right)  \text{.} \label{grassl1}%
\end{equation}
Therefore, applying now the S-work, the $5\times5$ symmetric coincidence
matrix $\Xi_{\left[  \left[  4,1\right]  \right]  }$ that characterizes the
graph with both input and output vertices becomes\textbf{,}%
\begin{equation}
\Xi_{\left[  \left[  4,1\right]  \right]  }\overset{\text{def}}{=}\left(
\begin{array}
[c]{ccccc}%
0 & 0 & 0 & 1 & 1\\
0 & 0 & 1 & 1 & 1\\
0 & 1 & 0 & 0 & 0\\
1 & 1 & 0 & 0 & 0\\
1 & 1 & 0 & 0 & 0
\end{array}
\right)  \text{.} \label{coincidencefour}%
\end{equation}
For the sake of completeness, we also remark that acting with a local
complementation with respect to the vertex $1$ of the graph with adjacency
matrix $\Gamma$ in Eq. (\ref{grassl1}), we obtain the fully connected graph
with adjacency matrix $\Gamma^{\prime}$,%
\begin{equation}
\Gamma\rightarrow\Gamma^{\prime}\equiv g_{1}\left(  \Gamma\right)
\overset{\text{def}}{=}\Gamma+\Gamma\Lambda_{1}\Gamma+\Lambda^{\left(
1\right)  }=\left(
\begin{array}
[c]{cccc}%
0 & 1 & 1 & 1\\
1 & 0 & 1 & 1\\
1 & 1 & 0 & 1\\
1 & 1 & 1 & 0
\end{array}
\right)  \text{,} \label{may7}%
\end{equation}
where,%
\begin{equation}
\Lambda_{1}\overset{\text{def}}{=}\left(
\begin{array}
[c]{cccc}%
1 & 0 & 0 & 0\\
0 & 0 & 0 & 0\\
0 & 0 & 0 & 0\\
0 & 0 & 0 & 0
\end{array}
\right)  \text{ and, }\Lambda^{\left(  1\right)  }\overset{\text{def}}%
{=}\left(
\begin{array}
[c]{cccc}%
0 & 0 & 0 & 0\\
0 & 1 & 0 & 0\\
0 & 0 & 1 & 0\\
0 & 0 & 0 & 1
\end{array}
\right)  \text{.}%
\end{equation}
We also point out that following the VdN-work, it turns out that the
$8\times8$ local unitary Clifford transformation $Q$ that links the
eight-dimensional binary vector representation of the operators in
$\mathcal{S}_{\text{CWS}}^{^{\prime}}$ and $\left[  \mathcal{S}_{\text{CWS}%
}^{^{\prime}}\right]  _{\text{new}}$ \textbf{(}associated with the graph with
adjacency matrix\textbf{\ }$\Gamma^{\prime}$\textbf{\ }in Eq. (\ref{may7}))
with,%
\begin{equation}
\mathcal{S}_{\text{CWS}}^{^{\prime}}\rightarrow\left[  \mathcal{S}%
_{\text{CWS}}^{^{\prime}}\right]  _{\text{new}}\overset{\text{def}}%
{=}\left\langle X^{1}Z^{2}Z^{3}Z^{4}\text{, }Z^{1}X^{2}Z^{3}Z^{4}\text{,
}Z^{1}Z^{2}X^{3}Z^{4}\text{, }Z^{1}Z^{2}Z^{3}X^{4}\right\rangle \text{,}%
\end{equation}
reads,%
\begin{equation}
Q\overset{\text{def}}{=}\left(
\begin{array}
[c]{cccccccc}%
1 & 0 & 0 & 0 & 0 & 0 & 0 & 0\\
0 & 1 & 0 & 0 & 0 & 1 & 0 & 0\\
0 & 0 & 1 & 0 & 0 & 0 & 1 & 0\\
0 & 0 & 0 & 1 & 0 & 0 & 0 & 1\\
1 & 0 & 0 & 0 & 1 & 0 & 0 & 0\\
0 & 0 & 0 & 0 & 0 & 1 & 0 & 0\\
0 & 0 & 0 & 0 & 0 & 0 & 1 & 0\\
0 & 0 & 0 & 0 & 0 & 0 & 0 & 1
\end{array}
\right)  \text{.}%
\end{equation}
To verify that $Q$ is indeed a local Clifford operation, we observe
that\textbf{\ }it exhibits the required block-diagonal structure and satisfies
the relation $Q^{\text{T}}PQ=P$ with%
\begin{equation}
P\overset{\text{def}}{=}\left(
\begin{array}
[c]{cc}%
0 & I_{4\times4}\\
I_{4\times4} & 0
\end{array}
\right)  \text{.}%
\end{equation}
Finally,\textbf{\ }it is fairly simple\textbf{\ }to use the graphical quantum
error correction conditions presented in the SW-work and verify that the
$\left[  \left[  4,1\right]  \right]  $ code with associated coincidence
matrix in\ Eq. (\ref{coincidencefour}) is a $1$-error detecting quantum code.

\begin{figure}[ptb]
\centering
\includegraphics[width=0.35\textwidth]{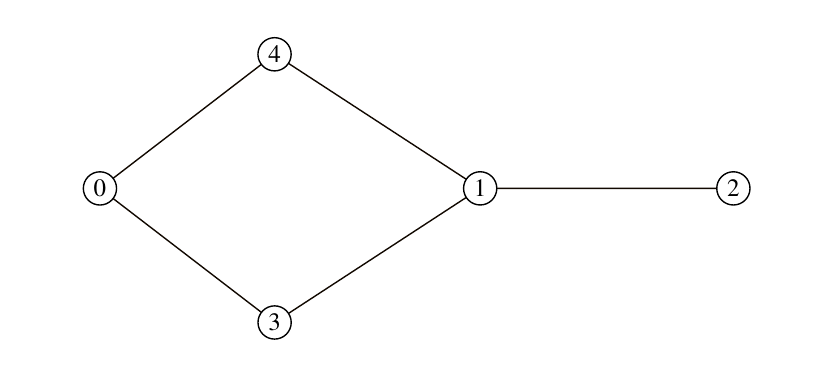}\caption{Graph for a quantum code
that is locally Clifford equivalent to the Leung et al. [[4,1]]-code.}%
\label{fig2}%
\end{figure}

\subsection{The $\left[  \left[  5,1,3\right]  \right]  $ stabilizer code}

The codespace of the perfect five-qubit stabilizer code is spanned by the
following codewords \cite{ray, charlie},%
\begin{equation}
\left\vert 0_{L}\right\rangle \overset{\text{def}}{=}\frac{1}{4}\left[
\begin{array}
[c]{c}%
\left\vert 00000\right\rangle +\left\vert 11000\right\rangle +\left\vert
01100\right\rangle +\left\vert 00110\right\rangle +\left\vert
00011\right\rangle +\left\vert 10001\right\rangle -\left\vert
01010\right\rangle -\left\vert 00101\right\rangle +\\
\\
-\left\vert 10010\right\rangle -\left\vert 01001\right\rangle -\left\vert
10100\right\rangle -\left\vert 11110\right\rangle -\left\vert
01111\right\rangle -\left\vert 10111\right\rangle -\left\vert
11011\right\rangle -\left\vert 11101\right\rangle
\end{array}
\right]  \text{,} \label{cd1}%
\end{equation}
and,%
\begin{equation}
\left\vert 1_{L}\right\rangle \overset{\text{def}}{=}\frac{1}{4}\left[
\begin{array}
[c]{c}%
\left\vert 11111\right\rangle +\left\vert 00111\right\rangle +\left\vert
10011\right\rangle +\left\vert 11001\right\rangle +\left\vert
11100\right\rangle +\left\vert 01110\right\rangle -\left\vert
10101\right\rangle -\left\vert 11010\right\rangle +\\
\\
-\left\vert 01101\right\rangle -\left\vert 10110\right\rangle -\left\vert
01011\right\rangle -\left\vert 00001\right\rangle -\left\vert
10000\right\rangle -\left\vert 01000\right\rangle -\left\vert
00100\right\rangle -\left\vert 00010\right\rangle
\end{array}
\right]  \text{.} \label{cd2}%
\end{equation}
Furthermore, the four stabilizer generators of the code are given by,%
\begin{equation}
g_{1}\overset{\text{def}}{=}X^{1}Z^{2}Z^{3}X^{4}\text{, }g_{2}\overset
{\text{def}}{=}X^{2}Z^{3}Z^{4}X^{5}\text{, }g_{3}\overset{\text{def}}{=}%
X^{1}X^{3}Z^{4}Z^{5}\text{ and, }g_{4}\overset{\text{def}}{=}Z^{1}X^{2}%
X^{4}Z^{5}\text{.}%
\end{equation}
A suitable choice of logical operations reads,%
\begin{equation}
\bar{X}\overset{\text{def}}{=}X^{1}X^{2}X^{3}X^{4}X^{5}\text{ and, }\bar
{Z}\overset{\text{def}}{=}Z^{1}Z^{2}Z^{3}Z^{4}Z^{5}\text{.}%
\end{equation}
We observe that the codespace of the $\left[  \left[  5,1,3\right]  \right]  $
code can be equally well-described by the following set of orthonormal
codewords,%
\begin{equation}
\left\vert 0_{L}^{\prime}\right\rangle \overset{\text{def}}{=}\frac{\left\vert
0_{L}\right\rangle +\left\vert 1_{L}\right\rangle }{\sqrt{2}}\text{ and,
}\left\vert 1_{L}^{\prime}\right\rangle \overset{\text{def}}{=}\frac
{\left\vert 0_{L}\right\rangle -\left\vert 1_{L}\right\rangle }{\sqrt{2}%
}\text{,}%
\end{equation}
with unchanged stabilizer and new logical operations given by,%
\begin{equation}
\bar{Z}^{\prime}=\bar{X}\overset{\text{def}}{=}X^{1}X^{2}X^{3}X^{4}X^{5}\text{
and, }\bar{X}^{\prime}=\bar{Z}\overset{\text{def}}{=}Z^{1}Z^{2}Z^{3}Z^{4}%
Z^{5}\text{.}%
\end{equation}
The codeword stabilizer $\mathcal{S}_{\text{CWS}}$ of the CWS code that
realizes the five-qubit code spanned by the codewords $\left\vert
0_{L}^{\prime}\right\rangle $ and $\left\vert 1_{L}^{\prime}\right\rangle $
reads\textbf{,}
\begin{equation}
\mathcal{S}_{\text{CWS}}\overset{\text{def}}{=}\left\langle g_{1}\text{,
}g_{2}\text{, }g_{3}\text{, }g_{4}\text{, }\bar{Z}^{\prime}\text{
}\right\rangle =\left\langle X^{1}Z^{2}Z^{3}X^{4}\text{, }X^{2}Z^{3}Z^{4}%
X^{5}\text{, }X^{1}X^{3}Z^{4}Z^{5}\text{, }Z^{1}X^{2}X^{4}Z^{5}\text{, }%
X^{1}X^{2}X^{3}X^{4}X^{5}\right\rangle \text{.}%
\end{equation}
The codeword stabilizer matrix $\mathcal{H}_{\mathcal{S}_{\text{CWS}}}$
associated with $\mathcal{S}_{\text{CWS}}$ is given by,%
\begin{equation}
\mathcal{H}_{\mathcal{S}_{\text{CWS}}}\overset{\text{def}}{=}\left(
Z\left\vert X\right.  \right)  =\left(
\begin{array}
[c]{ccccc}%
0 & 1 & 1 & 0 & 0\\
0 & 0 & 1 & 1 & 0\\
0 & 0 & 0 & 1 & 1\\
1 & 0 & 0 & 0 & 1\\
0 & 0 & 0 & 0 & 0
\end{array}
\left\vert
\begin{array}
[c]{ccccc}%
1 & 0 & 0 & 1 & 0\\
0 & 1 & 0 & 0 & 1\\
1 & 0 & 1 & 0 & 0\\
0 & 1 & 0 & 1 & 0\\
1 & 1 & 1 & 1 & 1
\end{array}
\right.  \right)  \text{.}%
\end{equation}
We observe $\det X\neq0$. Thus, using the VdN-work, the $5\times5$ adjacency
matrix $\Gamma$ becomes,%

\begin{figure}[ptb]
\centering
\includegraphics[width=0.35\textwidth]{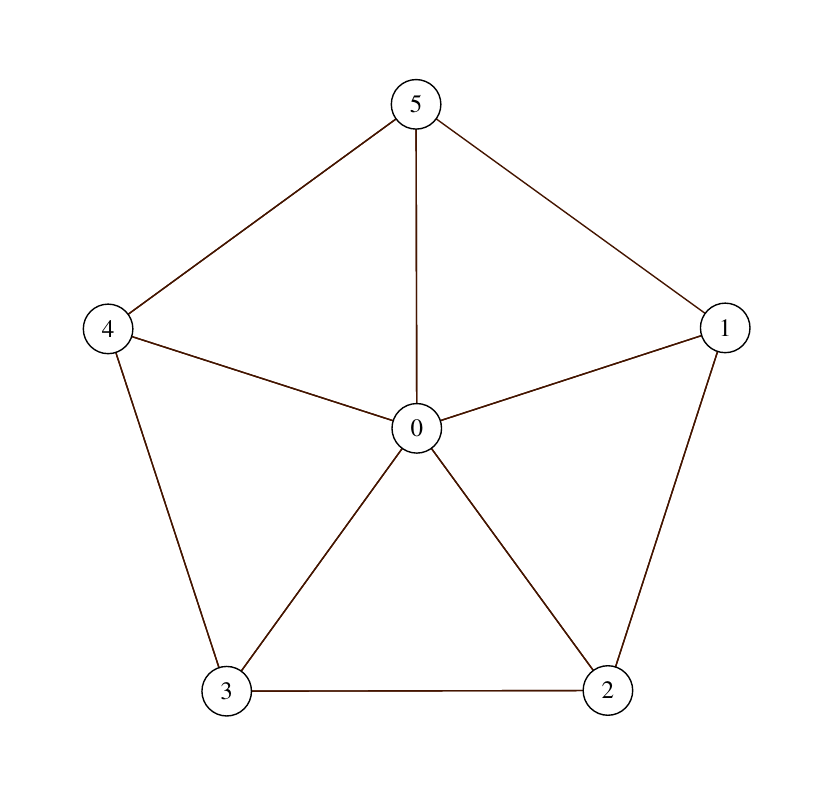}\caption{Graph for a quantum code
that is locally Clifford equivalent to the perfect [[5,1,3]]-code.}%
\label{fig3}%
\end{figure}

\begin{equation}
\Gamma\overset{\text{def}}{=}Z^{\text{T}}\cdot\left(  X^{\text{T}}\right)
^{-1}=\left(
\begin{array}
[c]{ccccc}%
0 & 1 & 0 & 0 & 1\\
1 & 0 & 1 & 0 & 0\\
0 & 1 & 0 & 1 & 0\\
0 & 0 & 1 & 0 & 1\\
1 & 0 & 0 & 1 & 0
\end{array}
\right)  \text{.}%
\end{equation}
Therefore, applying now the S-work, the $6\times6$ symmetric coincidence
matrix $\Xi_{\left[  \left[  5,1,3\right]  \right]  }$ characterizing the
graph with both input and output vertices is given by,%
\begin{equation}
\Xi_{\left[  \left[  5,1,3\right]  \right]  }\overset{\text{def}}{=}\left(
\begin{array}
[c]{cccccc}%
0 & 1 & 1 & 1 & 1 & 1\\
1 & 0 & 1 & 0 & 0 & 1\\
1 & 1 & 0 & 1 & 0 & 0\\
1 & 0 & 1 & 0 & 1 & 0\\
1 & 0 & 0 & 1 & 0 & 1\\
1 & 1 & 0 & 0 & 1 & 0
\end{array}
\right)  \text{.}%
\end{equation}
In order to show that the pentagon graph with five output vertices and one
input vertex with coincidence matrix $\Xi_{\left[  \left[  5,1,3\right]
\right]  }$ realizes a $1$-error correcting code, it is required to apply the
graph-theoretic error detection (correction) conditions of the SW-work to
$\binom{5}{2}=10$\textbf{\ }two-error configurations $E_{k}$ with
$k\in\left\{  1\text{,..., }10\right\}  $. These two-error configurations
read,%
\begin{align}
&  E_{1}\overset{\text{def}}{=}\left\{  0\text{, }1\text{, }2\right\}  \text{,
}E_{2}\overset{\text{def}}{=}\left\{  0\text{, }1\text{, }3\right\}  \text{,
}E_{3}\overset{\text{def}}{=}\left\{  0\text{, }1\text{, }4\right\}  \text{,
}E_{4}\overset{\text{def}}{=}\left\{  0\text{, }1\text{, }5\right\}  \text{,
}E_{5}\overset{\text{def}}{=}\left\{  0\text{, }2\text{, }3\right\}  \text{,
}\nonumber\\
& \nonumber\\
&  E_{6}\overset{\text{def}}{=}\left\{  0\text{, }2\text{, }4\right\}  \text{,
}E_{7}\overset{\text{def}}{=}\left\{  0\text{, }2\text{, }5\right\}  \text{,
}E_{8}\overset{\text{def}}{=}\left\{  0\text{, }3\text{, }4\right\}  \text{,
}E_{9}\overset{\text{def}}{=}\left\{  0\text{, }3\text{, }5\right\}  \text{,
}E_{10}\overset{\text{def}}{=}\left\{  0\text{, }4\text{, }5\right\}  \text{.}
\label{ec}%
\end{align}
For instance, the application of the SW-theorem to the error configuration
$E_{1}=\left\{  0\text{, }1\text{, }2\right\}  $ leads to the following set of
relations,%
\begin{equation}
d_{0}+d_{2}=0\text{, }d_{0}=0\text{ and, }d_{0}+d_{1}=0\text{.}%
\end{equation}

Solving this set of equations, we arrive at $d_{0}=d_{1}=d_{2}=0$. According
to the SW-theorem, this implies that the the error configuration
$E_{1}=\left\{  0\text{, }1\text{, }2\right\}  $ is a detectable
error-configuration. In other words, the detectability of $E_{1}$ is linked to
the non-singularity of the following $3\times3$ submatrix of the $6\times6$
coincidence matrix,%
\begin{equation}
E_{1}\overset{\text{def}}{=}\left\{  0\text{, }1\text{, }2\right\}
\leftrightarrow\left\{
\begin{array}
[c]{c}%
d_{0}+d_{2}=0\\
\text{ }d_{0}=0\\
\text{ }d_{0}+d_{1}=0
\end{array}
\right.  \leftrightarrow\det\left(
\begin{array}
[c]{ccc}%
1 & 0 & 1\\
1 & 0 & 0\\
1 & 1 & 0
\end{array}
\right)  \neq0\text{.}%
\end{equation}
Following this line of reasoning, it turns out that the remaining nine error
configurations in Eq. (\ref{ec}) are detectable as well. The detectability of
arbitrary error configurations with two nontrivial error operators leads to
the conclusion that the graph realizes a $1$-error correcting code.

\subsection{The $\left[  \left[  6,1,3\right]  \right]  $ stabilizer codes}

Calderbank et \textit{al}. discovered two distinct six-qubit quantum
degenerate codes which encode one logical qubit into six physical qubits
\cite{robert}. The first of these codes was discovered by trivially extending
the perfect five-qubit code and the other one through an exhaustive search of
the encoding space.

\subsubsection{Trivial case}

The first (trivial) degenerate six-qubit code that we consider can be obtained
from the $\left[  \left[  5,1,3\right]  \right]  $ code by appending an
ancilla qubit to the five-qubit code. Thus, we add a new qubit and a new
stabilizer generator which is $X$ for the new qubit \cite{daniel-phd}. The
other four stabilizer generators from the five-qubit code are tensored with
the identity on the new qubit to form the generators of the new code. To be
explicit, the codespace of this six-qubit code is spanned by the following two
orthonormal codewords,%
\begin{equation}
\left\vert 0_{L}^{\prime}\right\rangle \overset{\text{def}}{=}\left\vert
0_{L}\right\rangle \otimes\left\vert +\right\rangle _{6}\text{ and,
}\left\vert 1_{L}^{\prime}\right\rangle \overset{\text{def}}{=}\left\vert
1_{L}\right\rangle \otimes\left\vert +\right\rangle _{6}\text{,}%
\end{equation}
where $\left\vert 0_{L}\right\rangle $ and $\left\vert 1_{L}\right\rangle $
are defined in Eqs. (\ref{cd1}) and (\ref{cd2}), respectively. Following the
point of view adopted for the five-qubit code, let us choose a codespace for
the six-qubit code spanned by the new orthonormal codewords given by,%
\begin{equation}
\left\vert 0_{L}^{^{\prime\prime}}\right\rangle \overset{\text{def}}{=}%
\frac{\left\vert 0_{L}^{\prime}\right\rangle +\left\vert 1_{L}^{\prime
}\right\rangle }{\sqrt{2}}\text{ and, }\left\vert 1_{L}^{^{\prime\prime}%
}\right\rangle \overset{\text{def}}{=}\frac{\left\vert 0_{L}^{\prime
}\right\rangle -\left\vert 1_{L}^{\prime}\right\rangle }{\sqrt{2}}\text{.}%
\end{equation}
The five stabilizer generators of the code with codespace spanned by
$\left\vert 0_{L}^{^{\prime\prime}}\right\rangle $ and $\left\vert
1_{L}^{^{\prime\prime}}\right\rangle $ read,%
\begin{equation}
g_{1}\overset{\text{def}}{=}X^{1}Z^{2}Z^{3}X^{4}\text{, }g_{2}\overset
{\text{def}}{=}X^{2}Z^{3}Z^{4}X^{5}\text{, }g_{3}\overset{\text{def}}{=}%
X^{1}X^{3}Z^{4}Z^{5}\text{, }g_{4}\overset{\text{def}}{=}Z^{1}X^{2}X^{4}%
Z^{5}\text{ and, }g_{5}\overset{\text{def}}{=}X^{6}\text{.} \label{may-1}%
\end{equation}
A suitable choice of logical operations on $\left\vert 0_{L}^{^{\prime\prime}%
}\right\rangle $ and $\left\vert 1_{L}^{^{\prime\prime}}\right\rangle $ is
provided by,%
\begin{equation}
\bar{X}\overset{\text{def}}{=}Z^{1}Z^{2}Z^{3}Z^{4}Z^{5}\text{ and, }\bar
{Z}\overset{\text{def}}{=}X^{1}X^{2}X^{3}X^{4}X^{5}\text{.}%
\end{equation}
We remark that $\bar{X}$ and $\bar{Z}$ anticommute, and that each commutes
with all the five stabilizer generators in Eq. (\ref{may-1}). The codeword
stabilizer $\mathcal{S}_{\text{CWS}}$ of the CWS code that realizes the
six-qubit code spanned by the codewords $\left\vert 0_{L}^{^{\prime\prime}%
}\right\rangle $ and $\left\vert 1_{L}^{^{\prime\prime}}\right\rangle
$\textbf{\ }reads\textbf{,}%
\begin{equation}
\mathcal{S}_{\text{CWS}}\overset{\text{def}}{=}\left\langle g_{1}\text{,
}g_{2}\text{, }g_{3}\text{, }g_{4}\text{, }g_{5}\text{, }\bar{Z}\text{
}\right\rangle =\left\langle X^{1}Z^{2}Z^{3}X^{4}\text{, }X^{2}Z^{3}Z^{4}%
X^{5}\text{, }X^{1}X^{3}Z^{4}Z^{5}\text{, }Z^{1}X^{2}X^{4}Z^{5}\text{, }%
X^{6}\text{, }X^{1}X^{2}X^{3}X^{4}X^{5}\right\rangle \text{.}%
\end{equation}
The codeword stabilizer matrix $\mathcal{H}_{\mathcal{S}_{\text{CWS}}}$
associated with $\mathcal{S}_{\text{CWS}}$ is given by,%
\begin{equation}
\mathcal{H}_{\mathcal{S}_{\text{CWS}}}\overset{\text{def}}{=}\left(
Z\left\vert X\right.  \right)  =\left(
\begin{array}
[c]{cccccc}%
0 & 1 & 1 & 0 & 0 & 0\\
0 & 0 & 1 & 1 & 0 & 0\\
0 & 0 & 0 & 1 & 1 & 0\\
1 & 0 & 0 & 0 & 1 & 0\\
0 & 0 & 0 & 0 & 0 & 0\\
0 & 0 & 0 & 0 & 0 & 0
\end{array}
\left\vert
\begin{array}
[c]{cccccc}%
1 & 0 & 0 & 1 & 0 & 0\\
0 & 1 & 0 & 0 & 1 & 0\\
1 & 0 & 1 & 0 & 0 & 0\\
0 & 1 & 0 & 1 & 0 & 0\\
0 & 0 & 0 & 0 & 0 & 1\\
1 & 1 & 1 & 1 & 1 & 0
\end{array}
\right.  \right)  \text{.}%
\end{equation}
We observe $\det X\neq0$. Thus, using the VdN-work, the $6\times6$ adjacency
matrix $\Gamma$ becomes,%
\begin{equation}
\Gamma\overset{\text{def}}{=}Z^{\text{T}}\cdot\left(  X^{\text{T}}\right)
^{-1}=\left(
\begin{array}
[c]{cccccc}%
0 & 1 & 0 & 0 & 1 & 0\\
1 & 0 & 1 & 0 & 0 & 0\\
0 & 1 & 0 & 1 & 0 & 0\\
0 & 0 & 1 & 0 & 1 & 0\\
1 & 0 & 0 & 1 & 0 & 0\\
0 & 0 & 0 & 0 & 0 & 0
\end{array}
\right)  \text{.}%
\end{equation}
Therefore, applying now the S-work, the $7\times7$ symmetric coincidence
matrix $\Xi_{\left[  \left[  6,1,3\right]  \right]  }^{\text{trivial}}$
characterizing the graph with both input and output vertices reads,%
\begin{equation}
\Xi_{\left[  \left[  6,1,3\right]  \right]  }^{\text{trivial}}\overset
{\text{def}}{=}\left(
\begin{array}
[c]{ccccccc}%
0 & 1 & 1 & 1 & 1 & 1 & 1\\
1 & 0 & 1 & 0 & 0 & 1 & 0\\
1 & 1 & 0 & 1 & 0 & 0 & 0\\
1 & 0 & 1 & 0 & 1 & 0 & 0\\
1 & 0 & 0 & 1 & 0 & 1 & 0\\
1 & 1 & 0 & 0 & 1 & 0 & 0\\
1 & 0 & 0 & 0 & 0 & 0 & 0
\end{array}
\right)  \text{.}%
\end{equation}
In order to show that the graph with six output vertices and one input vertex
with coincidence matrix $\Xi_{\left[  \left[  6,1,3\right]  \right]
}^{\text{trivial}}$ realizes a $1$-error correcting degenerate code, we have
to apply the graph-theoretic error detection (correction) conditions of the
SW-work to $\binom{6}{2}=15$ error configurations $E_{k}$ with $k\in\left\{
1\text{,..., }15\right\}  $. It can be verified that any of the ten error
configurations $E_{k}\overset{\text{def}}{=}\left\{  0\text{, }e\text{,
}e^{\prime}\right\}  $ with $e$, $e^{\prime}\neq6$ satisfy the strong version
of the graph-theoretic error detection conditions. In addition, the five
two-error configurations $E_{k}$ with an error $e=6$ only satisfy the weak
form of the graph-theoretic error detection conditions. This fact is
consistent with the finding that concerns degenerate codes presented in the SW-work.

\begin{figure}[ptb]
\centering
\includegraphics[width=0.35\textwidth]{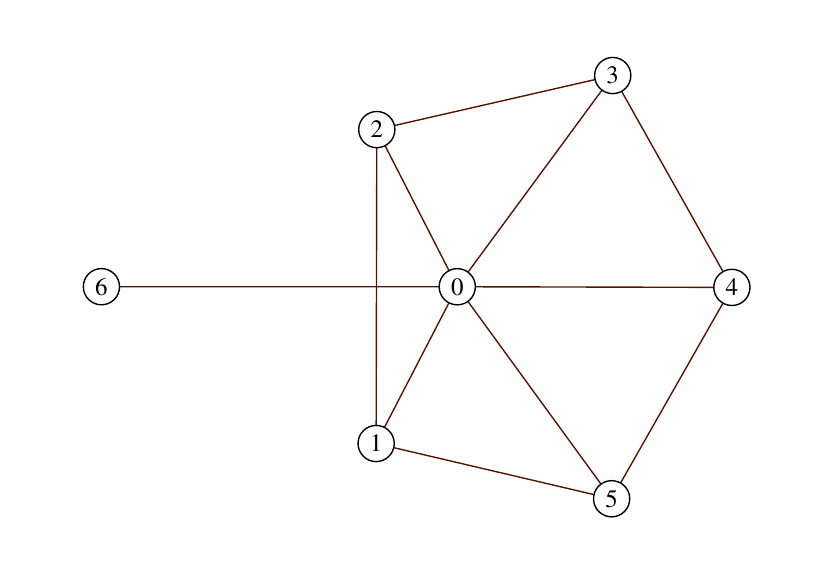}\caption{Graph for a quantum code
that is locally Clifford equivalent to the trivial [[6,1,3]]-code.}%
\label{fig4}%
\end{figure}

\subsubsection{Nontrivial case}

The second example of a six-qubit degenerate code provided by Calderbank et
\textit{al}. is a nontrivial six-qubit code \cite{robert}, which, according to
Calderbank et al., in unique up to equivalence. The example that we consider
was indeed introduced by Bilal et \textit{al}. in \cite{bilal}. They state
that since their example is not reducible to the trivial six-qubit code
because every one of its qubits is entangled with the others, their code is
equivalent to the (second) nontrivial six-qubit code according to the
arguments of Calderbank et \textit{al}. The codespace of this nontrivial
six-qubit code is spanned by the codewords $\left\vert 0_{L}\right\rangle $
and $\left\vert 1_{L}\right\rangle $ defined as \cite{bilal},%
\begin{equation}
\left\vert 0_{L}\right\rangle \overset{\text{def}}{=}\frac{1}{\sqrt{8}}\left[
\left\vert 000000\right\rangle -\left\vert 100111\right\rangle +\left\vert
001111\right\rangle -\left\vert 101000\right\rangle -\left\vert
010010\right\rangle +\left\vert 110101\right\rangle +\left\vert
011101\right\rangle -\left\vert 111010\right\rangle \right]  \text{,}%
\end{equation}
and,%
\begin{equation}
\left\vert 1_{L}\right\rangle \overset{\text{def}}{=}\frac{1}{\sqrt{8}}\left[
\left\vert 001010\right\rangle +\left\vert 101101\right\rangle +\left\vert
000101\right\rangle +\left\vert 1000010\right\rangle -\left\vert
011000\right\rangle -\left\vert 111111\right\rangle +\left\vert
010111\right\rangle +\left\vert 110000\right\rangle \right]  \text{,}%
\end{equation}
respectively. The five stabilizer generators for this code are given by,%
\begin{equation}
g_{1}\overset{\text{def}}{=}Y^{1}Z^{3}X^{4}X^{5}Y^{6}\text{, }g_{2}%
\overset{\text{def}}{=}Z^{1}X^{2}X^{5}Z^{6}\text{, }g_{3}\overset{\text{def}%
}{=}Z^{2}X^{3}X^{4}X^{5}X^{6}\text{, }g_{4}\overset{\text{def}}{=}Z^{4}%
Z^{6}\text{, }g_{5}\overset{\text{def}}{=}Z^{1}Z^{2}Z^{3}Z^{5}\text{.}%
\end{equation}
A suitable choice for the logical operations reads,%
\begin{equation}
\bar{X}\overset{\text{def}}{=}Z^{1}X^{3}X^{5}\text{ and, }\bar{Z}%
\overset{\text{def}}{=}Z^{2}Z^{5}Z^{6}\text{.}%
\end{equation}
In what follows, we shall consider the codespace spanned by the orthonormal
codewords%
\begin{equation}
\left\vert 0_{L}^{\prime}\right\rangle \overset{\text{def}}{=}\frac{\left\vert
0_{L}\right\rangle +\left\vert 1_{L}\right\rangle }{\sqrt{2}}\text{ and,
}\left\vert 1_{L}^{\prime}\right\rangle \overset{\text{def}}{=}\frac
{\left\vert 0_{L}\right\rangle -\left\vert 1_{L}\right\rangle }{\sqrt{2}%
}\text{.}%
\end{equation}
This way, the codeword stabilizer $\mathcal{S}_{\text{CWS}}$ of the CWS\ code
that realizes the six-qubit code with a codespace spanned by $\left\vert
0_{L}^{\prime}\right\rangle $ and $\left\vert 1_{L}^{\prime}\right\rangle $ is
given by,%
\begin{equation}
\mathcal{S}_{\text{CWS}}\overset{\text{def}}{=}\left\langle g_{1}\text{,
}g_{2}\text{, }g_{3}\text{, }g_{4}\text{, }g_{5}\text{, }\bar{Z}^{\prime
}\right\rangle \text{,}%
\end{equation}
with $\bar{Z}^{\prime}\equiv\bar{X}\overset{\text{def}}{=}Z^{1}X^{3}X^{5}$.
Therefore, $\mathcal{S}_{\text{CWS}}$ becomes\textbf{,}%
\begin{equation}
\mathcal{S}_{\text{CWS}}\overset{\text{def}}{=}\left\langle Y^{1}Z^{3}%
X^{4}X^{5}Y^{6}\text{, }Z^{1}X^{2}X^{5}Z^{6}\text{, }Z^{2}X^{3}X^{4}X^{5}%
X^{6}\text{, }Z^{4}Z^{6}\text{, }Z^{1}Z^{2}Z^{3}Z^{5}\text{, }Z^{1}X^{3}%
X^{5}\right\rangle \text{.}%
\end{equation}
Observe that $\mathcal{S}_{\text{CWS}}$ is locally Clifford equivalent to
$\mathcal{S}_{\text{CWS}}^{\prime}$ with $\mathcal{S}_{\text{CWS}}^{\prime
}\overset{\text{def}}{=}U\mathcal{S}_{\text{CWS}}U^{\dagger}$ and
$U\overset{\text{def}}{=}H^{1}H^{4}$. Therefore, we obtain\textbf{,}%
\begin{equation}
\mathcal{S}_{\text{CWS}}^{\prime}=\left\langle Y^{1}Z^{3}Z^{4}X^{5}%
Y^{6}\text{, }X^{1}X^{2}X^{5}Z^{6}\text{, }Z^{2}X^{3}Z^{4}X^{5}X^{6}\text{,
}X^{4}Z^{6}\text{, }X^{1}Z^{2}Z^{3}Z^{5}\text{, }X^{1}X^{3}X^{5}\right\rangle
\text{.}%
\end{equation}
The codeword stabilizer matrix $\mathcal{H}_{\mathcal{S}_{\text{CWS}}^{\prime
}}$ associated with $\mathcal{S}_{\text{CWS}}^{\prime}$ reads,%
\begin{equation}
\mathcal{H}_{\mathcal{S}_{\text{CWS}}^{\prime}}\overset{\text{def}}{=}\left(
Z^{\prime}\left\vert X^{\prime}\right.  \right)  =\left(
\begin{array}
[c]{cccccc}%
1 & 0 & 1 & 1 & 0 & 1\\
0 & 0 & 0 & 0 & 0 & 1\\
0 & 1 & 0 & 1 & 0 & 0\\
0 & 0 & 0 & 0 & 0 & 1\\
0 & 1 & 1 & 0 & 1 & 0\\
0 & 0 & 0 & 0 & 0 & 0
\end{array}
\left\vert
\begin{array}
[c]{cccccc}%
1 & 0 & 0 & 0 & 1 & 1\\
1 & 1 & 0 & 0 & 1 & 0\\
0 & 0 & 1 & 0 & 1 & 1\\
0 & 0 & 0 & 1 & 0 & 0\\
1 & 0 & 0 & 0 & 0 & 0\\
1 & 0 & 1 & 0 & 1 & 0
\end{array}
\right.  \right)  \text{.}%
\end{equation}
We observe that $\det X^{\prime\text{T}}\neq0$ and, applying the VdN-work, the
symmetric adjacency matrix $\Gamma$ becomes,%
\begin{equation}
\Gamma\overset{\text{def}}{=}Z^{\prime\text{T}}\cdot\left(  X^{\prime\text{T}%
}\right)  ^{-1}=\left(
\begin{array}
[c]{cccccc}%
0 & 1 & 1 & 0 & 1 & 0\\
1 & 0 & 0 & 0 & 1 & 0\\
1 & 0 & 0 & 0 & 1 & 1\\
0 & 0 & 0 & 0 & 0 & 1\\
1 & 1 & 1 & 0 & 0 & 1\\
0 & 0 & 1 & 1 & 1 & 0
\end{array}
\right)  \text{.}%
\end{equation}
Therefore, applying now the S-work, the $7\times7$ symmetric coincidence
matrix $\Xi_{\left[  \left[  6,1,3\right]  \right]  }^{\text{nontrivial}}$
characterizing the graph with both input and output vertices is given by,%
\begin{equation}
\Xi_{\left[  \left[  6,1,3\right]  \right]  }^{\text{nontrivial}}%
\overset{\text{def}}{=}\left(
\begin{array}
[c]{ccccccc}%
0 & 1 & 0 & 1 & 0 & 1 & 0\\
1 & 0 & 1 & 1 & 0 & 1 & 0\\
0 & 1 & 0 & 0 & 0 & 1 & 0\\
1 & 1 & 0 & 0 & 0 & 1 & 1\\
0 & 0 & 0 & 0 & 0 & 0 & 1\\
1 & 1 & 1 & 1 & 0 & 0 & 1\\
0 & 0 & 0 & 1 & 1 & 1 & 0
\end{array}
\right)  \text{.}%
\end{equation}
In order to show that the graph with six output vertices and one input vertex
with coincidence matrix $\Xi_{\left[  \left[  6,1,3\right]  \right]
}^{\text{nontrivial}}$ realizes a $1$-error correcting (degenerate) code, we
have to apply the graph-theoretic error detection (correction) conditions of
the SW-work to $\binom{6}{2}=15$ two-error configurations $E_{k}$ with
$k\in\left\{  1\text{,..., }15\right\}  $. It can be checked that the only
problematic error configuration is $E_{k}\overset{\text{def}}{=}\left\{
0\text{, }e\text{, }e^{\prime}\right\}  $ with $e=4$, $e^{\prime}=6$. The only
undetectable nontrivial error is represented by $X^{4}Z^{6}$. However, this
error operator belongs to the stabilizer of the code and therefore it will
have no impact on the encoded quantum state. Thus, the code considered has
indeed distance $d=3$. Furthermore, since a quantum stabilizer code with
distance $d$ is a degenerate code if and only if its stabilizer has an element
of weight less than $d$ (excluding the identity element), our code with $d=3$
and a stabilizer element of weight-$2$ is indeed a degenerate code.

\begin{figure}[ptb]
\centering
\includegraphics[width=0.35\textwidth]{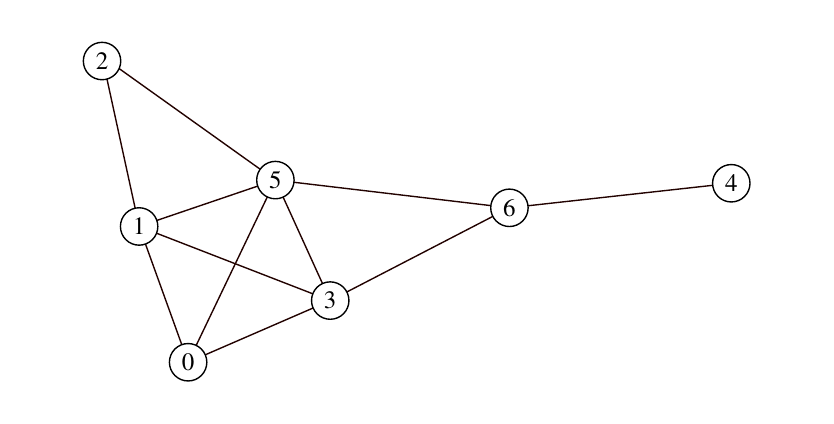}\caption{Graph for a quantum code
that is locally Clifford equivalent to the nontrivial [[6,1,3]]-code.}%
\label{fig5}%
\end{figure}

\subsection{The CSS $\left[  \left[  7,1,3\right]  \right]  $ stabilizer code}

The codespace of the CSS seven-qubit stabilizer code is spanned by the
following codewords \cite{steane, robert2},%
\begin{equation}
\left\vert 0_{L}\right\rangle \overset{\text{def}}{=}\frac{1}{\left(  \sqrt
{2}\right)  ^{3}}\left[
\begin{array}
[c]{c}%
\left\vert 0000000\right\rangle +\left\vert 0110011\right\rangle +\left\vert
1010101\right\rangle +\left\vert 1100110\right\rangle +\\
\\
+\left\vert 0001111\right\rangle +\left\vert 0111100\right\rangle +\left\vert
1011010\right\rangle +\left\vert 1101001\right\rangle
\end{array}
\right]  \text{,} \label{cd11}%
\end{equation}
and,%
\begin{equation}
\left\vert 1_{L}\right\rangle \overset{\text{def}}{=}\frac{1}{\left(  \sqrt
{2}\right)  ^{3}}\left[
\begin{array}
[c]{c}%
\left\vert 1111111\right\rangle +\left\vert 1001100\right\rangle +\left\vert
0101010\right\rangle +\left\vert 0011001\right\rangle +\\
\\
+\left\vert 1110000\right\rangle +\left\vert 1000011\right\rangle +\left\vert
0100101\right\rangle +\left\vert 0010110\right\rangle
\end{array}
\right]  \text{.} \label{cd22}%
\end{equation}
Furthermore, the six stabilizer generators of the code are given by%
\begin{equation}
g_{1}\overset{\text{def}}{=}X^{4}X^{5}X^{6}X^{7}\text{, }g_{2}\overset
{\text{def}}{=}X^{2}X^{3}X^{6}X^{7}\text{, }g_{3}\overset{\text{def}}{=}%
X^{1}X^{3}X^{5}X^{7}\text{, }g_{4}\overset{\text{def}}{=}Z^{4}Z^{5}Z^{6}%
Z^{7}\text{, }g_{5}\overset{\text{def}}{=}Z^{2}Z^{3}Z^{6}Z^{7}\text{, }%
g_{6}\overset{\text{def}}{=}Z^{1}Z^{3}Z^{5}Z^{7}\text{.}%
\end{equation}
A suitable choice of logical operations reads,%
\begin{equation}
\bar{X}\overset{\text{def}}{=}X^{1}X^{2}X^{3}\text{ and, }\bar{Z}%
\overset{\text{def}}{=}Z^{1}Z^{2}Z^{3}\text{.}%
\end{equation}
We observe that the codespace of the CSS code can be equally well-described by
the following set of orthonormal codewords,%
\begin{equation}
\left\vert 0_{L}^{\prime}\right\rangle \overset{\text{def}}{=}\frac{\left\vert
0_{L}\right\rangle +\left\vert 1_{L}\right\rangle }{\sqrt{2}}\text{ and,
}\left\vert 1_{L}^{\prime}\right\rangle \overset{\text{def}}{=}\frac
{\left\vert 0_{L}\right\rangle -\left\vert 1_{L}\right\rangle }{\sqrt{2}%
}\text{,}%
\end{equation}
with unchanged stabilizer and new logical operations given by,%
\begin{equation}
\bar{Z}^{\prime}=\bar{X}\overset{\text{def}}{=}X^{1}X^{2}X^{3}\text{ and,
}\bar{X}^{\prime}=\bar{Z}\overset{\text{def}}{=}Z^{1}Z^{2}Z^{3}\text{.}%
\end{equation}
The codeword stabilizer $\mathcal{S}_{\text{CWS}}$ of the CWS code that
realizes the seven-qubit code spanned by the codewords $\left\vert
0_{L}^{\prime}\right\rangle $ and $\left\vert 1_{L}^{\prime}\right\rangle $
reads,%
\begin{equation}
\mathcal{S}_{\text{CWS}}\overset{\text{def}}{=}\left\langle g_{1}\text{,
}g_{2}\text{, }g_{3}\text{, }g_{4}\text{, }g_{5}\text{, }g_{6}\text{, }\bar
{Z}^{\prime}\text{ }\right\rangle \text{,}%
\end{equation}
that is,%
\begin{equation}
\mathcal{S}_{\text{CWS}}=\left\langle X^{4}X^{5}X^{6}X^{7}\text{, }X^{2}%
X^{3}X^{6}X^{7}\text{, }X^{1}X^{3}X^{5}X^{7}\text{, }Z^{4}Z^{5}Z^{6}%
Z^{7}\text{, }Z^{2}Z^{3}Z^{6}Z^{7}\text{, }Z^{1}Z^{3}Z^{5}Z^{7}\text{, }%
X^{1}X^{2}X^{3}\right\rangle \text{.}%
\end{equation}
Observe that $\mathcal{S}_{\text{CWS}}$\ is local Clifford equivalent to
$\mathcal{S}_{\text{CWS}}^{\prime}$ with $\mathcal{S}_{\text{CWS}}^{\prime
}\overset{\text{def}}{=}U\mathcal{S}_{\text{CWS}}U^{\dagger}$ and
$U\overset{\text{def}}{=}H^{1}H^{2}H^{4}$. Therefore, $\mathcal{S}%
_{\text{CWS}}^{\prime}$ is given by,%
\begin{equation}
\mathcal{S}_{\text{CWS}}^{\prime}=\left\langle Z^{4}X^{5}X^{6}X^{7}\text{,
}Z^{2}X^{3}X^{6}X^{7}\text{, }Z^{1}X^{3}X^{5}X^{7}\text{, }X^{4}Z^{5}%
Z^{6}Z^{7}\text{, }X^{2}Z^{3}Z^{6}Z^{7}\text{, }X^{1}Z^{3}Z^{5}Z^{7}\text{,
}Z^{1}Z^{2}X^{3}\right\rangle \text{.}%
\end{equation}
The codeword stabilizer matrix $\mathcal{H}_{\mathcal{S}_{\text{CWS}}^{\prime
}}$ associated with $\mathcal{S}_{\text{CWS}}^{\prime}$ reads,%
\begin{equation}
\mathcal{H}_{\mathcal{S}_{\text{CWS}}^{\prime}}\overset{\text{def}}{=}\left(
Z^{\prime}\left\vert X^{\prime}\right.  \right)  =\left(
\begin{array}
[c]{ccccccc}%
0 & 0 & 0 & 1 & 0 & 0 & 0\\
0 & 1 & 0 & 0 & 0 & 0 & 0\\
1 & 0 & 0 & 0 & 0 & 0 & 0\\
0 & 0 & 0 & 0 & 1 & 1 & 1\\
0 & 0 & 1 & 0 & 0 & 1 & 1\\
0 & 0 & 1 & 0 & 1 & 0 & 1\\
1 & 1 & 0 & 0 & 0 & 0 & 0
\end{array}
\left\vert
\begin{array}
[c]{ccccccc}%
0 & 0 & 0 & 0 & 1 & 1 & 1\\
0 & 0 & 1 & 0 & 0 & 1 & 1\\
0 & 0 & 1 & 0 & 1 & 0 & 1\\
0 & 0 & 0 & 1 & 0 & 0 & 0\\
0 & 1 & 0 & 0 & 0 & 0 & 0\\
1 & 0 & 0 & 0 & 0 & 0 & 0\\
0 & 0 & 1 & 0 & 0 & 0 & 0
\end{array}
\right.  \right)  \text{.}%
\end{equation}
We observe $\det X^{\prime}\neq0$. Thus, using the VdN-work, the $7\times7 $
adjacency matrix $\Gamma$ reads,%
\begin{equation}
\Gamma\overset{\text{def}}{=}Z^{\prime\text{T}}\cdot\left(  X^{\prime\text{T}%
}\right)  ^{-1}=\left(
\begin{array}
[c]{ccccccc}%
0 & 0 & 1 & 0 & 1 & 0 & 1\\
0 & 0 & 1 & 0 & 0 & 1 & 1\\
1 & 1 & 0 & 0 & 0 & 0 & 0\\
0 & 0 & 0 & 0 & 1 & 1 & 1\\
1 & 0 & 0 & 1 & 0 & 0 & 0\\
0 & 1 & 0 & 1 & 0 & 0 & 0\\
1 & 1 & 0 & 1 & 0 & 0 & 0
\end{array}
\right)  \text{.}%
\end{equation}

\begin{figure}[ptb]
\centering
\includegraphics[width=0.35\textwidth]{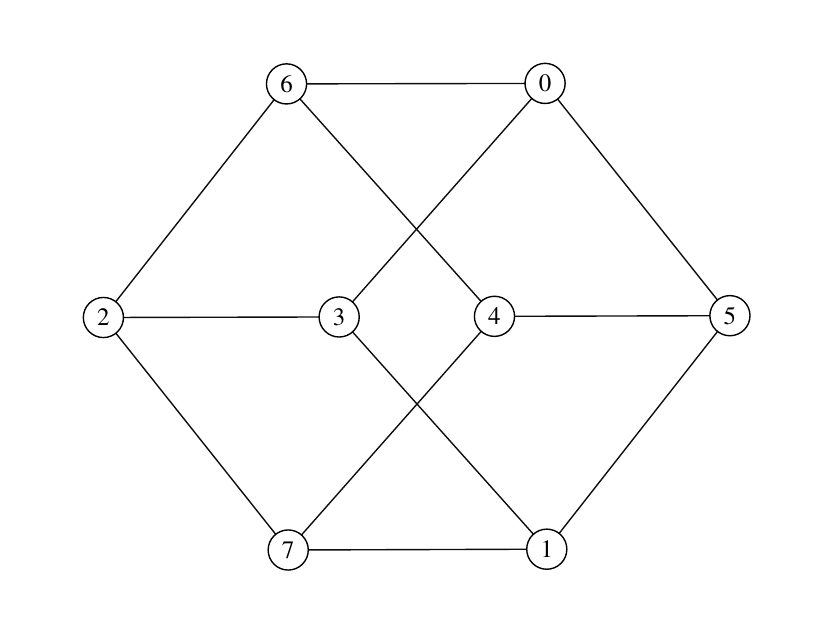}\caption{Graph for a quantum code
that is locally Clifford equivalent to the CSS [[7,1,3]]-code.}%
\label{fig6}%
\end{figure}

Therefore, applying now the S-work, the $8\times8$ symmetric coincidence
matrix $\Xi_{\text{CSS-}\left[  \left[  7,13\right]  \right]  }$
characterizing the graph with both input and output vertices\textbf{\ }%
becomes,%
\begin{equation}
\Xi_{\text{CSS-}\left[  \left[  7,13\right]  \right]  }\overset{\text{def}}%
{=}\left(
\begin{array}
[c]{cccccccc}%
0 & 0 & 0 & 1 & 0 & 1 & 1 & 0\\
0 & 0 & 0 & 1 & 0 & 1 & 0 & 1\\
0 & 0 & 0 & 1 & 0 & 0 & 1 & 1\\
1 & 1 & 1 & 0 & 0 & 0 & 0 & 0\\
0 & 0 & 0 & 0 & 0 & 1 & 1 & 1\\
1 & 1 & 0 & 0 & 1 & 0 & 0 & 0\\
1 & 0 & 1 & 0 & 1 & 0 & 0 & 0\\
0 & 1 & 1 & 0 & 1 & 0 & 0 & 0
\end{array}
\right)  \text{.}%
\end{equation}
It is straightforward to show that the cube graph with seven output vertices
and one input vertex with coincidence matrix $\Xi_{\text{CSS-}\left[  \left[
7,13\right]  \right]  }$ realizes a $1$-error correcting code. Namely, all the
$\binom{7}{2}=21$ two-error configurations $E_{k}$ with $k\in\left\{
1\text{,..., }21\right\}  $,%
\begin{align}
&  E_{1}\overset{\text{def}}{=}\left\{  0\text{, }1\text{, }2\right\}  \text{,
}E_{2}\overset{\text{def}}{=}\left\{  0\text{, }1\text{, }3\right\}  \text{,
}E_{3}\overset{\text{def}}{=}\left\{  0\text{, }1\text{, }4\right\}  \text{,
}E_{4}\overset{\text{def}}{=}\left\{  0\text{, }1\text{, }5\right\}  \text{,
}E_{5}\overset{\text{def}}{=}\left\{  0\text{, }1\text{, }6\right\}  \text{,
}E_{6}\overset{\text{def}}{=}\left\{  0\text{, }1\text{, }7\right\}  \text{,
}\nonumber\\
& \nonumber\\
&  E_{7}\overset{\text{def}}{=}\left\{  0\text{, }2\text{, }3\right\}  \text{,
}E_{8}\overset{\text{def}}{=}\left\{  0\text{, }2\text{, }4\right\}  \text{,
}E_{9}\overset{\text{def}}{=}\left\{  0\text{, }2\text{, }5\right\}  \text{,
}E_{10}\overset{\text{def}}{=}\left\{  0\text{, }2\text{, }6\right\}  \text{,
}E_{11}\overset{\text{def}}{=}\left\{  0\text{, }2\text{, }7\right\}  \text{,
}E_{12}\overset{\text{def}}{=}\left\{  0\text{, }3\text{, }4\right\}
\text{,}\nonumber\\
& \nonumber\\
&  \text{ }E_{13}\overset{\text{def}}{=}\left\{  0\text{, }3\text{,
}5\right\}  \text{, }E_{14}\overset{\text{def}}{=}\left\{  0\text{, }3\text{,
}6\right\}  \text{, }E_{15}\overset{\text{def}}{=}\left\{  0\text{, }3\text{,
}7\right\}  \text{, }E_{16}\overset{\text{def}}{=}\left\{  0\text{, }4\text{,
}5\right\}  \text{, }E_{17}\overset{\text{def}}{=}\left\{  0\text{, }4\text{,
}6\right\}  \text{,}\nonumber\\
& \nonumber\\
&  \text{ }E_{18}\overset{\text{def}}{=}\left\{  0\text{, }4\text{,
}7\right\}  \text{, }E_{19}\overset{\text{def}}{=}\left\{  0\text{, }5\text{,
}6\right\}  \text{, }E_{20}\overset{\text{def}}{=}\left\{  0\text{, }5\text{,
}7\right\}  \text{, }E_{21}\overset{\text{def}}{=}\left\{  0\text{, }6\text{,
}7\right\}  \text{,}%
\end{align}
satisfy the strong version of the graph-theoretic error detection (correction)
conditions of the SW-work in agreement with the fact that the code is nondegenerate.

\subsection{The Shor $\left[  \left[  9,1,3\right]  \right]  $ stabilizer
code}

\subsubsection{First case}

$\allowbreak$When realized as a CWS quantum code, the Shor nine-qubit code
\cite{shor} is characterized by the codeword stabilizer\textbf{\ }%
$\mathcal{S}_{\text{CWS}}$\textbf{\ }given by,%
\begin{equation}
\mathcal{S}_{\text{CWS}}\overset{\text{def}}{=}\left\langle g_{1}\text{,
}g_{2}\text{, }g_{3}\text{, }g_{4}\text{, }g_{5}\text{, }g_{6}\text{, }%
g_{7}\text{, }g_{8}\text{, }\bar{Z}\right\rangle \text{,} \label{sshor}%
\end{equation}
with codeword stabilizer generators given by,%
\begin{align}
&  g_{1}\overset{\text{def}}{=}Z^{1}Z^{2}\text{, }g_{2}\overset{\text{def}}%
{=}Z^{1}Z^{3}\text{, }g_{3}\overset{\text{def}}{=}Z^{4}Z^{5}\text{, }%
g_{4}\overset{\text{def}}{=}Z^{4}Z^{6}\text{, }g_{5}\overset{\text{def}}%
{=}Z^{7}Z^{8}\text{, }g_{6}\overset{\text{def}}{=}Z^{7}Z^{9}\text{,
}\nonumber\\
& \nonumber\\
&  g_{7}\overset{\text{def}}{=}X^{1}X^{2}X^{3}X^{4}X^{5}X^{6}\text{, }%
g_{8}\overset{\text{def}}{=}X^{1}X^{2}X^{3}X^{7}X^{8}X^{9}\text{, }\bar
{Z}\overset{\text{def}}{=}X^{1}X^{2}X^{3}X^{4}X^{5}X^{6}X^{7}X^{8}%
X^{9}\text{.}%
\end{align}
What is the graph that realizes the Shor code? The codeword stabilizer matrix
$\mathcal{H}_{\mathcal{S}_{\text{CWS}}}$ corresponding to $\mathcal{S}%
_{\text{CWS}}$ in Eq. (\ref{sshor}) can be formally written as,%
\begin{equation}
\mathcal{H}_{\mathcal{S}_{\text{CWS}}}\overset{\text{def}}{=}\left(
Z\left\vert X\right.  \right)  =\left(
\begin{array}
[c]{ccccccccc}%
1 & 1 & 0 & 0 & 0 & 0 & 0 & 0 & 0\\
1 & 0 & 1 & 0 & 0 & 0 & 0 & 0 & 0\\
0 & 0 & 0 & 1 & 1 & 0 & 0 & 0 & 0\\
0 & 0 & 0 & 1 & 0 & 1 & 0 & 0 & 0\\
0 & 0 & 0 & 0 & 0 & 0 & 1 & 1 & 0\\
0 & 0 & 0 & 0 & 0 & 0 & 1 & 0 & 1\\
0 & 0 & 0 & 0 & 0 & 0 & 0 & 0 & 0\\
0 & 0 & 0 & 0 & 0 & 0 & 0 & 0 & 0\\
0 & 0 & 0 & 0 & 0 & 0 & 0 & 0 & 0
\end{array}
\left\vert
\begin{array}
[c]{ccccccccc}%
0 & 0 & 0 & 0 & 0 & 0 & 0 & 0 & 0\\
0 & 0 & 0 & 0 & 0 & 0 & 0 & 0 & 0\\
0 & 0 & 0 & 0 & 0 & 0 & 0 & 0 & 0\\
0 & 0 & 0 & 0 & 0 & 0 & 0 & 0 & 0\\
0 & 0 & 0 & 0 & 0 & 0 & 0 & 0 & 0\\
0 & 0 & 0 & 0 & 0 & 0 & 0 & 0 & 0\\
1 & 1 & 1 & 1 & 1 & 1 & 0 & 0 & 0\\
1 & 1 & 1 & 0 & 0 & 0 & 1 & 1 & 1\\
1 & 1 & 1 & 1 & 1 & 1 & 1 & 1 & 1
\end{array}
\right.  \right)  \text{.}%
\end{equation}
Since $X^{\text{T}}$ is not invertible, the algorithmic procedure introduced
in the VdN-work cannot be applied. However, we notice that the codeword
stabilizer $\mathcal{S}_{\text{CWS}}$ in Eq. (\ref{sshor}) is locally Clifford
equivalent to $\mathcal{S}_{\text{CWS}}^{\prime}$ defined by,%
\begin{equation}
\mathcal{S}_{\text{CWS}}^{\prime}\overset{\text{def}}{=}U\mathcal{S}%
_{\text{CWS}}U^{\dagger}\text{ with, }U\overset{\text{def}}{=}I^{1}\otimes
H^{2}\otimes H^{3}\otimes I^{4}\otimes H^{5}\otimes H^{6}\otimes I^{7}\otimes
H^{8}\otimes H^{9}\text{.} \label{sshor1}%
\end{equation}
Using Eqs. (\ref{sshor}) and (\ref{sshor1}), it follows that%
\begin{equation}
\mathcal{S}_{\text{CWS}}^{\prime}\overset{\text{def}}{=}\left\langle
g_{1}^{\prime}\text{, }g_{2}^{\prime}\text{, }g_{3}^{\prime}\text{, }%
g_{4}^{\prime}\text{, }g_{5}^{\prime}\text{, }g_{6}^{\prime}\text{, }%
g_{7}^{\prime}\text{, }g_{8}^{\prime}\text{, }\bar{Z}^{\prime}\right\rangle
\text{,}%
\end{equation}
with,%
\begin{align}
&  g_{1}^{\prime}\overset{\text{def}}{=}Z^{1}X^{2}\text{, }g_{2}^{\prime
}\overset{\text{def}}{=}Z^{1}X^{3}\text{, }g_{3}^{\prime}\overset{\text{def}%
}{=}Z^{4}X^{5}\text{, }g_{4}^{\prime}\overset{\text{def}}{=}Z^{4}X^{6}\text{,
}g_{5}^{\prime}\overset{\text{def}}{=}Z^{7}X^{8}\text{, }g_{6}^{\prime
}\overset{\text{def}}{=}Z^{7}X^{9}\text{, }\nonumber\\
& \nonumber\\
&  g_{7}^{\prime}\overset{\text{def}}{=}X^{1}Z^{2}Z^{3}X^{4}Z^{5}Z^{6}\text{,
}g_{8}^{\prime}\overset{\text{def}}{=}X^{1}Z^{2}Z^{3}X^{7}Z^{8}Z^{9}\text{,
}\bar{Z}^{\prime}\overset{\text{def}}{=}X^{1}Z^{2}Z^{3}X^{4}Z^{5}Z^{6}%
X^{7}Z^{8}Z^{9}\text{.}%
\end{align}
We observe that the the codeword stabilizer matrix $\mathcal{H}_{\mathcal{S}%
_{\text{CWS}}^{\prime}}$ corresponding to $\mathcal{S}_{\text{CWS}}^{\prime}$
becomes,%
\begin{equation}
\mathcal{H}_{\mathcal{S}_{\text{CWS}}^{\prime}}\overset{\text{def}}{=}\left(
Z^{\prime}\left\vert X^{\prime}\right.  \right)  =\left(
\begin{array}
[c]{ccccccccc}%
1 & 0 & 0 & 0 & 0 & 0 & 0 & 0 & 0\\
1 & 0 & 0 & 0 & 0 & 0 & 0 & 0 & 0\\
0 & 0 & 0 & 1 & 0 & 0 & 0 & 0 & 0\\
0 & 0 & 0 & 1 & 0 & 0 & 0 & 0 & 0\\
0 & 0 & 0 & 0 & 0 & 0 & 1 & 0 & 0\\
0 & 0 & 0 & 0 & 0 & 0 & 1 & 0 & 0\\
0 & 1 & 1 & 0 & 1 & 1 & 0 & 0 & 0\\
0 & 1 & 1 & 0 & 0 & 0 & 0 & 1 & 1\\
0 & 1 & 1 & 0 & 1 & 1 & 0 & 1 & 1
\end{array}
\left\vert
\begin{array}
[c]{ccccccccc}%
0 & 1 & 0 & 0 & 0 & 0 & 0 & 0 & 0\\
0 & 0 & 1 & 0 & 0 & 0 & 0 & 0 & 0\\
0 & 0 & 0 & 0 & 1 & 0 & 0 & 0 & 0\\
0 & 0 & 0 & 0 & 0 & 1 & 0 & 0 & 0\\
0 & 0 & 0 & 0 & 0 & 0 & 0 & 1 & 0\\
0 & 0 & 0 & 0 & 0 & 0 & 0 & 0 & 1\\
1 & 0 & 0 & 1 & 0 & 0 & 0 & 0 & 0\\
1 & 0 & 0 & 0 & 0 & 0 & 1 & 0 & 0\\
1 & 0 & 0 & 1 & 0 & 0 & 1 & 0 & 0
\end{array}
\right.  \right)  \text{.}%
\end{equation}

\begin{figure}[ptb]
\centering
\includegraphics[width=0.35\textwidth]{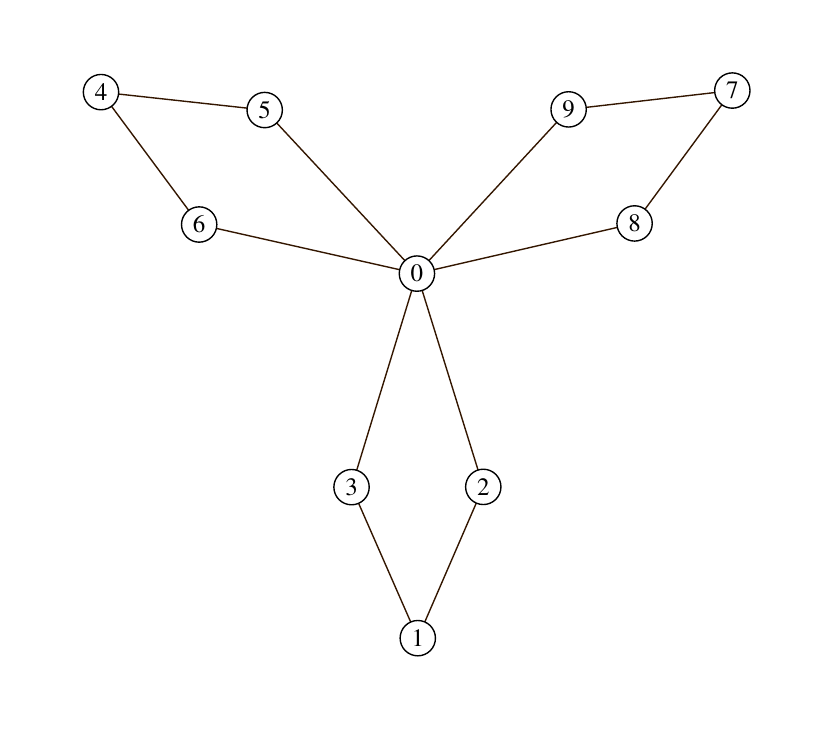}\caption{First example of a graph
for a quantum code that is locally Clifford equivalent to the Shor
[[9,1,3]]-code.}%
\label{fig7}%
\end{figure}

Omitting further details and applying the VdN-work, the symmetric $9\times9$
adjacency matrix $\Gamma$ for the Shor code becomes,%
\begin{equation}
\Gamma\overset{\text{def}}{=}Z^{\prime\text{T}}\cdot\left(  X^{\prime\text{T}%
}\right)  ^{-1}=\left(
\begin{array}
[c]{ccccccccc}%
0 & 1 & 1 & 0 & 0 & 0 & 0 & 0 & 0\\
1 & 0 & 0 & 0 & 0 & 0 & 0 & 0 & 0\\
1 & 0 & 0 & 0 & 0 & 0 & 0 & 0 & 0\\
0 & 0 & 0 & 0 & 1 & 1 & 0 & 0 & 0\\
0 & 0 & 0 & 1 & 0 & 0 & 0 & 0 & 0\\
0 & 0 & 0 & 1 & 0 & 0 & 0 & 0 & 0\\
0 & 0 & 0 & 0 & 0 & 0 & 0 & 1 & 1\\
0 & 0 & 0 & 0 & 0 & 0 & 1 & 0 & 0\\
0 & 0 & 0 & 0 & 0 & 0 & 1 & 0 & 0
\end{array}
\right)  \text{.} \label{gammashor1}%
\end{equation}
Finally, applying the S-work, the $10\times10$ symmetric coincidence matrix
$\Xi_{\left[  \left[  9,1,3\right]  \right]  }$ characterizing the graph with
both input and output vertices reads,%
\begin{equation}
\Xi_{\left[  \left[  9,1,3\right]  \right]  }\overset{\text{def}}{=}\left(
\begin{array}
[c]{cccccccccc}%
0 & 0 & 1 & 1 & 0 & 1 & 1 & 0 & 1 & 1\\
0 & 0 & 1 & 1 & 0 & 0 & 0 & 0 & 0 & 0\\
1 & 1 & 0 & 0 & 0 & 0 & 0 & 0 & 0 & 0\\
1 & 1 & 0 & 0 & 0 & 0 & 0 & 0 & 0 & 0\\
0 & 0 & 0 & 0 & 0 & 1 & 1 & 0 & 0 & 0\\
1 & 0 & 0 & 0 & 1 & 0 & 0 & 0 & 0 & 0\\
1 & 0 & 0 & 0 & 1 & 0 & 0 & 0 & 0 & 0\\
0 & 0 & 0 & 0 & 0 & 0 & 0 & 0 & 1 & 1\\
1 & 0 & 0 & 0 & 0 & 0 & 0 & 1 & 0 & 0\\
1 & 0 & 0 & 0 & 0 & 0 & 0 & 1 & 0 & 0
\end{array}
\right)  \text{.}%
\end{equation}
In what follows, we shall consider an alternative path leading to a graph for
the nine-qubit stabilizer code. Finally, we shall discuss the error-correcting
capability of the code in graph-theoretic terms as originally advocated in the SW-work.

\subsubsection{Second case}

Being within the CWS\ framework, consider a graph with nine vertices
characterized by the following canonical codeword stabilizer,%
\begin{equation}
\mathcal{S}_{\text{CWS}}\overset{\text{def}}{=}\left\langle g_{1}\text{,
}g_{2}\text{, }g_{3}\text{, }g_{4}\text{, }g_{5}\text{, }g_{6}\text{, }%
g_{7}\text{, }g_{8}\text{, }g_{9}\right\rangle \text{,} \label{s2}%
\end{equation}
with,%
\begin{align}
&  g_{1}\overset{\text{def}}{=}X^{1}Z^{2}Z^{3}Z^{4}Z^{5}\text{, }g_{2}%
\overset{\text{def}}{=}Z^{1}X^{2}\text{, }g_{3}\overset{\text{def}}{=}%
Z^{1}X^{3}\text{, }g_{4}\overset{\text{def}}{=}Z^{1}X^{4}Z^{5}Z^{6}%
Z^{7}\text{, }g_{5}\overset{\text{def}}{=}Z^{4}X^{5}\text{, }g_{6}%
\overset{\text{def}}{=}Z^{4}X^{6}\text{, }\nonumber\\
& \nonumber\\
&  g_{7}\overset{\text{def}}{=}Z^{1}Z^{4}X^{7}Z^{8}Z^{9}\text{, }g_{8}%
\overset{\text{def}}{=}Z^{7}X^{8}\text{, }g_{9}\equiv\bar{Z}\overset
{\text{def}}{=}Z^{7}X^{9}\text{.}%
\end{align}
The $9\times9$ adjacency matrix $\Gamma$ for this graph is given by,%
\begin{equation}
\Gamma\overset{\text{def}}{=}\left(
\begin{array}
[c]{ccccccccc}%
0 & 1 & 1 & 1 & 0 & 0 & 1 & 0 & 0\\
1 & 0 & 0 & 0 & 0 & 0 & 0 & 0 & 0\\
1 & 0 & 0 & 0 & 0 & 0 & 0 & 0 & 0\\
1 & 0 & 0 & 0 & 1 & 1 & 1 & 0 & 0\\
0 & 0 & 0 & 1 & 0 & 0 & 0 & 0 & 0\\
0 & 0 & 0 & 1 & 0 & 0 & 0 & 0 & 0\\
1 & 0 & 0 & 1 & 0 & 0 & 0 & 1 & 1\\
0 & 0 & 0 & 0 & 0 & 0 & 1 & 0 & 0\\
0 & 0 & 0 & 0 & 0 & 0 & 1 & 0 & 0
\end{array}
\right)  \text{.} \label{chist}%
\end{equation}
Applying the S-work, the $10\times10$ symmetric coincidence matrix
$\Xi_{\left[  \left[  9,1,3\right]  \right]  }^{\prime}$ characterizing the
graph with both input and output vertices becomes,%
\begin{equation}
\Xi_{\left[  \left[  9,1,3\right]  \right]  }^{\prime}\overset{\text{def}}%
{=}\left(
\begin{array}
[c]{cccccccccc}%
0 & 0 & 1 & 1 & 0 & 1 & 1 & 0 & 1 & 1\\
0 & 0 & 1 & 1 & 1 & 0 & 0 & 1 & 0 & 0\\
1 & 1 & 0 & 0 & 0 & 0 & 0 & 0 & 0 & 0\\
1 & 1 & 0 & 0 & 0 & 0 & 0 & 0 & 0 & 0\\
0 & 1 & 0 & 0 & 0 & 1 & 1 & 1 & 0 & 0\\
1 & 0 & 0 & 0 & 1 & 0 & 0 & 0 & 0 & 0\\
1 & 0 & 0 & 0 & 1 & 0 & 0 & 0 & 0 & 0\\
0 & 1 & 0 & 0 & 1 & 0 & 0 & 0 & 1 & 1\\
1 & 0 & 0 & 0 & 0 & 0 & 0 & 1 & 0 & 0\\
1 & 0 & 0 & 0 & 0 & 0 & 0 & 1 & 0 & 0
\end{array}
\right)  \text{.}%
\end{equation}
Does the graph \ associated with the adjacency matrix in Eq. (\ref{chist})
realize the Shor nine-qubit code? If we show that $\mathcal{S}_{\text{CWS}}$
in Eq. (\ref{s2}) is locally Clifford equivalent to a new stabilizer
$\mathcal{S}_{\text{CWS}}^{\prime}$ from which we can construct a graph that
realizes the Shor code, then we can reply with an affirmative answer. Observe
that $\mathcal{S}_{\text{CWS}}$ in Eq. (\ref{s2}) is locally Clifford
equivalent to $\mathcal{S}_{\text{CWS}}^{\prime}$ defined as,%
\begin{equation}
\mathcal{S}_{\text{CWS}}^{\prime}\overset{\text{def}}{=}U\mathcal{S}%
_{\text{CWS}}U^{\dagger}\text{ with, }U\overset{\text{def}}{=}P^{1}\otimes
H^{2}\otimes H^{3}\otimes P^{4}\otimes H^{5}\otimes H^{6}\otimes P^{7}\otimes
H^{8}\otimes H^{9}\text{.} \label{s3}%
\end{equation}
Using Eqs. (\ref{s2}) and (\ref{s3}), it follows that%
\begin{equation}
\mathcal{S}_{\text{CWS}}^{\prime}\overset{\text{def}}{=}\left\langle
g_{1}^{\prime}\text{, }g_{2}^{\prime}\text{, }g_{3}^{\prime}\text{, }%
g_{4}^{\prime}\text{, }g_{5}^{\prime}\text{, }g_{6}^{\prime}\text{, }%
g_{7}^{\prime}\text{, }g_{8}^{\prime}\text{, }\bar{Z}^{\prime}\right\rangle
\text{,}%
\end{equation}
with,%
\begin{align}
&  g_{1}^{\prime}\overset{\text{def}}{=}Z^{1}X^{2}\text{, }g_{2}^{\prime
}\overset{\text{def}}{=}Z^{1}X^{3}\text{, }g_{3}^{\prime}\overset{\text{def}%
}{=}Z^{4}X^{5}\text{, }g_{4}^{\prime}\overset{\text{def}}{=}Z^{4}X^{6}\text{,
}g_{5}^{\prime}\overset{\text{def}}{=}Z^{7}X^{8}\text{, }g_{6}^{\prime
}\overset{\text{def}}{=}Z^{7}X^{9}\text{, }\nonumber\\
& \nonumber\\
&  g_{7}^{\prime}\overset{\text{def}}{=}Y^{1}Z^{2}Z^{3}Y^{4}Z^{5}Z^{6}\text{,
}g_{8}^{\prime}\overset{\text{def}}{=}Y^{1}Z^{2}Z^{3}Y^{7}Z^{8}Z^{9}\text{,
}\bar{Z}^{\prime}\overset{\text{def}}{=}Y^{1}Z^{2}Z^{3}Y^{4}Z^{5}Z^{6}%
Y^{7}Z^{8}Z^{9}\text{.}%
\end{align}
We observe that the the codeword stabilizer matrix $\mathcal{H}_{\mathcal{S}%
_{\text{CWS}}^{\prime}}$ corresponding to $\mathcal{S}_{\text{CWS}}^{\prime}$
becomes,%
\begin{equation}
\mathcal{H}_{\mathcal{S}_{\text{CWS}}^{\prime}}\overset{\text{def}}{=}\left(
\begin{array}
[c]{ccccccccc}%
1 & 0 & 0 & 0 & 0 & 0 & 0 & 0 & 0\\
1 & 0 & 0 & 0 & 0 & 0 & 0 & 0 & 0\\
0 & 0 & 0 & 1 & 0 & 0 & 0 & 0 & 0\\
0 & 0 & 0 & 1 & 0 & 0 & 0 & 0 & 0\\
0 & 0 & 0 & 0 & 0 & 0 & 1 & 0 & 0\\
0 & 0 & 0 & 0 & 0 & 0 & 1 & 0 & 0\\
1 & 1 & 1 & 1 & 1 & 1 & 0 & 0 & 0\\
1 & 1 & 1 & 0 & 0 & 0 & 1 & 1 & 1\\
1 & 1 & 1 & 1 & 1 & 1 & 1 & 1 & 1
\end{array}
\left\vert
\begin{array}
[c]{ccccccccc}%
0 & 1 & 0 & 0 & 0 & 0 & 0 & 0 & 0\\
0 & 0 & 1 & 0 & 0 & 0 & 0 & 0 & 0\\
0 & 0 & 0 & 0 & 1 & 0 & 0 & 0 & 0\\
0 & 0 & 0 & 0 & 0 & 1 & 0 & 0 & 0\\
0 & 0 & 0 & 0 & 0 & 0 & 0 & 1 & 0\\
0 & 0 & 0 & 0 & 0 & 0 & 0 & 0 & 1\\
1 & 0 & 0 & 1 & 0 & 0 & 0 & 0 & 0\\
1 & 0 & 0 & 0 & 0 & 0 & 1 & 0 & 0\\
1 & 0 & 0 & 1 & 0 & 0 & 1 & 0 & 0
\end{array}
\right.  \right)  \text{.}%
\end{equation}
Omitting further details and applying the VdN-work, the $9\times9$ symmetric
adjacency matrix $\Gamma$ associated with the new graph reads,%
\begin{equation}
\Gamma\overset{\text{def}}{=}\left(
\begin{array}
[c]{ccccccccc}%
0 & 1 & 1 & 0 & 0 & 0 & 0 & 0 & 0\\
1 & 0 & 0 & 0 & 0 & 0 & 0 & 0 & 0\\
1 & 0 & 0 & 0 & 0 & 0 & 0 & 0 & 0\\
0 & 0 & 0 & 0 & 1 & 1 & 0 & 0 & 0\\
0 & 0 & 0 & 1 & 0 & 0 & 0 & 0 & 0\\
0 & 0 & 0 & 1 & 0 & 0 & 0 & 0 & 0\\
0 & 0 & 0 & 0 & 0 & 0 & 0 & 1 & 1\\
0 & 0 & 0 & 0 & 0 & 0 & 1 & 0 & 0\\
0 & 0 & 0 & 0 & 0 & 0 & 1 & 0 & 0
\end{array}
\right)  \text{.} \label{gammashor2}%
\end{equation}

\begin{figure}[ptb]
\centering
\includegraphics[width=0.35\textwidth]{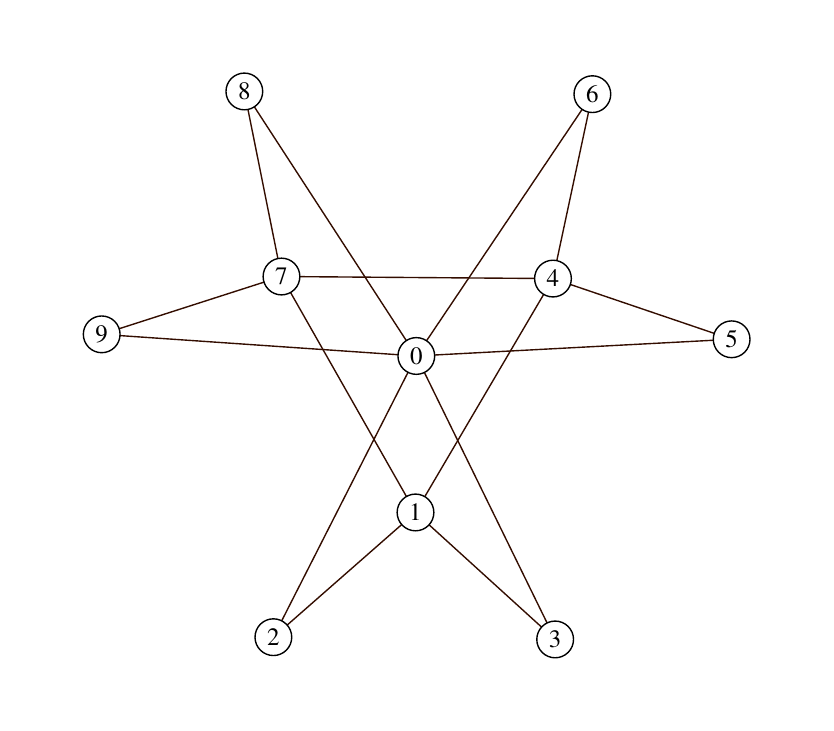}\caption{Second example of a graph
for a quantum code that is locally Clifford equivalent to the Shor
[[9,1,3]]-code.}%
\label{fig8}%
\end{figure}

Since the adjacency matrices in Eqs. (\ref{gammashor1}) and (\ref{gammashor2})
represent essentially the same graphs, we conclude that the graph with
canonical stabilizer (\ref{s2}) realizes\textbf{\ }the Shor code as well. In
addition, we point out that among all the possible $36$ two-element error
configurations, $27$ configurations satisfy the strong error correction
condition, $3$ satisfy the weak error correction condition and $6$ do not
satisfy neither of them but they are harmless as we shall show. The strongly
correctable $27$ configurations are given by,%
\begin{align}
&  \left\{  0\text{, }1\text{, }4\right\}  \text{, }\left\{  0\text{,
}1\text{, }5\right\}  \text{, }\left\{  0\text{, }1\text{, }6\right\}  \text{,
}\left\{  0\text{, }1\text{, }7\right\}  \text{, }\left\{  0\text{, }1\text{,
}8\right\}  \text{, }\left\{  0\text{, }1\text{, }9\right\}  \text{, }\left\{
0\text{, }2\text{, }4\right\}  \text{, }\left\{  0\text{, }2\text{,
}5\right\}  \text{, }\left\{  0\text{, }2\text{, }6\right\}  \text{, }\left\{
0\text{, }2\text{, }7\right\}  \text{,}\nonumber\\
& \nonumber\\
&  \text{ }\left\{  0\text{, }2\text{, }8\right\}  \text{, }\left\{  0\text{,
}2\text{, }9\right\}  \text{, }\left\{  0\text{, }3\text{, }4\right\}  \text{,
}\left\{  0\text{, }3\text{, }5\right\}  \text{, }\left\{  0\text{, }3\text{,
}6\right\}  \text{, }\left\{  0\text{, }3\text{, }7\right\}  \text{, }\left\{
0\text{, }3\text{, }8\right\}  \text{, }\left\{  0\text{, }3\text{,
}9\right\}  \text{, }\left\{  0\text{, }4\text{, }7\right\}  \text{, }\left\{
0\text{, }4\text{, }8\right\}  \text{,}\nonumber\\
& \nonumber\\
&  \text{ }\left\{  0\text{, }4\text{, }9\right\}  \text{, }\left\{  0\text{,
}5\text{, }7\right\}  \text{, }\left\{  0\text{, }5\text{, }8\right\}  \text{,
}\left\{  0\text{, }5\text{, }9\right\}  \text{, }\left\{  0\text{, }6\text{,
}7\right\}  \text{, }\left\{  0\text{, }6\text{, }8\right\}  \text{, }\left\{
0\text{, }6\text{, }9\right\}  \text{,}%
\end{align}
while the weakly correctable $3$ configurations read,%
\begin{equation}
\left\{  0\text{, }2\text{, }3\right\}  \text{, }\left\{  0\text{, }5\text{,
}6\right\}  \text{, }\left\{  0\text{, }8\text{, }9\right\}  \text{.}%
\end{equation}
Finally, the potentially dangerous $6$ error configurations are,%
\begin{equation}
\left\{  0\text{, }1\text{, }2\right\}  \text{, }\left\{  0\text{, }1\text{,
}3\right\}  \text{, }\left\{  0\text{, }4\text{, }5\right\}  \text{, }\left\{
0\text{, }4\text{, }6\right\}  \text{, }\left\{  0\text{, }7\text{,
}8\right\}  \text{, }\left\{  0\text{, }7\text{, }9\right\}  \text{.}
\label{badcon}%
\end{equation}
Each of the $6$ two-error configurations in Eq. (\ref{badcon}) generates $9$
weight-$2$ error operators for a total of $54$ errors. It turns out that in
each set of errors of cardinality $9$, there is $1$ weight-$2$ nondetectable
nontrivial error. However, this single error operator belongs to the
stabilizer $\mathcal{S}_{\text{stabilizer}}$ of the code and therefore it will
have no impact on the encoded quantum state. Thus, the code considered has
indeed distance $d=3$. To be explicit, consider the set $\left\{  0\text{,
}1\text{, }2\right\}  $. This sets generates the following $9$ weight-$2$
error operators,%
\begin{equation}
X^{1}X^{2}\text{, }X^{1}Y^{2}\text{, }X^{1}Z^{2}\text{, }Y^{1}X^{2}\text{,
}Y^{1}Y^{2}\text{, }Y^{1}Z^{2}\text{, }Z^{1}X^{2}\text{, }Z^{1}Y^{2}\text{,
}Z^{1}Z^{2}\text{.}%
\end{equation}
The only nontrivial error with vanishing error syndrome is $Z^{1}Z^{2}$ which,
however, belongs to the stabilizer.

\subsection{The $\left[  \left[  11,1,5\right]  \right]  $ stabilizer code}

$\allowbreak$The smallest possible code protecting against two arbitrary
errors maps one logical qubit into eleven physical qubits. The existence of
such a code was proven in \cite{robert} while its stabilizer structure was
constructed in \cite{daniel-phd}. When\ realized as a CWS code, the
eleven-qubit quantum stabilizer code is characterized by the codeword
stabilizer,%
\begin{equation}
\mathcal{S}_{\text{CWS}}\overset{\text{def}}{=}\left\langle g_{1}\text{,
}g_{2}\text{, }g_{3}\text{, }g_{4}\text{, }g_{5}\text{, }g_{6}\text{, }%
g_{7}\text{, }g_{8}\text{, }g_{9}\text{, }g_{10}\text{, }\bar{Z}\right\rangle
\text{,} \label{eleven}%
\end{equation}
with \cite{daniel-phd},%
\begin{align}
&  g_{1}\overset{\text{def}}{=}Z^{1}Z^{2}Z^{3}Z^{4}Z^{5}Z^{6}\text{, }%
g_{2}\overset{\text{def}}{=}X^{1}X^{2}X^{3}X^{4}X^{5}X^{6}\text{, }%
g_{3}\overset{\text{def}}{=}Z^{4}X^{5}Y^{6}Y^{7}Y^{8}Y^{9}X^{10}Z^{11}\text{,
}g_{4}\overset{\text{def}}{=}X^{4}Y^{5}Z^{6}Z^{7}Z^{8}Z^{9}Y^{10}%
X^{11}\text{,}\nonumber\\
& \nonumber\\
&  g_{5}\overset{\text{def}}{=}Z^{1}Y^{2}X^{3}Z^{7}Y^{8}X^{9}\text{, }%
g_{6}\overset{\text{def}}{=}X^{1}Z^{2}Y^{3}X^{7}Z^{8}Y^{9}\text{, }%
g_{7}\overset{\text{def}}{=}Z^{4}Y^{5}X^{6}X^{7}Y^{8}Z^{9}\text{, }%
g_{8}\overset{\text{def}}{=}X^{4}Z^{5}Y^{6}Z^{7}X^{8}Y^{9}\text{,}\nonumber\\
& \nonumber\\
&  g_{9}\overset{\text{def}}{=}Z^{1}X^{2}Y^{3}Z^{7}Z^{8}Z^{9}X^{10}%
Y^{11}\text{, }g_{10}\overset{\text{def}}{=}Y^{1}Z^{2}X^{3}Y^{7}Y^{8}%
Y^{9}Z^{10}X^{11}\text{, }\bar{Z}\overset{\text{def}}{=}Z^{7}Z^{8}Z^{9}%
Z^{10}Z^{11}\text{.}%
\end{align}
What is the graph that realizes such eleven-qubit code? The codeword
stabilizer matrix $\mathcal{H}_{\mathcal{S}_{\text{CWS}}}$ corresponding to
$\mathcal{S}_{\text{CWS}}$ in Eq. (\ref{eleven}) can be formally written as,%
\begin{equation}
\mathcal{H}_{\mathcal{S}_{\text{CWS}}}\overset{\text{def}}{=}\left(
Z\left\vert X\right.  \right)  =\left(
\begin{array}
[c]{ccccccccccc}%
1 & 1 & 1 & 1 & 1 & 1 & 0 & 0 & 0 & 0 & 0\\
0 & 0 & 0 & 0 & 0 & 0 & 0 & 0 & 0 & 0 & 0\\
0 & 0 & 0 & 1 & 0 & 1 & 1 & 1 & 1 & 0 & 1\\
0 & 0 & 0 & 0 & 1 & 1 & 1 & 1 & 1 & 1 & 0\\
1 & 1 & 0 & 0 & 0 & 0 & 1 & 1 & 0 & 0 & 0\\
0 & 1 & 1 & 0 & 0 & 0 & 0 & 1 & 1 & 0 & 0\\
0 & 0 & 0 & 1 & 1 & 0 & 0 & 1 & 1 & 0 & 0\\
0 & 0 & 0 & 0 & 1 & 1 & 1 & 0 & 1 & 0 & 0\\
1 & 0 & 1 & 0 & 0 & 0 & 1 & 1 & 1 & 0 & 1\\
1 & 1 & 0 & 0 & 0 & 0 & 1 & 1 & 1 & 1 & 0\\
0 & 0 & 0 & 0 & 0 & 0 & 1 & 1 & 1 & 1 & 1
\end{array}
\left\vert
\begin{array}
[c]{ccccccccccc}%
0 & 0 & 0 & 0 & 0 & 0 & 0 & 0 & 0 & 0 & 0\\
1 & 1 & 1 & 1 & 1 & 1 & 0 & 0 & 0 & 0 & 0\\
0 & 0 & 0 & 0 & 1 & 1 & 1 & 1 & 1 & 1 & 0\\
0 & 0 & 0 & 1 & 1 & 0 & 0 & 0 & 0 & 1 & 1\\
0 & 1 & 1 & 0 & 0 & 0 & 0 & 1 & 1 & 0 & 0\\
1 & 0 & 1 & 0 & 0 & 0 & 1 & 0 & 1 & 0 & 0\\
0 & 0 & 0 & 0 & 1 & 1 & 1 & 1 & 0 & 0 & 0\\
0 & 0 & 0 & 1 & 0 & 1 & 0 & 1 & 1 & 0 & 0\\
0 & 1 & 1 & 0 & 0 & 0 & 0 & 0 & 0 & 1 & 1\\
1 & 0 & 1 & 0 & 0 & 0 & 1 & 1 & 1 & 0 & 1\\
0 & 0 & 0 & 0 & 0 & 0 & 0 & 0 & 0 & 0 & 0
\end{array}
\right.  \right)  \text{.}%
\end{equation}
Since $X^{\text{T}}$ is not invertible, the algorithmic procedure introduced
in the VdN-work cannot be applied. However, we notice that the stabilizer
$\mathcal{S}_{\text{CWS}}$ in Eq. (\ref{eleven}) is locally Clifford
equivalent to $\mathcal{S}_{\text{CWS}}^{\prime}$ defined by,%
\begin{equation}
\mathcal{S}_{\text{CWS}}^{\prime}\overset{\text{def}}{=}U\mathcal{S}%
_{\text{CWS}}U^{\dagger}\text{,}%
\end{equation}
where the unitary operator $U$ is defined as,%
\begin{equation}
U\overset{\text{def}}{=}\left(  H^{1}P^{1}H^{1}\right)  \otimes H^{2}\otimes
H^{3}\otimes H^{4}\otimes H^{5}\otimes H^{6}\otimes H^{7}\otimes H^{8}\otimes
H^{9}\otimes H^{10}\otimes\left(  H^{11}P^{11}H^{11}\right)  \text{.}
\label{eleven1}%
\end{equation}
The operator $U$ can be regarded as the composition of three unitary operators
$U\overset{\text{def}}{=}U_{3}\circ U_{2}\circ U_{1}$ with,%
\begin{align}
&  U_{1}\overset{\text{def}}{=}H^{1}\otimes I^{2}\otimes I^{3}\otimes
I^{4}\otimes I^{5}\otimes I^{6}\otimes I^{7}\otimes I^{8}\otimes I^{9}\otimes
I^{10}\otimes H^{11}\text{,}\nonumber\\
& \nonumber\\
&  U_{2}\overset{\text{def}}{=}P^{1}\otimes I^{2}\otimes I^{3}\otimes
I^{4}\otimes I^{5}\otimes I^{6}\otimes I^{7}\otimes I^{8}\otimes I^{9}\otimes
I^{10}\otimes P^{11}\text{,}\nonumber\\
& \nonumber\\
&  U_{3}\overset{\text{def}}{=}H^{1}\otimes H^{2}\otimes H^{3}\otimes
H^{4}\otimes H^{5}\otimes H^{6}\otimes H^{7}\otimes H^{8}\otimes H^{9}\otimes
H^{10}\otimes H^{11}\text{.}%
\end{align}
Using Eqs. (\ref{eleven}) and (\ref{eleven1}), it follows that%
\begin{equation}
\mathcal{S}_{\text{CWS}}^{\prime}\overset{\text{def}}{=}\left\langle
g_{1}^{\prime}\text{, }g_{2}^{\prime}\text{, }g_{3}^{\prime}\text{, }%
g_{4}^{\prime}\text{, }g_{5}^{\prime}\text{, }g_{6}^{\prime}\text{, }%
g_{7}^{\prime}\text{, }g_{8}^{\prime}\text{, }g_{9}^{\prime}\text{, }%
g_{10}^{\prime}\text{, }\bar{Z}^{\prime}\right\rangle \text{,}%
\end{equation}
with,%
\begin{align}
&  g_{1}^{\prime}\overset{\text{def}}{=}Y^{1}X^{2}X^{3}X^{4}X^{5}X^{6}\text{,
}g_{2}^{\prime}\overset{\text{def}}{=}X^{1}Z^{2}Z^{3}Z^{4}Z^{5}Z^{6}\text{,
}g_{3}^{\prime}\overset{\text{def}}{=}X^{4}Z^{5}Y^{6}Y^{7}Y^{8}Y^{9}%
Z^{10}Y^{11}\text{, }g_{4}^{\prime}\overset{\text{def}}{=}Z^{4}Y^{5}X^{6}%
X^{7}X^{8}X^{9}Y^{10}X^{11}\text{,}\nonumber\\
& \nonumber\\
&  g_{5}^{\prime}\overset{\text{def}}{=}Y^{1}Y^{2}Z^{3}X^{7}Y^{8}Z^{9}\text{,
}g_{6}^{\prime}\overset{\text{def}}{=}X^{1}X^{2}Y^{3}Z^{7}X^{8}Y^{9}\text{,
}g_{7}^{\prime}\overset{\text{def}}{=}X^{4}Y^{5}Z^{6}Z^{7}Y^{8}X^{9}\text{,
}g_{8}^{\prime}\overset{\text{def}}{=}Z^{4}X^{5}Y^{6}X^{7}Z^{8}Y^{9}%
\text{,}\nonumber\\
& \nonumber\\
&  g_{9}^{\prime}\overset{\text{def}}{=}Y^{1}Z^{2}Y^{3}X^{7}X^{8}X^{9}%
Z^{10}Z^{11}\text{, }g_{10}^{\prime}\overset{\text{def}}{=}Z^{1}X^{2}%
Z^{3}Y^{7}Y^{8}Y^{9}X^{10}X^{11}\text{, }\bar{Z}^{\prime}\overset{\text{def}%
}{=}X^{7}X^{8}X^{9}X^{10}Y^{11}\text{.}%
\end{align}
We observe that the codeword stabilizer matrix $\mathcal{H}_{\mathcal{S}%
_{\text{CWS}}^{\prime}}$ corresponding to $\mathcal{S}_{\text{CWS}}^{\prime}$
becomes,%
\begin{equation}
\mathcal{H}_{\mathcal{S}_{\text{CWS}}^{\prime}}\overset{\text{def}}{=}\left(
X^{\prime}\left\vert Z^{\prime}\right.  \right)  =\left(
\begin{array}
[c]{ccccccccccc}%
1 & 0 & 0 & 0 & 0 & 0 & 0 & 0 & 0 & 0 & 0\\
0 & 1 & 1 & 1 & 1 & 1 & 0 & 0 & 0 & 0 & 0\\
0 & 0 & 0 & 0 & 1 & 1 & 1 & 1 & 1 & 1 & 1\\
0 & 0 & 0 & 1 & 1 & 0 & 0 & 0 & 0 & 1 & 0\\
1 & 1 & 1 & 0 & 0 & 0 & 0 & 1 & 1 & 0 & 0\\
0 & 0 & 1 & 0 & 0 & 0 & 1 & 0 & 1 & 0 & 0\\
0 & 0 & 0 & 0 & 1 & 1 & 1 & 1 & 0 & 0 & 0\\
0 & 0 & 0 & 1 & 0 & 1 & 0 & 1 & 1 & 0 & 0\\
1 & 1 & 1 & 0 & 0 & 0 & 0 & 0 & 0 & 1 & 1\\
1 & 0 & 1 & 0 & 0 & 0 & 1 & 1 & 1 & 0 & 0\\
0 & 0 & 0 & 0 & 0 & 0 & 0 & 0 & 0 & 0 & 1
\end{array}
\left\vert
\begin{array}
[c]{ccccccccccc}%
1 & 1 & 1 & 1 & 1 & 1 & 0 & 0 & 0 & 0 & 0\\
1 & 0 & 0 & 0 & 0 & 0 & 0 & 0 & 0 & 0 & 0\\
0 & 0 & 0 & 1 & 0 & 1 & 1 & 1 & 1 & 0 & 1\\
0 & 0 & 0 & 0 & 1 & 1 & 1 & 1 & 1 & 1 & 1\\
1 & 1 & 0 & 0 & 0 & 0 & 1 & 1 & 0 & 0 & 0\\
1 & 1 & 1 & 0 & 0 & 0 & 0 & 1 & 1 & 0 & 0\\
0 & 0 & 0 & 1 & 1 & 0 & 0 & 1 & 1 & 0 & 0\\
0 & 0 & 0 & 0 & 1 & 1 & 1 & 0 & 1 & 0 & 0\\
1 & 0 & 1 & 0 & 0 & 0 & 1 & 1 & 1 & 0 & 0\\
0 & 1 & 0 & 0 & 0 & 0 & 1 & 1 & 1 & 1 & 1\\
0 & 0 & 0 & 0 & 0 & 0 & 1 & 1 & 1 & 1 & 1
\end{array}
\right.  \right)  \text{.}%
\end{equation}

\begin{figure}[ptb]
\centering
\includegraphics[width=0.35\textwidth]{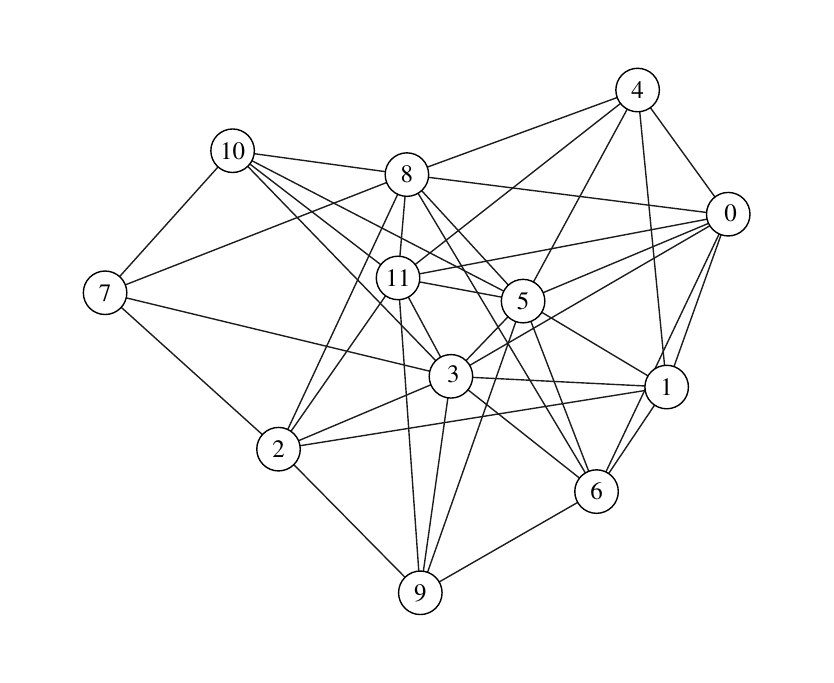}\caption{Graph for a quantum code
that is locally Clifford equivalent to the Gottesman [[11,1,5]]-code.}%
\label{fig9}%
\end{figure}

Omitting further details and applying the VdN-work, the $11\times11$ symmetric
adjacency matrix $\Gamma$ for the eleven-qubit code becomes,%
\begin{equation}
\Gamma\overset{\text{def}}{=}\left(
\begin{array}
[c]{ccccccccccc}%
0 & 1 & 1 & 1 & 1 & 1 & 0 & 0 & 0 & 0 & 0\\
1 & 0 & 1 & 0 & 0 & 0 & 1 & 1 & 1 & 0 & 1\\
1 & 1 & 0 & 0 & 1 & 1 & 1 & 0 & 1 & 1 & 1\\
1 & 0 & 0 & 0 & 1 & 0 & 0 & 1 & 0 & 0 & 1\\
1 & 0 & 1 & 1 & 0 & 1 & 0 & 1 & 1 & 1 & 1\\
1 & 0 & 1 & 0 & 1 & 0 & 0 & 1 & 1 & 0 & 0\\
0 & 1 & 1 & 0 & 0 & 0 & 0 & 1 & 0 & 1 & 0\\
0 & 1 & 0 & 1 & 1 & 1 & 1 & 0 & 0 & 1 & 1\\
0 & 1 & 1 & 0 & 1 & 1 & 0 & 0 & 0 & 0 & 1\\
0 & 0 & 1 & 0 & 1 & 0 & 1 & 1 & 0 & 0 & 1\\
0 & 1 & 1 & 1 & 1 & 0 & 0 & 1 & 1 & 1 & 0
\end{array}
\right)  \text{.}%
\end{equation}
Employing the S-work, the $12\times12$ symmetric coincidence matrix
$\Xi_{\left[  \left[  11,1,5\right]  \right]  }$ can be written as,%
\begin{equation}
\Xi_{\left[  \left[  11,1,5\right]  \right]  }\overset{\text{def}}{=}\left(
\begin{array}
[c]{cccccccccccc}%
0 & a_{1} & a_{2} & a_{3} & a_{4} & a_{5} & a_{6} & a_{7} & a_{8} & a_{9} &
a_{10} & a_{11}\\
a_{1} & 0 & 1 & 1 & 1 & 1 & 1 & 0 & 0 & 0 & 0 & 0\\
a_{2} & 1 & 0 & 1 & 0 & 0 & 0 & 1 & 1 & 1 & 0 & 1\\
a_{3} & 1 & 1 & 0 & 0 & 1 & 1 & 1 & 0 & 1 & 1 & 1\\
a_{4} & 1 & 0 & 0 & 0 & 1 & 0 & 0 & 1 & 0 & 0 & 1\\
a_{5} & 1 & 0 & 1 & 1 & 0 & 1 & 0 & 1 & 1 & 1 & 1\\
a_{6} & 1 & 0 & 1 & 0 & 1 & 0 & 0 & 1 & 1 & 0 & 0\\
a_{7} & 0 & 1 & 1 & 0 & 0 & 0 & 0 & 1 & 0 & 1 & 0\\
a_{8} & 0 & 1 & 0 & 1 & 1 & 1 & 1 & 0 & 0 & 1 & 1\\
a_{9} & 0 & 1 & 1 & 0 & 1 & 1 & 0 & 0 & 0 & 0 & 1\\
a_{10} & 0 & 0 & 1 & 0 & 1 & 0 & 1 & 1 & 0 & 0 & 1\\
a_{11} & 0 & 1 & 1 & 1 & 1 & 0 & 0 & 1 & 1 & 1 & 0
\end{array}
\right)  \text{,}%
\end{equation}
where the eleven matrix coefficients $a_{k}$ with $k\in\left\{  1\text{,...,
}11\right\}  $ satisfy the following eleven constraints,%
\begin{align}
a_{2}+a_{3}+a_{4}+a_{5}+a_{6}  &  =0\text{, }\nonumber\\
a_{1}+a_{3}+a_{7}+a_{8}+a_{9}+a_{11}  &  =0\text{,}\nonumber\\
\text{ }a_{1}+a_{2}+a_{5}+a_{6}+a_{7}+a_{9}+a_{10}+a_{11}  &  =0\text{,}%
\nonumber\\
a_{1}+a_{5}+a_{8}+a_{11}  &  =0\text{, }\nonumber\\
a_{1}+a_{3}+a_{4}+a_{6}+a_{8}+a_{9}+a_{10}+a_{11}  &  =0\text{, }\nonumber\\
a_{1}+a_{3}+a_{5}+a_{8}+a_{9}  &  =0\text{, }\nonumber\\
a_{2}+a_{3}+a_{8}+a_{10}  &  =0\text{,}\nonumber\\
a_{2}+a_{4}+a_{5}+a_{6}+a_{7}+a_{10}+a_{11}  &  =0\text{,}\nonumber\\
\text{ }a_{2}+a_{3}+a_{5}+a_{6}+a_{11}  &  =0\text{, }\nonumber\\
a_{3}+a_{5}+a_{7}+a_{8}+a_{11}  &  =0\text{, }\nonumber\\
a_{2}+a_{3}+a_{4}+a_{5}+a_{8}+a_{9}+a_{10}  &  =0\text{.} \label{sys}%
\end{align}
It turns out that a suitable solution of the system of equations in
(\ref{sys}) reads,%
\begin{equation}
\mathbf{a}=\left(  a_{1}\text{, }a_{2}\text{, }a_{3}\text{, }a_{4}\text{,
}a_{5}\text{, }a_{6}\text{, }a_{7}\text{, }a_{8}\text{, }a_{9}\text{, }%
a_{10}\text{, }a_{11}\right)  =\left(  1\text{, }0\text{, }1\text{, }1\text{,
}1\text{, }1\text{, }0\text{, }1\text{, }0\text{, }0\text{, }1\right)
\text{.}%
\end{equation}
Finally, the coincidence matrix $\Xi_{\left[  \left[  11,1,5\right]  \right]
}$ for a graph\textbf{ }associated with the eleven-qubit code reads,%
\begin{equation}
\Xi_{\left[  \left[  11,1,5\right]  \right]  }\overset{\text{def}}{=}\left(
\begin{array}
[c]{cccccccccccc}%
0 & 1 & 0 & 1 & 1 & 1 & 1 & 0 & 1 & 0 & 0 & 1\\
1 & 0 & 1 & 1 & 1 & 1 & 1 & 0 & 0 & 0 & 0 & 0\\
0 & 1 & 0 & 1 & 0 & 0 & 0 & 1 & 1 & 1 & 0 & 1\\
1 & 1 & 1 & 0 & 0 & 1 & 1 & 1 & 0 & 1 & 1 & 1\\
1 & 1 & 0 & 0 & 0 & 1 & 0 & 0 & 1 & 0 & 0 & 1\\
1 & 1 & 0 & 1 & 1 & 0 & 1 & 0 & 1 & 1 & 1 & 1\\
1 & 1 & 0 & 1 & 0 & 1 & 0 & 0 & 1 & 1 & 0 & 0\\
0 & 0 & 1 & 1 & 0 & 0 & 0 & 0 & 1 & 0 & 1 & 0\\
1 & 0 & 1 & 0 & 1 & 1 & 1 & 1 & 0 & 0 & 1 & 1\\
0 & 0 & 1 & 1 & 0 & 1 & 1 & 0 & 0 & 0 & 0 & 1\\
0 & 0 & 0 & 1 & 0 & 1 & 0 & 1 & 1 & 0 & 0 & 1\\
1 & 0 & 1 & 1 & 1 & 1 & 0 & 0 & 1 & 1 & 1 & 0
\end{array}
\right)  \text{.}%
\end{equation}
It is straightforward, though tedious, to check that all the $\binom{11}%
{2}=330$ four-error configurations satisfy the graph-theoretic error detection
conditions in their strong version in agreement with the SW-work for
nondegenerate codes. However, we also remark that checking out $330$ graphical
error detection conditions is always better that checking out $529$
Knill-Laflamme error correction conditions,%
\begin{equation}
3^{0}\binom{11}{0}+3^{1}\binom{11}{1}+3^{2}\binom{11}{2}=529>330\text{.}%
\end{equation}
In the next section, we shall consider few graphical constructions of
stabilizer codes characterized by multi-qubit encoding operators.

\section{Multi-qubit encoding}

\subsection{The $\left[  \left[  4,2,2\right]  \right]  $ stabilizer code}

In what follows, we shall consider the graphical construction of two
non-equivalent quantum stabilizer codes encoding two logical qubits into four
physical qubits.

\subsubsection{First case}

The $\left[  \left[  4,2,2\right]  \right]  $ code is the simplest example of
a class of $\left[  \left[  n-1,k+1,d-1\right]  \right]  $ codes that are
derivable from pure (or, nondegenerate) codes $\left[  \left[  n,k,d\right]
\right]  $ with $n\geq2$ (for more details, we refer to \cite{robert}) and is
an explicit example of multi-qubit encoding. It is derivable from the perfect
five-qubit code and can detect a single qubit error. The stabilizer generators
of the code are defined by \cite{gaitan},%
\begin{equation}
g_{1}\overset{\text{def}}{=}X^{1}Z^{2}Z^{3}X^{4}\text{ and, }g_{2}%
\overset{\text{def}}{=}Y^{1}X^{2}X^{3}Y^{4}\text{.} \label{please}%
\end{equation}
Each encoded qubit $i$ with $i\in\left\{  1\text{, }2\right\}  $ has its own
of logical operations $\bar{X}_{i}$ and $\bar{Z}_{i}$. A convenient choice is,%
\begin{equation}
\bar{X}_{1}\overset{\text{def}}{=}X^{1}Y^{3}Y^{4}\text{, }\bar{X}_{2}%
\overset{\text{def}}{=}X^{1}X^{3}Z^{4}\text{, }\bar{Z}_{1}\overset{\text{def}%
}{=}Y^{1}Z^{2}Y^{3}\text{ and, }\bar{Z}_{2}\overset{\text{def}}{=}X^{2}%
Z^{3}Z^{4}\text{.}%
\end{equation}
The codeword stabilizer $\mathcal{S}_{\text{CWS}}$ associated with the CWS
code that realizes this stabilizer code reads,%
\begin{equation}
\mathcal{S}_{\text{CWS}}\overset{\text{def}}{=}\left\langle g_{1}\text{,
}g_{2}\text{, }\bar{Z}_{1}\text{, }\bar{Z}_{2}\text{ }\right\rangle
=\left\langle X^{1}Z^{2}Z^{3}X^{4}\text{, }Y^{1}X^{2}X^{3}Y^{4}\text{, }%
Y^{1}Z^{2}Y^{3}\text{, }X^{2}Z^{3}Z^{4}\text{ }\right\rangle \text{.}
\label{SBcase1}%
\end{equation}
The codeword stabilizer matrix $\mathcal{H}_{\mathcal{S}_{\text{CWS}}}$
associated with $\mathcal{S}_{\text{CWS}}$ reads,%
\begin{equation}
\mathcal{H}_{\mathcal{S}_{\text{CWS}}}\overset{\text{def}}{=}\left(
Z\left\vert X\right.  \right)  =\left(
\begin{array}
[c]{cccc}%
0 & 1 & 1 & 0\\
1 & 0 & 0 & 1\\
1 & 1 & 1 & 0\\
0 & 0 & 1 & 1
\end{array}
\left\vert
\begin{array}
[c]{cccc}%
1 & 0 & 0 & 1\\
1 & 1 & 1 & 1\\
1 & 0 & 1 & 0\\
0 & 1 & 0 & 0
\end{array}
\right.  \right)  \text{.}%
\end{equation}
We observe $\det X\neq0$. Thus, using the VdN-work, the $4\times4$ adjacency
matrix $\Gamma$ becomes,%
\begin{equation}
\Gamma\overset{\text{def}}{=}\left(
\begin{array}
[c]{cccc}%
0 & 0 & 1 & 0\\
0 & 0 & 1 & 1\\
1 & 1 & 0 & 0\\
0 & 1 & 0 & 0
\end{array}
\right)  \text{.} \label{422g}%
\end{equation}
We remark that the graph with symmetric adjacency matrix $\Gamma$ in Eq.
(\ref{422g}) is in the local unitary equivalence class of the square graph
(see Figure $7$ in \cite{hein}). Therefore, an alternative graph (with only
output vertices) for our stabilizer code can be characterized by the
alternative $4\times4$ symmetric adjacency matrix $\Gamma^{\prime}$,%
\begin{equation}
\Gamma^{\prime}\overset{\text{def}}{=}\left(
\begin{array}
[c]{cccc}%
0 & 1 & 0 & 1\\
1 & 0 & 1 & 0\\
0 & 1 & 0 & 1\\
1 & 0 & 1 & 0
\end{array}
\right)  \text{.}%
\end{equation}
Finally, applying the S-work and considering $\Gamma^{\prime}$, the $6\times6$
symmetric coincidence matrix $\Xi_{\left[  \left[  4,2,2\right]  \right]
}^{\text{Beigi}}$ characterizing the graph with both input and output vertices
is given by,%
\begin{equation}
\Xi_{\left[  \left[  4,2,2\right]  \right]  }^{\text{Beigi}}\overset
{\text{def}}{=}\left(
\begin{array}
[c]{cccccc}%
0 & 0 & 1 & 0 & 0 & 1\\
0 & 0 & 0 & 1 & 1 & 0\\
1 & 0 & 0 & 1 & 0 & 1\\
0 & 1 & 1 & 0 & 1 & 0\\
0 & 1 & 0 & 1 & 0 & 1\\
1 & 0 & 1 & 0 & 1 & 0
\end{array}
\right)  \text{.} \label{bei}%
\end{equation}
Using the SW-work, it is simple to verify that any graphical single-error
configuration $\left\{  0\text{, }0^{\prime}\text{, }e\right\}  $ with
$e\in\left\{  1\text{, }2\text{, }3\text{, }4\right\}  $ is detectable. Thus,
the code detects any single-qubit error. As a final remark, we emphasize that
the\textbf{\ }graph\textbf{ }associated with $\Xi_{\left[  \left[
4,2,2\right]  \right]  }^{\text{Beigi}}$ is identical to the one appeared in
\cite{beigi}.

\begin{figure}[ptb]
\centering
\includegraphics[width=0.35\textwidth]{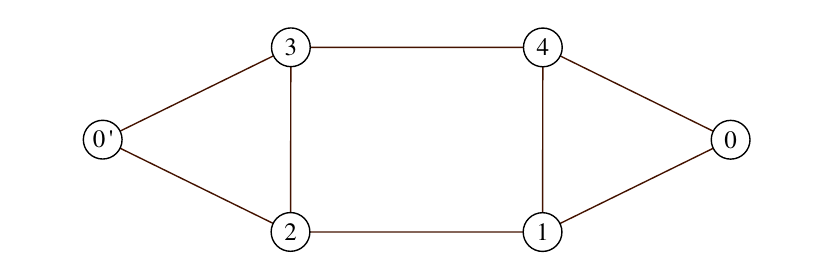}\caption{Graph for a quantum code
that is locally Clifford equivalent to the Beigi et al. [[4,2,2]]-code.}%
\label{fig10}%
\end{figure}

\subsubsection{Second case}

Let us consider a different\textbf{\ }$\left[  \left[  4,2,2\right]  \right]
$ stabilizer code with stabilizer generators defined by \cite{gaitan},
\begin{subequations}
\begin{equation}
g_{1}\overset{\text{def}}{=}X^{1}X^{2}X^{3}X^{4}\text{ and, }g_{2}%
\overset{\text{def}}{=}Z^{1}Z^{2}Z^{3}Z^{4}\text{.} \tag{B8}%
\end{equation}
Each encoded qubit $i$ with $i\in\left\{  1\text{, }2\right\}  $ has its own
of logical operations $\bar{X}_{i}$ and $\bar{Z}_{i}$. A convenient choice
is,
\end{subequations}
\begin{equation}
\bar{X}_{1}\overset{\text{def}}{=}X^{1}X^{2}\text{, }\bar{X}_{2}%
\overset{\text{def}}{=}X^{1}X^{3}\text{, }\bar{Z}_{1}\overset{\text{def}}%
{=}Z^{2}Z^{4}\text{ and, }\bar{Z}_{2}\overset{\text{def}}{=}Z^{3}Z^{4}\text{.}%
\end{equation}
The codeword stabilizer $\mathcal{S}_{\text{CWS}}$ associated with the CWS
code that realizes this stabilizer code reads,%
\begin{equation}
\mathcal{S}_{\text{CWS}}\overset{\text{def}}{=}\left\langle g_{1}\text{,
}g_{2}\text{, }\bar{Z}_{1}\text{, }\bar{Z}_{2}\text{ }\right\rangle
=\left\langle X^{1}X^{2}X^{3}X^{4}\text{, }Z^{1}Z^{2}Z^{3}Z^{4}\text{, }%
Z^{2}Z^{4}\text{, }Z^{3}Z^{4}\text{ }\right\rangle \text{.} \label{SBcase2}%
\end{equation}
The codeword stabilizer matrix $\mathcal{H}_{\mathcal{S}_{\text{CWS}}}$
associated with $\mathcal{S}_{\text{CWS}}$ reads,%
\begin{equation}
\mathcal{H}_{\mathcal{S}_{\text{CWS}}}\overset{\text{def}}{=}\left(
Z\left\vert X\right.  \right)  =\left(
\begin{array}
[c]{cccc}%
0 & 0 & 0 & 0\\
1 & 1 & 1 & 1\\
0 & 1 & 0 & 1\\
0 & 0 & 1 & 1
\end{array}
\left\vert
\begin{array}
[c]{cccc}%
1 & 1 & 1 & 1\\
0 & 0 & 0 & 0\\
0 & 0 & 0 & 0\\
0 & 0 & 0 & 0
\end{array}
\right.  \right)  \text{.}%
\end{equation}
We observe $\det X=0$ and the VdN-work cannot be applied. However, we also
notice that $\mathcal{S}_{\text{CWS}}$ is locally Clifford equivalent to
$\mathcal{S}_{\text{CWS}}^{\prime}$ with $\mathcal{S}_{\text{CWS}}^{\prime
}=U\mathcal{S}_{\text{CWS}}U^{\dagger}$ and $U\overset{\text{def}}{=}%
H^{2}H^{3}H^{4}$. Thus, $\mathcal{S}_{\text{CWS}}^{\prime}$ becomes,%
\begin{equation}
\mathcal{S}_{\text{CWS}}^{\prime}\overset{\text{def}}{=}\left\langle
g_{1}^{\prime}\text{, }g_{2}^{\prime}\text{, }\bar{Z}_{1}^{\prime}\text{,
}\bar{Z}_{2}^{\prime}\text{ }\right\rangle =\left\langle X^{1}Z^{2}Z^{3}%
Z^{4}\text{, }Z^{1}X^{2}X^{3}X^{4}\text{, }X^{2}X^{4}\text{, }X^{3}X^{4}\text{
}\right\rangle \text{.}%
\end{equation}
The codeword stabilizer matrix $\mathcal{H}_{\mathcal{S}_{\text{CWS}}^{\prime
}}$ associated with $\mathcal{S}_{\text{CWS}}^{\prime}$ is given by,%
\begin{equation}
\mathcal{H}_{\mathcal{S}_{\text{CWS}}^{\prime}}\overset{\text{def}}{=}\left(
Z^{\prime}\left\vert X^{\prime}\right.  \right)  =\left(
\begin{array}
[c]{cccc}%
0 & 1 & 1 & 1\\
1 & 0 & 0 & 0\\
0 & 0 & 0 & 0\\
0 & 0 & 0 & 0
\end{array}
\left\vert
\begin{array}
[c]{cccc}%
1 & 0 & 0 & 0\\
0 & 1 & 1 & 1\\
0 & 1 & 0 & 1\\
0 & 0 & 1 & 1
\end{array}
\right.  \right)  \text{.}%
\end{equation}
We now have $\det X^{\prime}\neq0$. Therefore, using the VdN-work, the
$4\times4$ adjacency matrix $\Gamma$ becomes,%
\begin{equation}
\Gamma\overset{\text{def}}{=}\left(
\begin{array}
[c]{cccc}%
0 & 1 & 1 & 1\\
1 & 0 & 0 & 0\\
1 & 0 & 0 & 0\\
1 & 0 & 0 & 0
\end{array}
\right)  \text{.} \label{422gg}%
\end{equation}
Finally, applying the S-work and considering $\Gamma$ in Eq. (\ref{422gg}),
the $6\times6$ symmetric coincidence matrix $\Xi_{\left[  \left[
4,2,2\right]  \right]  }^{\text{Schlingemann}}$ characterizing the graph with
both input and output vertices reads,%
\begin{equation}
\Xi_{\left[  \left[  4,2,2\right]  \right]  }^{\text{Schlingemann}}%
\overset{\text{def}}{=}\left(
\begin{array}
[c]{cccccc}%
0 & 0 & 0 & 1 & 1 & 0\\
0 & 0 & 0 & 0 & 1 & 1\\
0 & 0 & 0 & 1 & 1 & 1\\
1 & 0 & 1 & 0 & 0 & 0\\
1 & 1 & 1 & 0 & 0 & 0\\
0 & 1 & 1 & 0 & 0 & 0
\end{array}
\right)  \text{.}%
\end{equation}
Applying the SW-work, it is simple to verify that any graphical single-error
configuration $\left\{  0\text{, }0^{\prime}\text{, }e\right\}  $ with
$e\in\left\{  1\text{, }2\text{, }3\text{, }4\right\}  $ is detectable.
Therefore, the code detects any single-qubit error. As a final remark, we
emphasize that the graph\textbf{ }associated with\textbf{\ }$\Xi_{\left[
\left[  4,2,2\right]  \right]  }^{\text{Schlingemann}}$\textbf{\ }is identical
to the one appeared in \cite{dirk}.

We stress that the stabilizer generated by the stabilizers in Eq.
(\ref{please}) for the first code can be obtained from the stabilizer
generated by the stabilizers in Eq. (B8) for the second code by applying a
local unitary transformation $U\overset{\text{def}}{=}Q^{1}H^{2}H^{3}Q^{4}$
where $Q\overset{\text{def}}{=}PHP$. However, the codeword stabilizer in Eq.
(\ref{SBcase1}) cannot be obtained from the codeword stabilizer in Eq.
(\ref{SBcase2}) via a local unitary transformation. This feature is consistent
with the fact that graphs associated with adjacency matrices in Eqs.
(\ref{422g}) and (\ref{422gg}) are inequivalent. In other words, these two
matrices characterize graphs that belong to different orbits \cite{hein}.

\begin{figure}[ptb]
\centering
\includegraphics[width=0.35\textwidth]{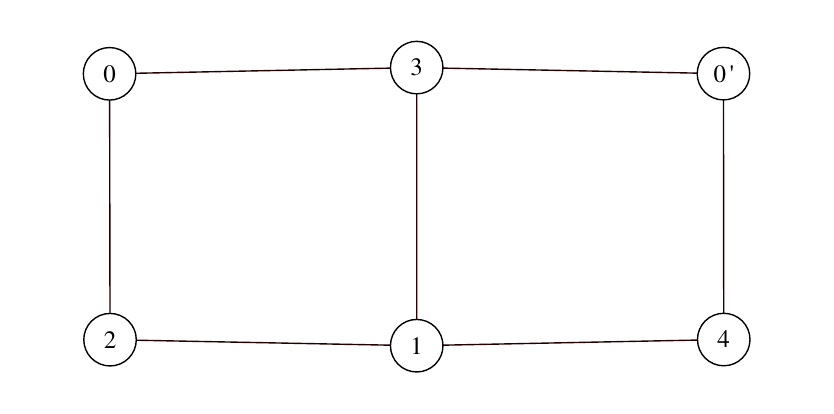}\caption{Graph for a quantum code
that is locally Clifford equivalent to the Schlingemann [[4,2,2]]-code.}%
\label{fig11}%
\end{figure}

\subsection{The $\left[  \left[  8,3,3\right]  \right]  $ stabilizer code}

The $\left[  \left[  8,3,3\right]  \right]  $ code is a special case of a
class of $\left[  \left[  2^{j},2^{j}-j-2,3\right]  \right]  $ codes
\cite{danielpra}. It encodes three logical qubits into eight physical qubits
and corrects all single-qubit errors. The five stabilizer generators are given
by \cite{gaitan},%
\begin{align}
&  g_{1}\overset{\text{def}}{=}X^{1}X^{2}X^{3}X^{4}X^{5}X^{6}X^{7}X^{8}\text{,
}g_{2}\overset{\text{def}}{=}Z^{1}Z^{2}Z^{3}Z^{4}Z^{5}Z^{6}Z^{7}Z^{8}\text{,
}g_{3}\overset{\text{def}}{=}X^{2}X^{4}Y^{5}Z^{6}Y^{7}Z^{8}\text{,
}\nonumber\\
& \nonumber\\
&  g_{4}\overset{\text{def}}{=}X^{2}Z^{3}Y^{4}X^{6}Z^{7}Y^{8}\text{, }%
g_{5}\overset{\text{def}}{=}Y^{2}X^{3}Z^{4}X^{5}Z^{6}Y^{8}\text{,}%
\end{align}
and a suitable choice for the logical operations $\bar{X}_{i}$ and $\bar
{Z}_{i}$ with $i\in\left\{  1\text{, }2\text{, }3\right\}  $ reads,
\begin{equation}
\bar{X}_{1}\overset{\text{def}}{=}X^{1}X^{2}Z^{6}Z^{8}\text{, }\bar{X}%
_{2}\overset{\text{def}}{=}X^{1}X^{3}Z^{4}Z^{7}\text{, }\bar{X}_{3}%
\overset{\text{def}}{=}X^{1}Z^{4}X^{5}Z^{6}\text{, }\bar{Z}_{1}\overset
{\text{def}}{=}Z^{2}Z^{4}Z^{6}Z^{8}\text{, }\bar{Z}_{2}\overset{\text{def}}%
{=}Z^{3}Z^{4}Z^{7}Z^{8}\text{, }\bar{Z}_{3}\overset{\text{def}}{=}Z^{5}%
Z^{6}Z^{7}Z^{8}\text{.}%
\end{equation}
The codeword stabilizer $\mathcal{S}_{\text{CWS}}$ of the CWS code that
realizes this stabilizer code is given by,%
\begin{equation}
\mathcal{S}_{\text{CWS}}\overset{\text{def}}{=}\left\langle g_{1}\text{,
}g_{2}\text{, }g_{3}\text{, }g_{4}\text{, }g_{5}\text{, }\bar{Z}_{1}\text{,
}\bar{Z}_{2}\text{, }\bar{Z}_{3}\right\rangle \text{.}%
\end{equation}
We observe that $\mathcal{S}_{\text{CWS}}$ is locally Clifford equivalent to
$\mathcal{S}_{\text{CWS}}^{\prime}\overset{\text{def}}{=}U\mathcal{S}%
_{\text{CWS}}U^{\dagger}$ with $U\overset{\text{def}}{=}H^{1}H^{2}H^{3}H^{5}$.
Therefore, $\mathcal{S}_{\text{CWS}}^{\prime}$ reads\textbf{,}%
\begin{equation}
\mathcal{S}_{\text{CWS}}^{\prime}\overset{\text{def}}{=}\left\langle
g_{1}^{\prime}\text{, }g_{2}^{\prime}\text{, }g_{3}^{\prime}\text{, }%
g_{4}^{\prime}\text{, }g_{5}^{\prime}\text{, }\bar{Z}_{1}^{\prime}\text{,
}\bar{Z}_{2}^{\prime}\text{, }\bar{Z}_{3}^{\prime}\right\rangle \text{,}%
\end{equation}
with,%
\begin{align}
&  g_{1}^{\prime}\overset{\text{def}}{=}Z^{1}Z^{2}Z^{3}X^{4}Z^{5}X^{6}%
X^{7}X^{8}\text{, }g_{2}^{\prime}\overset{\text{def}}{=}X^{1}X^{2}X^{3}%
Z^{4}X^{5}Z^{6}Z^{7}Z^{8}\text{, }g_{3}^{\prime}\overset{\text{def}}{=}%
Z^{2}X^{4}Y^{5}Z^{6}Y^{7}Z^{8}\text{, }\nonumber\\
& \nonumber\\
&  g_{4}^{\prime}\overset{\text{def}}{=}Z^{2}X^{3}Y^{4}X^{6}Z^{7}Y^{8}\text{,
}g_{5}^{\prime}\overset{\text{def}}{=}Y^{2}Z^{3}Z^{4}Z^{5}Z^{6}Y^{8}\text{,}%
\end{align}
and,%
\begin{equation}
\text{ }\bar{Z}_{1}^{\prime}\overset{\text{def}}{=}X^{2}Z^{4}Z^{6}Z^{8}\text{,
}\bar{Z}_{2}^{\prime}\overset{\text{def}}{=}X^{3}Z^{4}Z^{7}Z^{8}\text{, }%
\bar{Z}_{3}^{\prime}\overset{\text{def}}{=}X^{5}Z^{6}Z^{7}Z^{8}\text{.}%
\end{equation}
The codeword stabilizer matrix $\mathcal{H}_{\mathcal{S}_{\text{CWS}}^{\prime
}}$ associated with $\mathcal{S}_{\text{CWS}}^{\prime}$ is given by,%
\begin{equation}
\mathcal{H}_{\mathcal{S}_{\text{CWS}}^{\prime}}\overset{\text{def}}{=}\left(
Z^{\prime}\left\vert X^{\prime}\right.  \right)  =\left(
\begin{array}
[c]{cccccccc}%
1 & 1 & 1 & 0 & 1 & 0 & 0 & 0\\
0 & 0 & 0 & 1 & 0 & 1 & 1 & 1\\
0 & 1 & 0 & 0 & 1 & 1 & 1 & 1\\
0 & 1 & 0 & 1 & 0 & 0 & 1 & 1\\
0 & 1 & 1 & 1 & 1 & 1 & 0 & 1\\
0 & 0 & 0 & 1 & 0 & 1 & 0 & 1\\
0 & 0 & 0 & 1 & 0 & 0 & 1 & 1\\
0 & 0 & 0 & 0 & 0 & 1 & 1 & 1
\end{array}
\left\vert
\begin{array}
[c]{cccccccc}%
0 & 0 & 0 & 1 & 0 & 1 & 1 & 1\\
1 & 1 & 1 & 0 & 1 & 0 & 0 & 0\\
0 & 0 & 0 & 1 & 1 & 0 & 1 & 0\\
0 & 0 & 1 & 1 & 0 & 1 & 0 & 1\\
0 & 1 & 0 & 0 & 0 & 0 & 0 & 1\\
0 & 1 & 0 & 0 & 0 & 0 & 0 & 0\\
0 & 0 & 1 & 0 & 0 & 0 & 0 & 0\\
0 & 0 & 0 & 0 & 1 & 0 & 0 & 0
\end{array}
\right.  \right)  \text{.}%
\end{equation}

\begin{figure}[ptb]
\centering
\includegraphics[width=0.35\textwidth]{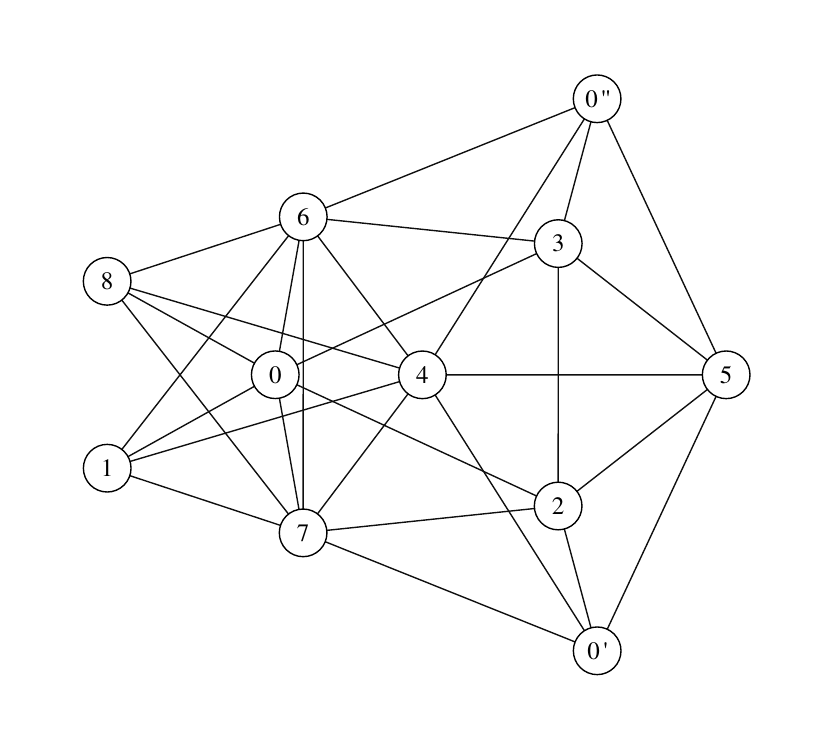}\caption{Graph for a quantum code
that is locally Clifford equivalent to the Gottesman [[8,3,3]]-code.}%
\label{fig12}%
\end{figure}

Since $\det X^{\prime}\neq0$, we can use the VdN-work and the $8\times8$
adjacency matrix $\Gamma$ becomes,%
\begin{equation}
\Gamma\overset{\text{def}}{=}Z^{\prime\text{T}}\cdot\left(  X^{\prime\text{T}%
}\right)  ^{-1}=\left(
\begin{array}
[c]{cccccccc}%
0 & 0 & 0 & 1 & 0 & 1 & 1 & 0\\
0 & 0 & 0 & 1 & 0 & 1 & 0 & 1\\
0 & 0 & 0 & 1 & 0 & 0 & 1 & 1\\
1 & 1 & 1 & 0 & 0 & 0 & 0 & 0\\
0 & 0 & 0 & 0 & 0 & 1 & 1 & 1\\
1 & 1 & 0 & 0 & 1 & 0 & 0 & 0\\
1 & 0 & 1 & 0 & 1 & 0 & 0 & 0\\
0 & 1 & 1 & 0 & 1 & 0 & 0 & 0
\end{array}
\right)  \text{.} \label{ccube}%
\end{equation}
We observe that the graph associated with the adjacency matrix $\Gamma$ (with
$\det\Gamma\neq0$) in Eq. (\ref{ccube})\ is the cube. Acting with a local
complementation with respect to the vertex $1$, $\Gamma$ becomes
$\Gamma^{\prime}$ (with $\det\Gamma^{\prime}=0$).
\begin{equation}
\Gamma^{\prime}\overset{\text{def}}{=}\left(
\begin{array}
[c]{cccccccc}%
0 & 0 & 0 & 1 & 0 & 1 & 1 & 0\\
0 & 0 & 1 & 0 & 1 & 0 & 1 & 0\\
0 & 1 & 0 & 0 & 1 & 1 & 0 & 0\\
1 & 0 & 0 & 0 & 1 & 1 & 1 & 1\\
0 & 1 & 1 & 1 & 0 & 0 & 0 & 0\\
1 & 0 & 1 & 1 & 0 & 0 & 1 & 1\\
1 & 1 & 0 & 1 & 0 & 1 & 0 & 1\\
0 & 0 & 0 & 1 & 0 & 1 & 1 & 0
\end{array}
\right)  \label{833a}%
\end{equation}
Finally, applying the S-work and considering $\Gamma^{\prime}$ in Eq.
(\ref{833a}), the $11\times11$ symmetric coincidence matrix $\Xi_{\left[
\left[  8,3,3\right]  \right]  }$ associated\textbf{\ }with the graph with
both input and output vertices becomes,%
\begin{equation}
\Xi_{\left[  \left[  8,3,3\right]  \right]  }\overset{\text{def}}{=}\left(
\begin{array}
[c]{ccccccccccc}%
0 & 0 & 0 & 1 & 1 & 1 & 0 & 0 & 1 & 1 & 1\\
0 & 0 & 0 & 0 & 1 & 0 & 1 & 1 & 0 & 1 & 0\\
0 & 0 & 0 & 0 & 0 & 1 & 1 & 1 & 1 & 0 & 0\\
1 & 0 & 0 & 0 & 0 & 0 & 1 & 0 & 1 & 1 & 0\\
1 & 1 & 0 & 0 & 0 & 1 & 0 & 1 & 0 & 1 & 0\\
1 & 0 & 1 & 0 & 1 & 0 & 0 & 1 & 1 & 0 & 0\\
0 & 1 & 1 & 1 & 0 & 0 & 0 & 1 & 1 & 1 & 1\\
0 & 1 & 1 & 0 & 1 & 1 & 1 & 0 & 0 & 0 & 0\\
1 & 0 & 1 & 1 & 0 & 1 & 1 & 0 & 0 & 1 & 1\\
1 & 1 & 0 & 1 & 1 & 0 & 1 & 0 & 1 & 0 & 1\\
1 & 0 & 0 & 0 & 0 & 0 & 1 & 0 & 1 & 1 & 0
\end{array}
\right)  \text{.}%
\end{equation}
Using the SW-work, it can be finally verified that any of the $\binom{8}{2}$
graphical\ two-error configuration $\left\{  0\text{, }0^{\prime}\text{,
}0^{\prime\prime}\text{, }e_{1}\text{, }e_{2}\right\}  $ with $e_{1,2}%
\in\left\{  1\text{,..., }8\right\}  $ is detectable. Thus, the code corrects
any single-qubit error.



\end{document}